\documentclass{aastex631}

\usepackage{amsmath}

\submitjournal{ApJ}

\shorttitle{Diversity of low-mass planet atmospheres}
\shortauthors{Bower et al.}
\graphicspath{{./}{figures/}}
\begin{document}

\title{Diversity of Low-mass Planet Atmospheres in the C--H--O--N--S--Cl System with Interior Dissolution, Nonideality, and Condensation: Application to TRAPPIST-1e and Sub-Neptunes}

\correspondingauthor{Dan J. Bower}
\email{dbower@eaps.ethz.ch}

\author[0000-0002-0673-4860]{Dan J. Bower}
\affiliation{Institute of Geochemistry and Petrology \\
Department of Earth and Planetary Sciences \\
ETH Zurich \\
Clausiusstrasse 25 \\
8092 Zurich, Switzerland}

\author[0000-0002-6178-9055]{Maggie A. Thompson}
\affiliation{Institute of Geochemistry and Petrology \\
Department of Earth and Planetary Sciences \\
ETH Zurich \\
Clausiusstrasse 25 \\
8092 Zurich, Switzerland}

\author[0000-0003-4815-2874]{Kaustubh Hakim}
\affiliation{Royal Observatory of Belgium \\
Ringlaan 3 \\
1180 Brussels, Belgium}
\affiliation{KU Leuven \\
Institute of Astronomy \\
Celestijnenlaan 200D bus 2401 \\
3001 Leuven, Belgium}

\author[0000-0002-7384-8577]{Meng Tian}
\affiliation{University Observatory Munich \\
Faculty of Physics \\
Ludwig Maximilian University \\
Scheinerstrasse 1 \\
D-81679 Munich, Germany}

\author[0000-0002-1462-1882]{Paolo A. Sossi}
\affiliation{Institute of Geochemistry and Petrology \\
Department of Earth and Planetary Sciences \\
ETH Zurich \\
Clausiusstrasse 25 \\
8092 Zurich, Switzerland}



% User-defined commands
\newcommand{\atmodeller}{\texttt{Atmodeller}}

%% Mark off the abstract in the ``abstract'' environment. 250 words max.
\begin{abstract}
% NOTE: This is at the maximum of 250 words
A quantitative understanding of the nature and composition of low-mass rocky (exo)planet atmospheres during their evolution is needed to interpret observations. The magma ocean stage of terrestrial- and sub-Neptune planets permits mass exchange between their interiors and atmospheres, during which the mass and speciation of the atmosphere is dictated by the planet's volatile budget, chemical equilibria, and gas/fluid solubility in molten rock. As the atmosphere cools, it is modified by gas-phase reactions and condensation. We combine these processes into an open-source Python package built using JAX called \atmodeller{}, and perform calculations for planet sizes and conditions analogous to TRAPPIST-1e and K2-18b. For TRAPPIST-1e-like planets, our simulations indicate that CO-dominated atmospheres are prevalent during the magma ocean stage, which, upon isochemical cooling, predominantly evolve into CO$_2$-rich atmospheres of a few hundred bar at 280 K. Around 40\% of our simulations predict the coexistence of liquid water, graphite, $\alpha$-sulfur, and ammonium chloride---key ingredients for surface habitability. For sub-Neptune gas dwarfs, pressures are sufficiently high ($\sim$GPa) that gas fugacities deviate from ideality, thereby drastically enhancing solubilities. This buffers the total atmospheric pressure to lower values than for the ideal case. These effects conspire to produce CH$_4$-rich sub-Neptune atmospheres for total pressures exceeding $\sim$3.5~GPa, provided H/C is $\sim$100$\times$ solar and $f$O$_2$ moderately reducing (3 log$_{10}$ units below the iron-w\"ustite buffer). Otherwise, molecular hydrogen remains the predominant species at lower total pressures and/or higher H/C. For all planets at high temperature, solubility enriches C/H in the atmosphere relative to the initial composition.
\end{abstract}




\section{Introduction}
\label{section:intro}

Whilst the discovery of exoplanets has proliferated in recent years, a significant challenge lies in unraveling how their bulk interior compositions relate to atmospheric speciation. Based on current demographics studies, the most abundant kinds of planets are those with masses in between those of Earth and Neptune and are divided into "super-Earths," which are rocky planets with radii less than $\sim$1.5 $R_{\oplus}$, and "sub-Neptunes,"' which are planets with radii larger than $\sim$2 $R_{\oplus}$ that likely have H$_2$-rich atmospheres \citep{Bean2021}. Characterizing planets with possible molten surfaces (magma oceans)---such as close-in rocky and super-Earth planets in addition to sub-Neptunes with thick hydrogen (H$_2$) envelopes---demands a comprehensive treatment of thermodynamics at the melt-atmosphere interface \citep{CDL21,charnoz2023,misener2023,rigby2024,seo2024,WDS25} because the mass of the melt reservoirs always predominates over that contained in the atmosphere, such that the physicochemical state of the liquid (i.e., the chemical potentials of components set by pressure, temperature, and bulk composition) controls the near-surface atmospheric speciation, thereby setting a lower boundary condition for the atmosphere \citep{hirschmann2012,SBB20}. Above this boundary, the atmospheric speciation may be modified by dynamics, chemistry, photochemistry, and atmospheric escape \citep{CatlingKasting2017, wordsworth2022, selsis2023cool, krissansen2024erosion, owen2024, rogers2024fleeting, cherubim2025oxidation, nicholls2025convective}. Ultimately, the spectral signal in the uppermost atmosphere is most accessible for interrogation by space-based observatories such as the James Webb Space Telescope (JWST). Hence, adequately modellng physical and chemical processes at the melt-atmosphere interface is foundational to interpreting current and upcoming observations.

During planet formation, volatiles delivered in planetary building blocks \cite[e.g.,][]{alexander2012provenances,marty2012origins} and by capture of the nebular gas \cite[e.g.,][]{sharp2022multi,young2023earth} likely equilibrated with the growing planet's rocky magma ocean, whose capacity to store volatile species varies according to their solubilities. 
A solubility law for any given volatile species relates its dissolved concentration in magma to its fugacity, which is, in turn, related to partial pressure through a fugacity coefficient. These laws are determined either experimentally or theoretically via ab initio simulations for a set of conditions, including total pressure, temperature, oxygen fugacity, and melt composition, that affect the stable melt and/or gas species. Solubility studies have focused on deriving relations for major atmosphere-forming species (e.g., H$_2$O, H$_2$, CO$_2$, CO, CH$_4$, N$_2$, SO$_2$, Cl$_2$, and the noble gases) in silicate melts \citep[e.g.,][]{DSH95, NBB17, STB23, HWA12, AHW13, LMH03, BW22, BGF21, OM22, TW21, daviscaracas2024, JWB86}, and were largely conducted at moderate to high-pressure conditions (hundreds of bar to several GPa) since they were originally devised to cover the conditions of Earth's mantle and crust. These conditions, however, are also comparable to the melt-atmosphere interface of large rocky planets and hence these solubility relations are now finding new applications in exoplanet research.

For interface pressures greater than $\sim$1~GPa, gases can no longer be reasonably treated as obeying the ideal gas law, depending on temperature and the nature of the gas species \citep{bridgman1964, DZ06}. Fugacity coefficients quantify how strongly a gas departs from ideal behavior and are often expressed in terms of temperature ($T$), pressure ($P$), and composition ($X$) \citep[e.g.,][]{RK49, TH24}. Deviations from unity pertain physically to differences in volume at a given pressure and temperature relative to the ideal gas law, which are products of (1) the finite size of gas particles and (2) attractive/repulsive interactions between them \citep{vanderwaals1873} and are themselves a function of $T$, $P$ and $X$. They can be derived from real gas equations of state (EOS) \citep[e.g.,][]{SF88, HP91}, thermodynamic tables and charts \citep[e.g.,][]{JANAF}, or directly from experimental data and measurements \citep[e.g.,][]{P69, jakobsson1994system, FW97}. Nonideality is of direct relevance to interpreting the formation and evolution of rocky planets, which are mostly rock by mass and gas by volume. Both gas-phase reactions and gas solubilities are regulated by fugacities, which are only equivalent to partial pressures in the ideal limit. Meanwhile, pressure (not fugacity) is the operative quantity in satisfying volatile mass balance \citep[e.g., Appendix C,][]{BHS21}. In essence, nonideality demands that "thermodynamic pressure," relevant to reactions and solubility, and "mechanical pressure," relevant to mass balance, are treated as distinct entities. This may lead to exoplanet demographics that are notably different from those derived by assuming ideal gas behavior and chemically inert rocky cores \citep[e.g.,][]{KFS19}.

For planets that subsequently cool and crystallize their mantles, such as the solar system bodies, volatiles exsolve to form an early outgassed atmosphere that sets the initial state for the subsequent long-term evolution of the atmosphere \citep[e.g.,][]{BKW19, BHS21, NKT19, KOG20, LBH21, GBF21, liggins2022, krissansen2024erosion, maurice2024volatile}. The gaseous species in these atmospheres, in addition to any condensed phases that form upon cooling, may be the earliest ingredients necessary for prebiotic chemical reactions to take place \citep{SBB20,zahnle2020, benner2020,preiner2021life,johansen2024,korenaga2025tectonics}. However, no systematic study of how volcanically degassed atmospheres cool and condense under a range of conditions distinct from that of the Earth has been undertaken. Therefore, whether the raw chemical ingredients for the origin of life were endemic to Earth, and to what extent they occur in other planetary systems, remains unanswered. Moreover, super-Earths and sub-Neptunes may sustain a molten interior, either due to close proximity to their host star or an insulating envelope, and hence volatiles could remain sequestered in the molten interior, giving rise to a wide variety of atmospheres. Hence, quantifying both the volatile budget of rocky worlds that potentially become habitable (e.g., Earth and Earth twins), as well as the interior--atmosphere exchange occurring on super-Earths and sub-Neptunes, requires compositionally consistent and accurate volatile solubility relations applicable to the unique conditions at the interior--atmosphere interface.

We have therefore devised a versatile and modular toolkit (\atmodeller{}) that the community can utilize and extend to compute volatile partitioning between planetary reservoirs with solubility, nonideality, and condensation. In particular, this will enable new theoretically and experimentally derived solubility laws and fugacity coefficients to be efficiently and accurately incorporated into models devised by (exo)planetary scientists. Ultimately, \atmodeller{} provides a transparent and robust framework for planetary outgassing calculations. After presenting the theory and method on which \atmodeller{} is based (Section~\ref{sect:method}), we then showcase its features by calculating the earliest atmosphere of TRAPPIST-1e (Section~\ref{sect:trappist1e}) as well as the atmosphere of a gas dwarf sub-Neptune (Section~\ref{sect:gasdwarf}).

\section{Method}
\label{sect:method}
\subsection{Formulation and numerical solution}
We focus on providing an overview of the specific design considerations of \atmodeller{}, noting that a review of the general approaches to formulate and solve chemical equilibrium calculations are presented in \cite{LKS17}. Gas and (optionally) condensed species are specified by the user and are used to determine an independent reaction set via Gaussian elimination \citep[see Appendix A,][]{LKK16}. \atmodeller{} then applies the extended law of mass-action (xLMA) equations \citep{LKK16} to create a closed set of nonlinear equations that combines equilibrium chemistry with elemental mass balance, cast as a root-finding problem. The xLMA has an intuitive formulation using equilibrium constants and allows for the simultaneous solution of all (user-specified) species in the system without a priori knowledge of the species' thermodynamic stability, which is particularly important to facilitate solutions with condensates \citep[see also][]{KSP24}. Solubility laws and real gas EOS provide optional closure conditions, where, if desired, solubility can be neglected and the system assumed ideal. In this regard, \atmodeller{} can also model planetary atmospheres above chemically inert mantles or cool solidified surfaces; there is no inherent requirement for magma to be present nor the gas and condensed matter to chemically interact.

If included, solubility relations ensure chemical equilibrium between a species' fugacity and its dissolved abundance in melt (Section~\ref{section:solubility}), whereas real gas EOS relate a species' partial pressure to its fugacity (Section~\ref{sect:eos}). The mass balance is formulated by considering elemental abundances in condensates, in the molten mantle, and in the atmosphere \citep[e.g., Appendix A,][]{BKW19}, which requires the specification of some planetary properties (e.g., surface temperature, mantle mass, melt fraction, etc.). The mantle mass (and hence the gravitational acceleration on the planet) remains constant, irrespective of the mass of volatiles dissolved in it. However, insofar as the masses of the combined volatile elements added never exceed 3\%--4\% of the total planetary mass, even in the case of the most volatile-rich sub-Neptunes (Section \ref{sect:gasdwarf}), this approximation introduces minimal error. Hence, in this formulation, the mass of volatiles in the atmosphere directly determines the magma-atmosphere surface pressure, and therefore is inextricably tied to the partitioning of gaseous species between the atmosphere and molten mantle. \atmodeller{} allows the specification of mixed constraints, such that fugacity constraints can be combined with elemental mass constraints as long as the system of equations is closed. This facilitates a common, desirable scenario in which oxygen fugacity is defined, in absolute terms as $f$O$_2$, or relative to a mineralogical buffer (e.g., the iron-w\"{u}stite buffer; IW) while other elements have fixed abundance.

\atmodeller{} is written in the Python programming language for seamless integration with existing modeling frameworks and leverages JAX \citep{jax2018github} for high-performance computation. It consists of three subpackages that contain JAX-compliant solubility laws, real gas EOS, and thermodynamic data, all of which can be accessed independently and used as components of Python-based modeling frameworks if desired (with or without JAX support). Notably, the automatic differentiation feature of JAX is convenient for its ability to propagate derivative information through solubility laws and real gas EOS, which often present complex functional forms that may themselves require iterative solution during the overall solution procedure. Another feature, automatic vectorization, enables \atmodeller{} to easily spawn batch calculations to efficiently explore parameter space in a single vectorized solve. Finally, just-in-time compilation enables a model to be pre-compiled, which facilitates fast \atmodeller{} calculations as part of a repeating workflow such as a time integration in evolution models. \atmodeller{} uses Optimistix \citep{optimistix2024} to solve the root-finding problem and additionally leverages Lineax for linear solves \citep{lineax2023} and Equinox \citep{equinox}. As with all systems of nonlinear equations, obtaining a solution requires that applied constraints (fugacity, mass) can be satisfied simultaneously with the given system of equations (mass balance, equilibrium chemistry) and closure conditions (solubility laws, free energies of individual gas species, real gas EOS). We independently verified the results of \atmodeller{} by comparison to FactSage 8.2 \citep{bale2016} calculations assuming ideal gas behavior (Appendix~\ref{app:factsage}), and the Python package has additional comparisons with condensates as well as FastChem 3.1.1 \citep{KSP24}. Furthermore, each solubility law and real gas EOS implemented in \atmodeller{} includes an associated test that verifies its output against tabulated data or published figures from the original source, ensuring the accuracy and reproducibility of the model components.

\subsection{Solubility laws}
\label{section:solubility}

\begin{deluxetable}{lllrrl}
\tablecaption{Some solubility laws included in \atmodeller{}. Each solubility law relates the dissolved concentration (ppm by weight) of a given volatile in the melt to its fugacity in the gas phase. Solubility laws can depend on the total pressure, temperature, and oxygen fugacity ($f$O$_2$). \label{table:solubilitylaws}} 
\tablewidth{0pt}
\tablehead{\multicolumn3c{} & \multicolumn3c{Experimental Calibration$^\dagger$}\\
\colhead{Species} & \colhead{Composition$^*$} & \colhead{Reference} & \colhead{Pressure} & \colhead{Temperature} & \colhead{$f$O$_2$ rel. IW}\\
\multicolumn3c{} & \colhead{(kbar)} & \colhead{(K)} & \colhead{(log$_{10}$ units)}}
\startdata
H$_2$ & Basalt, Andesite (Fe-free) & \citet[their Table~2]{HWA12} & 7--30 & 1673--1773 & -1 to 3.8 \\ %  % \maggie{Check the other buffers NNO (IW+3.8), MMC, and IWC for IWC (IW-1), MMC (IW-0.45) checked with online factsage} \\
H$_2$ & Silicate glass (obsidian) & \citet[their Table~4]{GSM03} & 0.001--0.265 & 1073 & -0.36 to 0.6 \\
H$_2$O & Basalt &  \citet[their Fig.~4]{DSH95} & 0.2--0.72 & 1473 & 4.2 to 5.5 \\ 
H$_2$O & Peridotite & \citet{STB23} & 0.001 & 2173 & -1.9 to 6.0 \\
H$_2$O & Lunar basalt, An-Di & \citet[their Fig.~5]{NBB17} & 0.001 & 1623 & -3.0 to 4.8 \\
H$_2$O & Basalt & \citet[their Fig.~8]{MGO17} & 10 & 1473 & 4.2 \\ 
He & Basalt (tholeiitic) & \citet{JWB86} & 0.001 & 1523--1873 & 7.3 to 10.7 \\ % (fO2 is given by the range for air between 1523-1873 K) \\
Ne & Basalt (tholeiitic) & \cite{JWB86} & 0.001 & 1523--1873 & 7.3 to 10.7 \\
Ar & Basalt (tholeiitic) & \cite{JWB86} & 0.001 & 1523--1873 & 7.3 to 10.7 \\
Kr & Basalt (tholeiitic) & \cite{JWB86} & 0.001 & 1523--1873 & 7.3 to 10.7 \\
Xe & Basalt (tholeiitic) & \cite{JWB86} & 0.001 & 1523--1873 & 7.3 to 10.7 \\
CO & Basalt, Rhyolite &  \citet{YNN19} & 2--30 & 1473--1773 & 1 to 3.8 \\
CO & Basalt & \citet[their Eq.~10]{AHS15} & 10--12 & 1673 & -3.65 to 1.46 \\
CO$_2$ & Basalt & \citet[their Eq.~6]{DSH95} & 0.21--0.98 & 1473 & 4.2 to 5.5 \\
CH$_4$ & Basalt (Fe-free) & \citet[their Eq. 7a and 8]{AHW13} & 7--30 & 1673--1723 & -9.50 to -1.36 \\
N$_2$ & Basalt (tholeiitic) & \citet[their Eq.~23]{LMH03} & 0.001 & 1673--1698 & -8.3 to 8.7 \\
N$_2$ & Basalt & \citet{DFP22} & 0.001--82 & 1323--2600 & -8.3 to 8.7 \\ % Dan - I just took the most extreme bounds including the other studies
N$_2$ & Basalt & \citet{BGF21} & 0.8--10 & 1473--1573 & -4.7 to 4.9 \\
{S$_2$}$^\ddagger$ & Basalt, Andesite & \citet{BW22, BW23} & 0.001 & 1473--1773 & -0.14 to 10.9 \\ % Dan combined the ranges of sulfide and sulfate
%S$_2$ & Mafic silicate melts & \citet[Eq.~10]{NCH16} & 0.001-40 & 1473-2023 & -9.4 to -1.5 \\
Cl$_2$ & Basalt, An-Di-Fo & \citet[their Fig.~4]{TW21} & 15 & 1673 & 2.03 \\ % (CCO) Note the full range of experiments was P: 5-20 kbar; T: 1473-1773 K; fO2: CCO buffer (equation 4 gives expression for CCO buffer)\\
\enddata
\tablecomments{$^*$Anorthite (An), Diopside (Di), Forsterite (Fo). $\dagger$: Range of experimental conditions used to determine the solubility laws. The oxygen fugacity, $f$O$_2$ is given in log$_{10}$ units relative to the IW buffer. $\ddagger$: Combines the solubility of sulfur as sulfide (S$^{2-}$) and sulfate (SO$_4^{2-}$/S$^{6+}$).}
\end{deluxetable}

The atmospheric compositions of planets with molten interiors are modulated by both the volatile endowment of the planet and the solubility of these volatiles in the melt. The solubility of a gas species depends on its fugacity (or partial pressure under ideal gas conditions) in the gas phase, the composition of the melt, and sometimes additional parameters, such as the oxygen fugacity ($f$O$_2$), temperature, and total pressure. Solubility laws for major atmosphere-forming species are determined from experiments that, so far, have largely been focused on Earth-like melt compositions and conducted under $P$, $T$ conditions motivated by Earth's mantle. These empirically determined laws often take the form of power-law or linear fits to the experimental data to relate the dissolved concentration of a species in the melt to its fugacity in the gas phase. Importantly, solubility experiments are performed for certain melt compositions under specific temperature and pressure conditions, meaning that extrapolating these laws beyond their calibrated conditions can be erroneous. The composition of the gas phase also affects the solubility by expanding or shrinking the stability of both gas and melt species in the system. For example, the dissolution of N may occur as NH$_2^-$, NH$_3$, CN$^-$, N$^{3-}$ and N$_2$ moieties in silicate melts depending on the fugacities of N-, C- and H-bearing species at a given $P$, $T$ \citep{dalou2019,grewal2020}, such that solubilities are not independent of atmospheric composition.  \atmodeller{} includes a library of solubility laws from the literature for H$_2$, H$_2$O, CO, CO$_2$, CH$_4$, N$_2$, S$_2$, Cl$_2$ and the noble gases He, Ne, Ar, Kr, Xe for various melt compositions and chemical systems (Table~\ref{table:solubilitylaws}).  We note the need for more solubility experiments of major volatiles for diverse melt compositions beyond those of the modern Earth, and under a wide range of $P$, $T$ conditions as well as gas-phase compositions.

\subsection{Real gas equations of state}
\label{sect:eos}
\atmodeller{} includes a library of real gas EOS (Table~\ref{table:realgas}), which yield expressions for the volume of any given species as a function of pressure and temperature. With this information, it is possible to quantify the relationship between a species' fugacity and its partial pressure, expressed as the fugacity coefficient $\phi=f/P$, where $f$ is fugacity and $P$ is pressure:
\begin{equation}
\ln \phi = \frac{1}{RT} \int_{P^0}^{P} (V_\text{m} - V_\text{m}^\text{ideal})\ dP = \int_{P^0}^{P} \left( \frac{Z-1}{P} \right)\ dP,
\label{eq:fugacity}
\end{equation}
where $R$ is the gas constant, $T$ temperature, $V_\text{m}$ molar volume, $V_\text{m}^\text{ideal}=RT/P$, and the compressibility factor $Z=V_\text{m}/V_\text{m}^\text{ideal}$. The lower bound of the integration $P^0$ is the standard state pressure (usually 1 bar or 1 atmosphere, depending on the thermodynamic data source), and conventionally $P^0$ defines an ideal state, i.e. $\phi(P=P^0)=1$. Typically, experimental data or numerical calculations determine $V_\text{m}$ or $Z$, which are then fit to an empirical or physically motivated function of temperature and pressure, such as the compensated Redlich-Kwong EOS \citep[CORK,][]{HP91} or the virial EOS \citep{SS92}. Because the behavior of gases under high $P--T$ is related to their critical temperature and pressure, often, the pressure and temperature dependence of volume is expressed relative to these critical quantities ($T_c$, $P_c$), i.e., the law of corresponding states \citep{vanderwaals1873}. This enables regression of experimental data across similarly behaving species and thereby improves the average fit. For known pressure and temperature, fugacity coefficients and fugacities can then be computed by evaluating the integral in Equation~\ref{eq:fugacity}. Some functional forms have convenient closed-form solutions of the fugacity, whereas others require numerical solution \citep[see][for a discussion]{HP91}. It is also beneficial for numerical stability to ensure the EOS are bounded and smooth, so we extrapolate the compressibility factor, $Z$, linearly according to a virial-type expression (Appendix~\ref{app:boundedeos}).

% For Saxena and Fei (1988), ranges from Fig 1 and Table 1.
% For Holland and Powell (1991), ranges from Fig. 4
% For Holland and Powell (1998) the extended pressure range of CO2 and H2O is given in the text
% For Holland and Powell (2011) the corresponding states model is the same as HP91
\begin{deluxetable}{llllrr}
\tablecaption{Some real gas EOS included in \atmodeller{}. \label{table:realgas}}
\tablewidth{0pt}
\tablehead{\multicolumn4c{} & \multicolumn2c{Range$^\dagger$}\\
\colhead{Species} & \colhead{EOS$^*$} & \colhead{Reference} & \colhead{Note} & \colhead{Pressure} & \colhead{Temperature}\\
\multicolumn4c{} & \colhead{(kbar)} & \colhead{(K)}}
\startdata
Ar & Virial & \cite{SF87} & Corresponding states & 0-1000 & 200-3000 \\
CH$_4$ & CORK & \cite{HP91} & Corresponding states & 0-50 & 400-1900 \\
CH$_4$ & Virial & \cite{SS92} & Corresponding states & 0-1000 & 200-3000 \\
CH$_4$ & B-B & \cite{HWZ58} & & 0-1 & 200-500 \\
CO & CORK & \cite{HP91} & Corresponding states & 0-50 & 400-1900 \\
CO & Virial & \cite{SS92} & Corresponding states & 0-1000 & 200-3000 \\
CO$_2$ & CORK & \cite{HP91} & Corresponding states & 0-50 & 400-1900 \\ 
CO$_2$ & CORK & \cite{HP91} & & 0-50 & 400-1900 \\
CO$_2$ & CORK & \cite{HP98} & Updated virial coefficients & 0-120 & 400-1900  \\
CO$_2$ & Virial & \cite{SS92} & Corresponding states & 0-1000 & 200-3000 \\
CO$_2$ & B-B & \cite{HWZ58} & & 0-0.1 & 300-1000 \\
COS & Virial & \cite{SS92} & Corresponding states & 0-1000 & 200-3000 \\
H$_2$ & CORK & \cite{HP91} & Corresponding states & 0-50 & 400-1900 \\
H$_2$ & Virial & \cite{SF87} & Corresponding states & 0-1000 & 200-3000 \\
H$_2$ & Virial & \cite{SS92} & Corresponding states & 0-1000 & 200-3000 \\
H$_2$ & B-B & \cite{HWZ58} & & 0-1 & 20-600 \\
H$_2$ & ab initio & \cite{CD21} & Also mixtures with He & 0-10$^{14}$ & 100-$10^8$ \\
H$_2$O & CORK & \cite{HP91} & & 0-50 & 400-1700 \\ 
H$_2$O & CORK & \cite{HP98} & Updated virial coefficients & 0-120 & 400-1700 \\
H$_2$S & CORK & \cite{HP11} & Corresponding states & 0-50 & 400-1900 \\
H$_2$S & Virial & \cite{SS92} & & 0-10 & 400-1100 \\ % Fig. 3
He & B-B & \cite{HWZ58} & & 0-1 & 200-400 \\
He & ab initio & \cite{CD21} & Also mixtures with H$_2$ & 0-10$^{14}$ & 100-$10^8$ \\
N$_2$ & CORK & \cite{HP91} & Corresponding states & 0-50 & 400-1900 \\
N$_2$ & Virial & \cite{SF87} & Corresponding states & 0-1000 & 200-3000 \\
N$_2$ & B-B & \cite{HWZ58} & & 0-0.1 & 100-1000 \\
NH$_3$ & B-B & \cite{HWZ58} & & 0-0.1 & 300-500 \\
O$_2$ & Virial & \cite{SS92} & Corresponding states & 0-1000 & 200-3000 \\
O$_2$ & B-B & \cite{HWZ58} & & 0-0.1 & 100-1000 \\
S$_2$ & CORK & \cite{HP11} & Corresponding states & 0-50 & 400-1900 \\
S$_2$ & Virial & \cite{SS92} & Corresponding states & 0-1000 & 200-3000 \\
SO$_2$ & Virial & \cite{SS92} & & 0.5-10 & 500-1100 \\ % Fig. 2
\enddata
\tablecomments{$^*$Compensated-Redlich Kwong (CORK), Beattie-Bridgeman (B-B). $\dagger$: Approximate range of thermodynamic conditions, accessed by experiments or ab initio calculations, used to inform the real gas EOS.}
\end{deluxetable}

\subsection{Condensation}
\label{sect:condensation}
The condition for condensation is met when the addition of a condensed phase, liquid or solid, to the system of gas species satisfies the xLMA equations. We recall that by construction the xLMA accommodates both stable and unstable condensates in the same framework through the specification of stability variables, which are independent for each condensed species and solved self-consistently as part of the solution process. For implementation details, the reader is referred to \cite{LKK16} and \cite{KSP24}. For numerical stability, it is necessary to define a parameter $\tau$ that controls the minimum amount that a stable species can achieve. Following \cite{LKK16}, we adopt $\tau=10^{-25} \mathrm{min}(b)$, where $b$ is the minimum amount of an element in the system that composes a given species. In principle, stability variables could also be applied to gas species, although since gas-phase chemistry is notably sensitive to $f$O$_2$---even at very low abundance---introducing a minimum cut-off for the amounts of gas species could distort results.

In \atmodeller{}, only pure condensed phases are treated, and, as such, the phase is stable when its activity, $a=1$. The activity is defined as the ratio of the fugacity of any component at $P$, $T$, relative to that of the pure phase at standard state conditions; $T$ and $P^0=f^0=1$ bar:
\begin{equation}
    a = f/f^0,
    \label{eq:activity}
\end{equation}
where, by definition, $a=1$ for a pure phase. We do not account for the compressibilities of condensed phases (i.e., the Poynting correction is negligible). The stable condensates included in the base package are H$_2$O(l), C(cr), $\alpha$-S(cr), and ClH$_4$N(cr,l)---ammonium chloride, because these are of primary importance for assessing planetary habitability. Thermodynamic data for condensates as well as gas species are sourced from \cite{MZG02} and \cite{JANAF}.

\section{Results}
\subsection{TRAPPIST-1e and similar Earth-sized planets}
\label{sect:trappist1e}
\subsubsection{Context and parameters}
\label{sect:T1e_context}
There is significant interest in probing the atmospheres of temperate Earth-sized rocky planets, such as the TRAPPIST-1 planets \citep{GTD17}, \added{whose atmospheres are now being probed by JWST \citep{zieba2023,ducrot2025,piaulet2025}}, to understand the nature of potentially habitable environments beyond Earth and the solar system. The earliest outgassed atmospheres form during and immediately following planet assembly, whereby one or several epochs of magma ocean formation is thought to have distributed volatile elements readily between the atmosphere, mantle, and, potentially the metallic core \citep[e.g.,][]{hirschmann2012,LMC13,SBB20,chen2022impact,gu2024,dasgupta2024, huang2024nitrogen}. This establishes the initial budget of volatiles in the atmosphere and interior, which may subsequently participate in geological volatile cycling that further modulates their atmospheric abundances \citep[e.g.,][]{sleep2001,KAC2018,2021PSJ.....2...49H}. Here, we apply \atmodeller{} to a young TRAPPIST-1e analog as an example of a small rocky planet during its magma ocean phase following iron-nickel-alloy core formation, although the results are applicable broadly for any Earth-sized rocky planet. These results are used to inform subsequent \atmodeller{} calculations to determine the postformation atmosphere and surface abundances of TRAPPIST-1e at its present-day equilibrium temperature (Section~\ref{sect:trappist1e_cool}).

TRAPPIST-1e has a total mass of 4.13$\times 10^{24}$ kg, a mantle mass of 2.91$\times 10^{24}$ kg (assuming the same core-mantle mass proportion as Earth), and surface radius of 5861 km (0.9 R$_{\oplus}$) \citep{ADG21}. To compute the properties of the high-temperature atmosphere, we assume a surface temperature of 1800 K, \added{which lies above the liquidus for an Earth-like mantle and is compatible with either a fully or partially molten magma ocean and a melt--gas interface at thermodynamic equilibrium, a state likely to have been approximated if not achieved \citep[see][]{salvador2023,walbecq2025}.} We consider 14 gas species: \added{H$_2$O, H$_2$, O$_2$, CO, CO$_2$, CH$_4$, N$_2$, NH$_3$, S$_2$, H$_2$S, SO$_2$, SO, Cl$_2$, and HCl.} This encompasses the majority of the stable, gaseous molecules of the so-called CHONS group of elements, in addition to Cl, that constitute the majority of living matter. \added{Species of moderately volatile elements such as Na, K and P are excluded because they are likely to be present in far lower abundances than are the major volatile species \citep[see, for example][]{fegley2016,sossifegley2018}}.

Of the gas species listed above, the following solubility laws are used to distribute them between the atmosphere and the magma ocean: H$_2$O \citep{DSH95}, H$_2$ \citep{HWA12}, CO \citep{YNN19}, CO$_2$ \citep{DSH95}, CH$_4$ \citep{AHW13}, N$_2$ \citep{LMH03}, S-bearing species \citep{BW22, BW23}, Cl-bearing species \citep{TW21}. Both the EOS of pure species and the mixing between them are assumed to be ideal. Condensed carbon (graphite) is also allowed to exist depending on thermodynamic stability, since this high-temperature condensate, if present, can buffer the fugacities of carbon-bearing species at the most reducing conditions.

We do not explicitly model the metallic core as a reservoir for volatiles, but rather calibrate the initial elemental abundances relative to the bulk silicate Earth (BSE), which is best constrained from natural samples \citep[e.g.,][]{palmeoneill2014}. This allows us to model the exchange of volatiles between the silicate mantle and atmosphere without necessitating comparatively uncertain high $P$, $T$ partitioning data that governs equilibrium between silicate and metallic phases. Nevertheless, our approach does not mandate that the metallic core is volatile free, but rather that it is isolated during the period of atmosphere-mantle exchange that we consider. Alternatively, the core could be thought of as volatile free (relatively, compared to the mantle), if the prescribed abundances are to be correlated 1:1 with the bulk composition.

Therefore, we fix the total abundances of N, S, and Cl based on BSE estimates, scaled to the mass of TRAPPIST-1e. The average BSE estimates for N, S, and Cl are 7.01$\times 10^{18}$ kg, 9.47$\times 10^{20}$ kg, and 1.09$\times 10^{20}$ kg, respectively \citep{SKG21, H16, KHK17}, and, when scaled to the mantle mass of TRAPPIST-1e, they are 4.85$\times 10^{18}$ kg, 6.55$\times 10^{20}$ kg, and 7.57$\times 10^{19}$ kg, respectively. In a Monte Carlo simulation with 10,000 realizations we uniformly sample, in log$_{10}$ space, the total amount of hydrogen between 0.1 and 10 Earth oceans \added{(one present-day Earth ocean contains $1.55 \times 10^{20}$ kg of hydrogen)}, C/H (by mass) between 0.1 and 10, and the oxygen fugacity by 5 log$_{10}$ units either side of the IW buffer (Appendix~\ref{app:oxygen}). \added{The rationale for this sampling strategy and the choice of parameter ranges is detailed in Appendix~\ref{app:sampling}.} All simulation data are available to download \citep{atmodeller_data}.

\subsubsection{Volatile partitioning during the magma ocean stage}
\label{sect:early}
\begin{figure}
\includegraphics[width=1\textwidth]{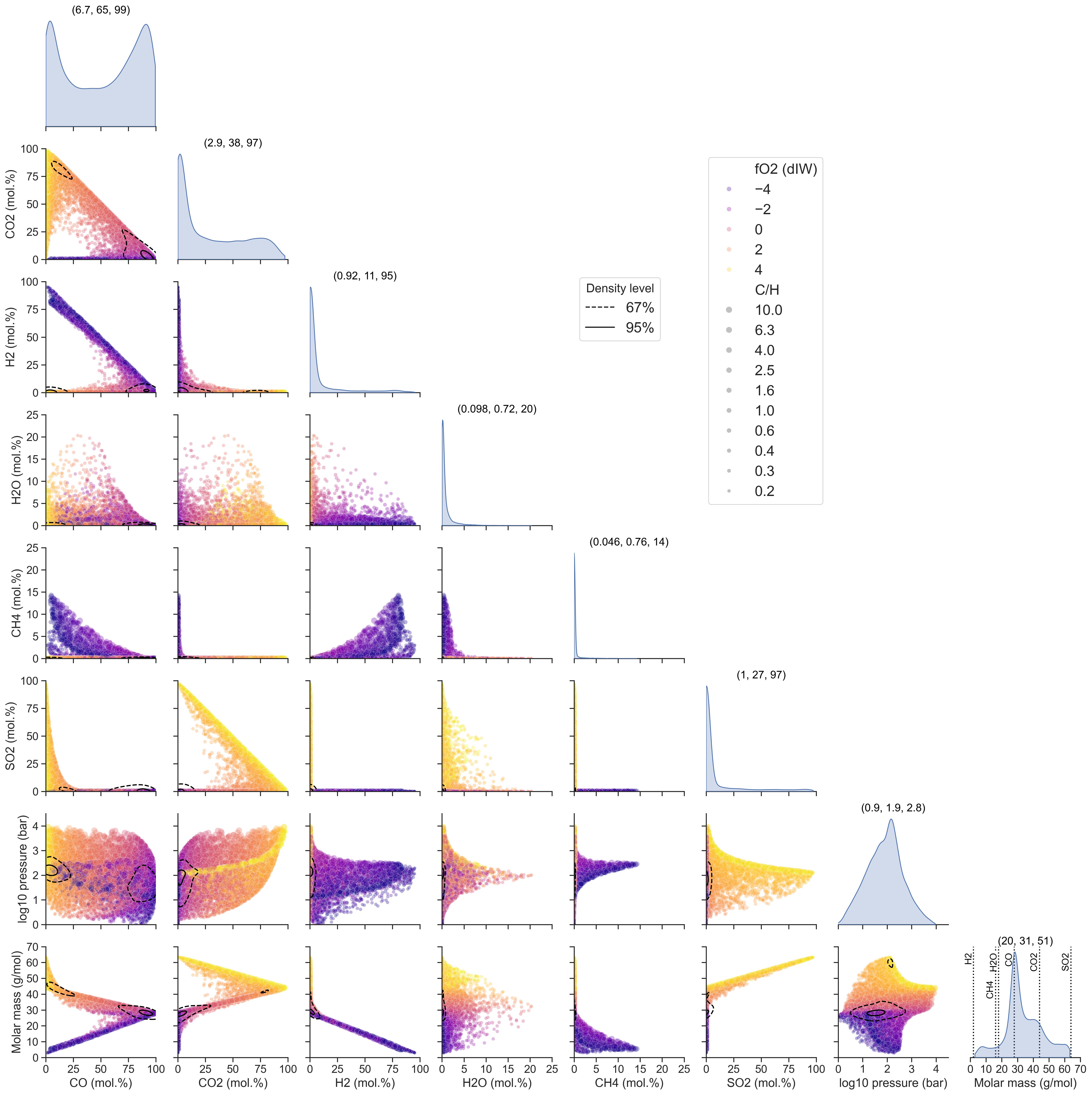}
\caption{For an early TRAPPIST-1e with a fully molten mantle (surface temperature of 1800 K), scatter plots of atmospheric molar abundance, total pressure, and molar mass for major species. Points are colored by oxygen fugacity expressed relative to the IW buffer and sized in proportion to log$_{10}$ C/H. Density levels indicate areas with high likelihood in the scatter plots, while marginal distributions are shown on the diagonal. The 10th, 50th (median), and 90th percentiles of these distributions are also annotated above the marginal distributions as (p10, p50, and p90), respectively. Compare to a partially molten mantle in Figure~\ref{fig:T1e_partially_molten}, \added{noting that the extent of the axes for H$_2$O and CH$_4$ is different.}}
\label{fig:T1e_fully_molten}
\end{figure}

\begin{figure}
\includegraphics[width=1\textwidth]{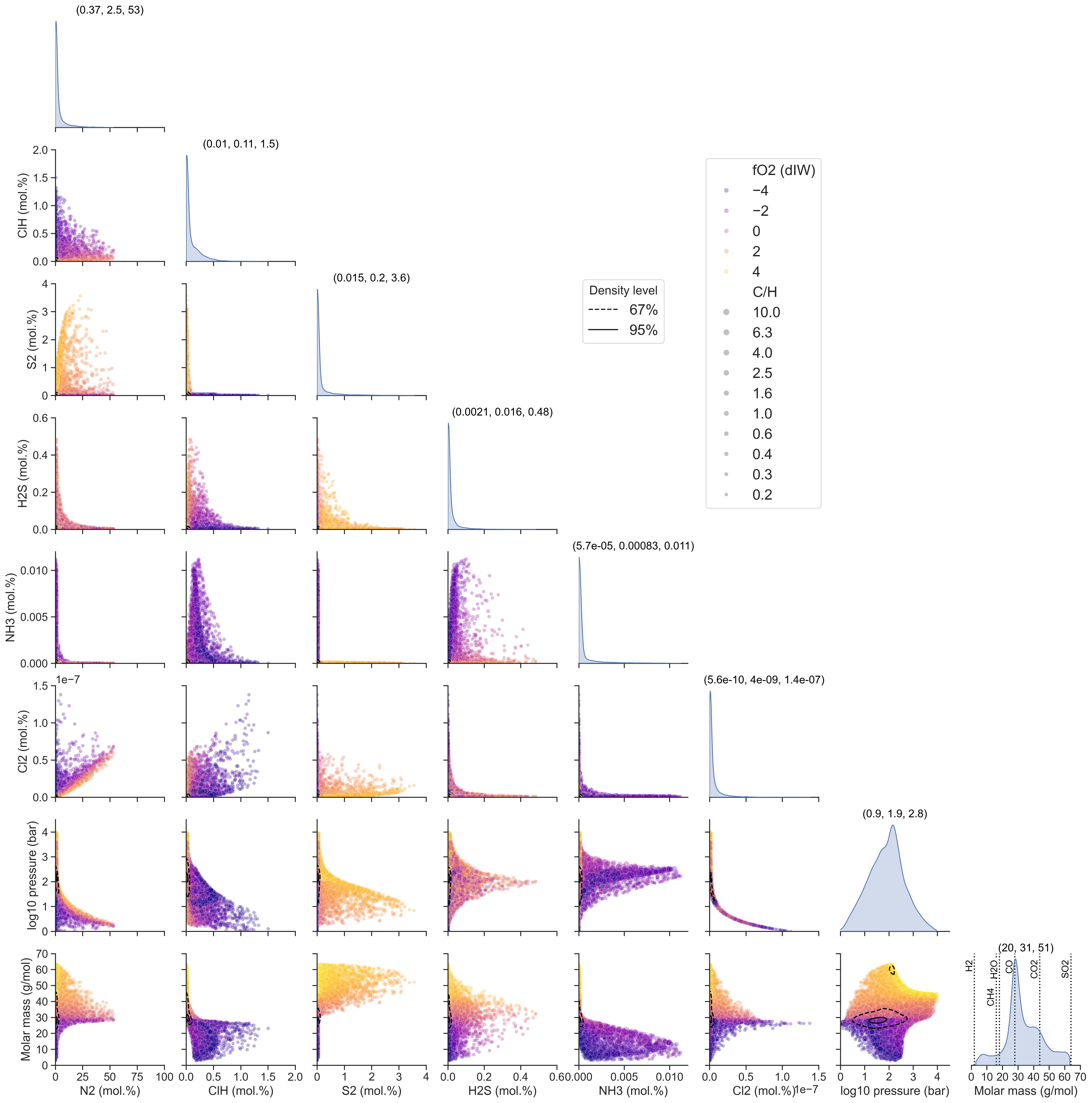}
\caption{For an early TRAPPIST-1e with a fully molten mantle (surface temperature of 1800 K), scatter plots of atmospheric molar abundance, total pressure, and molar mass, for N$_2$ and minor species: ClH, S$_2$, H$_2$S, NH$_3$, and Cl$_2$. Points are colored by oxygen fugacity expressed relative to the IW buffer and sized in proportion to log$_{10}$ C/H. Density levels indicate areas with high likelihood in the scatter plots, while marginal distributions are shown on the diagonal. The 10th, 50th (median), and 90th percentiles of these distributions are also annotated above the marginal distributions as (p10, p50, and p90), respectively. Compare to a partially molten mantle in Figure~\ref{fig:T1e_partially_molten_minor}.}
\label{fig:T1e_fully_molten_minor}
\end{figure}

We consider TRAPPIST-1e with a fully molten mantle, for which solubility and gas-phase equilibria are both important in determining the atmospheric mass and composition. For comparison, Appendix~\ref{app:trappist1e} presents the same analysis assuming a partially molten mantle with 10\% melt fraction, with major and minor species shown in Figure~\ref{fig:T1e_partially_molten} and Figure~\ref{fig:T1e_partially_molten_minor}, respectively. For a fully molten mantle, solubility conspires to restrict the diversity of atmospheres (Figure~\ref{fig:T1e_fully_molten}) compared to the low melt fraction case (Figure~\ref{fig:T1e_partially_molten}). The data are colored according to the oxygen fugacity relative to the IW buffer ($\Delta$IW), which defines whether the atmosphere is reduced (lower values) or oxidized (higher values). The scatter plots capture the diversity of atmospheres, whereas the contoured density levels and marginal distributions reveal the most probable range of values \added{and are dependent on the sampling strategy (Appendix~\ref{app:sampling})}. \added{Total pressure represents the sum of the partial pressures of the gaseous species according to Dalton's law.} Figure~\ref{fig:T1e_fully_molten} and Figure~\ref{fig:T1e_fully_molten_minor} show major and minor species, respectively.

%%%
%%%

%%%
%%%

Reduced atmospheres are dominated (volume mixing ratio, VMR $>0.5$) by either CO at low total H budgets or by H$_2$ at high total H budgets, because neither carbon monoxide nor molecular hydrogen are particularly soluble in silicate melts \citep{HWA12,YNN19}. Furthermore, H$_2$ can be accompanied by modest amounts (up to 15 mol\%) of CH$_4$, where $f$CH$_4$ is buffered by graphite precipitation:

\begin{equation}
    \mathrm{CH}_4 \mathrm{(g)} = \mathrm{C(cr)} + \mathrm{2H}_2\mathrm{(g)}.
\end{equation}

At $f$O$_2$s higher than the IW buffer, the mixing ratio of CH$_4$ drops precipitously as it is mostly replaced by H$_2$O and CO$_2$, as described by the homogeneous equilibrium:

\begin{equation}
    \mathrm{CH}_4\mathrm{(g)} + \mathrm{2O}_2\mathrm{(g)} = \mathrm{CO}_2\mathrm{(g)} + \mathrm{2H}_2\mathrm{O(g)}.
\end{equation}

However, because H$_2$O(g) is more soluble in silicate melt than H$_2$(g) or CO$_2$(g), it never becomes a dominant species in the atmosphere, even under oxidizing conditions \citep[see also][]{BHS21,STB23}. Consequently, the abundances of hydrogen-bearing species in the atmosphere remain low, with H$_2$O(g) and H$_2$(g) comprising just a few percent of the total. Therefore, owing to its relatively lower solubility, CO$_2$(g) assumes the role of the prevailing gas species in atmospheres between $\Delta$IW = +1 and $\Delta$IW = +3. At the highest $f$O$_2$s above $\sim \Delta$IW = +3, SO$_2$(g) overtakes CO$_2$(g) as the prevailing gas species for BSE-like S abundances \citep[see also][]{GBR22,maurice2024volatile, gillmann2024}. Because no other gaseous species form in appreciable quantities at these conditions, the mixing ratios of CO$_2$(g) and SO$_2$(g) are anticorrelated in oxidizing atmospheres.

\begin{figure}
\plotone{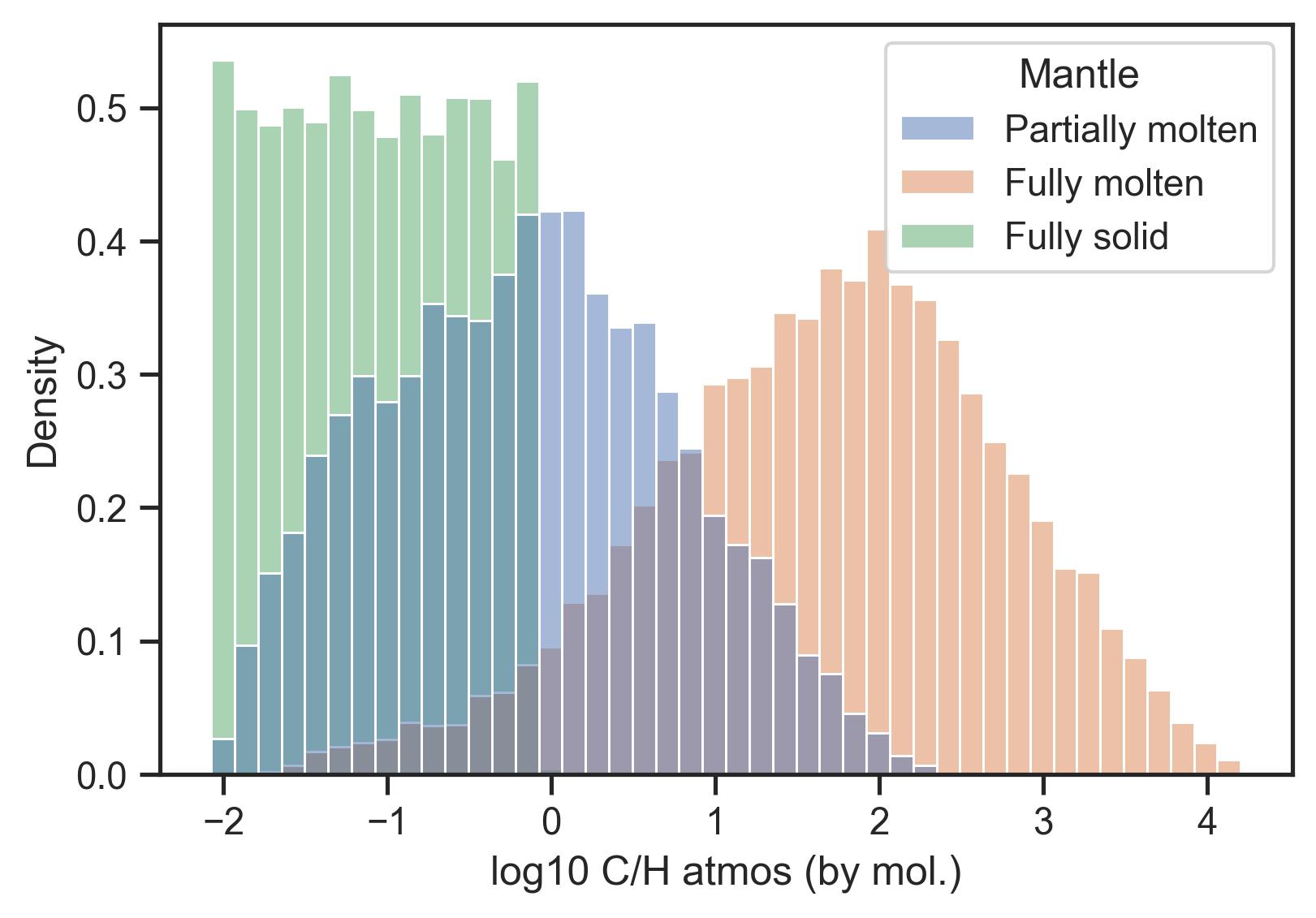}
\caption{Histogram of log$_{10}$ C/H in the atmosphere, computed by moles, for a partially molten mantle (10\% melt) (Appendix~\ref{app:trappist1e}) and fully molten mantle (main text), both at 1800 K and each based on 10,000 simulations. For reference, the `fully solid' case shows the total C/H imposed as an initial condition that is available to partition between the silicate mantle and atmosphere. Interior dissolution both increases C/H in the atmosphere and broadens its distribution, and for a fully molten mantle there is a long tail to low C/H.}
\label{fig:T1e_starting_CH_ratio}
\end{figure}

\begin{figure}
\includegraphics[width=1\textwidth]{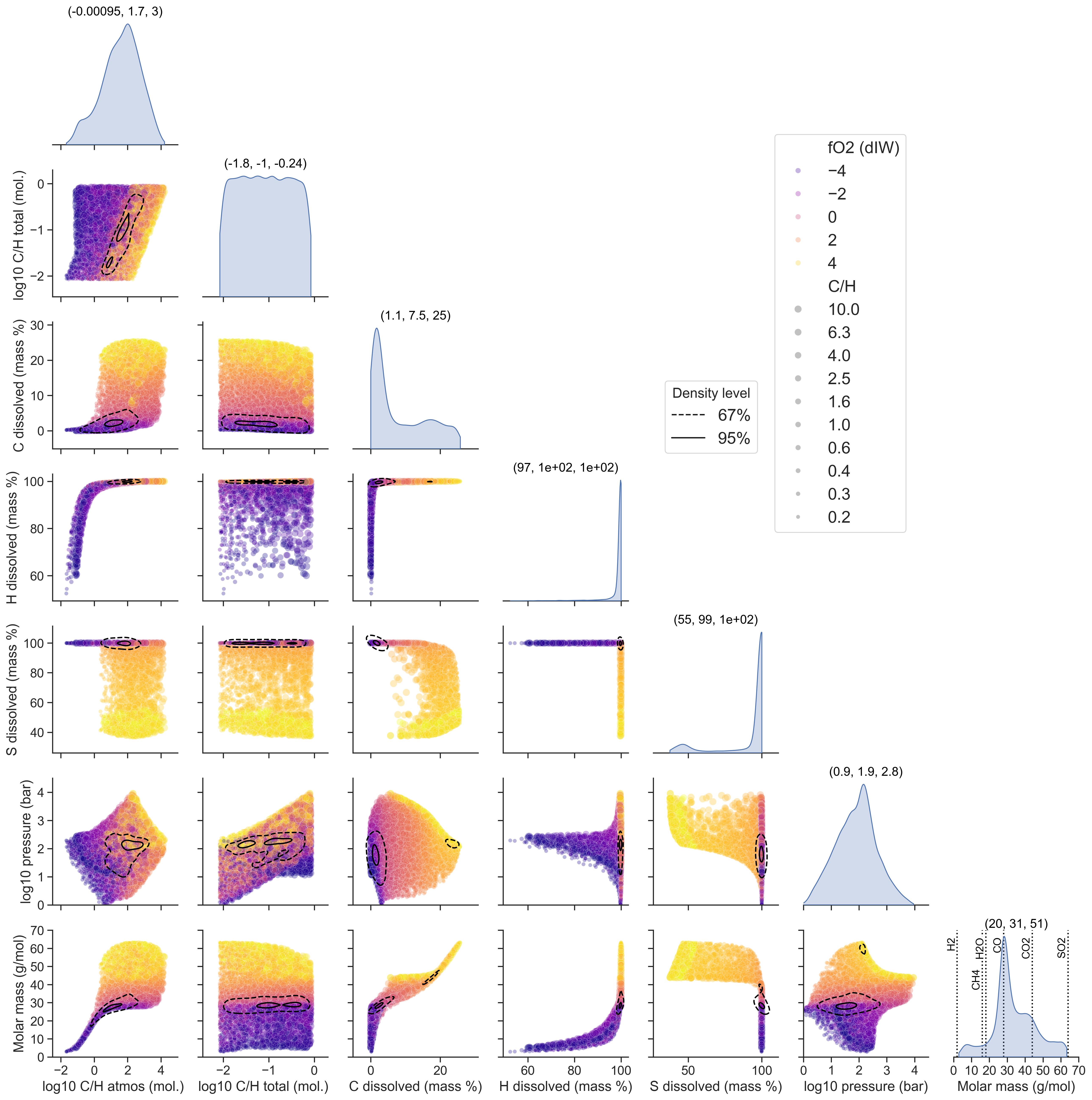}
\caption{For an early TRAPPIST-1e with a fully molten mantle (surface temperature of 1800 K), scatter plots of atmospheric and total C/H, C, H, and S dissolved in the interior relative to the total inventory of the element by mass, pressure, and molar mass. Points are colored by oxygen fugacity expressed relative to the IW buffer and sized in proportion to log$_{10}$ C/H. Density levels indicate areas with high likelihood in the scatter plots, while marginal distributions are shown on the diagonal. The 10th, 50th (median), and 90th percentiles of these distributions are also annotated above the marginal distributions as (p10, p50, and p90), respectively. Compare to a partially molten mantle in Figure~\ref{fig:T1e_partially_molten_ratios}.}
\label{fig:T1e_fully_molten_ratios}
\end{figure}

The dominant outcome of preferential H$_2$O dissolution is to enhance the relative abundances of carbon-bearing species in the atmosphere (Figure~\ref{fig:T1e_starting_CH_ratio}). This is illustrated by the mean molar mass ($\bar{\mu}$) frequency distribution (Figure~\ref{fig:T1e_fully_molten}, bottom right), which shows a low plateau between the molar masses corresponding to H$_2$ and H$_2$O, a dominant peak around that of CO, a smaller peak around the molar mass of CO$_2$, and a tail extending to SO$_2$. The small peak at $\bar{\mu} \sim$7--8 g mol$^{-1}$ corresponds to the most reducing, CH$_4$--H$_2$ atmospheres. For "neutral" $f$O$_2$ atmospheres 1--2 log units either side of the IW buffer, the near-quantitative removal of H$_2$O from the atmosphere means that peaks in the molar mass frequency distribution can be ascribed to a singular, dominant gas species (Figure~\ref{fig:T1e_fully_molten}, bottom row). Therefore, an atmosphere with $\bar{\mu}=28$ g mol$^{-1}$ must be CO-rich, rather than a mixture of lighter hydrogen-bearing and heavier carbon-bearing species, because H$_2$O(g) is never a major gas species and CO$_2$--H$_2$ atmospheres are excluded thermodynamically \citep[see also][]{STB23}. The most oxidizing atmospheres have molar masses that are defined by binary mixtures of CO$_2$ and SO$_2$. As such, the mean molar mass of an atmosphere is broadly positively correlated with its $f$O$_2$, where atmospheres with $\bar{\mu}<16$ g mol$^{-1}$ must be both reducing and H-rich (Figure~\ref{fig:T1e_fully_molten_ratios}, bottom left).

%%%
%%%

A fully molten mantle always dissolves at least $\sim$60\% of the total hydrogen budget. At $\Delta$IW close to $-4$, high C/H atmospheres have marginally less total H dissolved, all else being equal, due to the stability of CH$_4$(g) over H$_2$(g). For all but these most reduced conditions, the dissolved fraction of H approaches 100\% (Figure~\ref{fig:T1e_fully_molten_ratios}). This is evident by the sharp peak around 100\% in the marginal distribution of dissolved hydrogen where even the 10th percentile exceeds 97\% \citep[see also][]{SBB20}. As a result, atmospheric C/H ratios can be around 4 orders of magnitude larger than total C/H, with the disparity increasing to more oxidized conditions due to increasing $f$H$_2$O/$f$H$_2$. Nearly no C is dissolved in the mantle under reducing conditions (owing to the low solubilities of CO(g) and CH$_4$(g)), but this \replaced{figure}{value} increases up to about 25\% of the total carbon inventory at the most oxidized conditions where CO$_2$(g) is stable (Figure~\ref{fig:T1e_fully_molten_ratios}). This proportion decreases for correspondingly lower melt fractions (Figure~\ref{fig:T1e_partially_molten_ratios}).

%%%
%%%

Total pressures range between $\sim$10$^0$ and 10$^4$ bar (90\% lie within $\sim$5--500 bar), with atmospheres produced in equilibrium at intermediate $f$O$_2$s near the IW buffer delimiting both the high- and low ends of the range. The total pressure is largely governed by the number of ocean masses of H, together with the C/H ratio, with high C/H resulting in the highest atmospheric pressures at a given $f$O$_2$. Even though the mantles in equilibrium with reducing atmospheres store proportionally less H and C than their more oxidizing counterparts, their lower mean molecular masses ($\bar{\mu}$) result in only modest total pressures (Equation~\ref{eq:volmasssingle}). At high $f$O$_2$s near $\Delta$IW = +4, the increasing degree of CO$_2$ dissolution into the mantle restricts the total pressure of SO$_2$-rich atmospheres to around $\sim$200 bar. Sulfur is almost entirely dissolved below $\Delta$IW = +3, but above this value up to 60\% of the sulfur inventory can exist in the atmosphere (Figure~\ref{fig:T1e_fully_molten_ratios}). The nonmonotonic solubility of S arises due to the occurrence of two moieties in silicate liquids; S$^{2-}$ below $\sim \Delta$IW = +4 and SO$_4^{2-}$ above this value \citep{OM22,BW23}. Because the S content of the silicate liquid is proportional to $f$O$_2^{-1/2}$ (for constant temperature, $f$S$_2$ and FeO content) when S$^{2-}$ is stable, and to $f$O$_2^{3/2}$ when SO$_4^{2-}$ is stable, the solubility of S exhibits a minimum at $x$S$^{2-}$/$x$SO$_4^{2-}$ = 0.5 (where $x$ is mole fraction), or roughly $\Delta$IW = +4 \citep[see also][]{hughes2023sulfur}. Therefore, $f$SO$_2$ should decrease in atmospheres that are more oxidizing and more reducing than those formed at the solubility minimum.

Minor species for a 100\% molten mantle are shown in Figure~\ref{fig:T1e_fully_molten_minor} (see Figure~\ref{fig:T1e_partially_molten_minor} for the 10\% molten equivalent). Of these, N$_2$(g) can become a major species (i.e., its mixing ratio exceeds 0.5) in a handful of cases. These atmospheres correspond to those with the lowest total pressures. That is, since the abundances of N, Cl, and S are held constant (while those of H, C, and O vary) in the Monte Carlo simulations, N$_2$-rich atmospheres arise when the prescribed initial abundances of H and C are low. This behavior is also reflected in the inverse correlation between Cl$_2$(g) mixing ratio (which never exceeds 1.5$\times$10$^{-7}$) and total pressure (Figure~\ref{fig:T1e_fully_molten_minor}). Indeed, the mixing ratios of N$_2$(g) and Cl$_2$(g) are positively correlated\added{, except in reduced, H-rich atmospheres due to the formation of abundant HCl(g)}. The other minor species diverge from this expectation because their partial pressures are sensitive to the presence of other volatile species, whose fugacities depend on $f$O$_2$ and/or the bulk chemical composition of the atmosphere.

The second most abundant minor species is S$_2$(g), reaching mixing ratios up to $\sim$3\% (0.03), with these cases corresponding exclusively to oxidizing atmospheres. As detailed above, more reducing conditions promote the dissolution of S as S$^{2-}$ in the silicate liquid, thereby restricting the S budget of the atmosphere and thus its ability to stabilize S$_2$(g). A third S-bearing gas species, H$_2$S(g), is less abundant than S$_2$(g) and SO$_2$(g) in atmospheres more oxidized than $\Delta$IW=+2, yet it becomes the dominant S-bearing gas species (mixing ratios up to 0.005) at $f$O$_2$s below this value. This reflects the increase in $f$H$_2$ at the expense of $f$H$_2$O, promoting the formation of H$_2$S(g) by:

\begin{equation}
\mathrm{H_2(g)} + \frac{1}{2}\mathrm{S_2(g)} = \mathrm{H_2S(g)}.
\label{eq:H2S}
\end{equation}

The higher mixing ratios of NH$_3$(g) (up to 0.0001) and HCl(g) (up to 0.015) in reduced atmospheres relative to their oxidized counterparts (Figure~\ref{fig:T1e_fully_molten_minor}) are explained by reactions analogous to Equation~\ref{eq:H2S} occurring when substituting S$_2$ for N$_2$ and for Cl$_2$, respectively. Since the abundance of H$>>$Cl on a molar basis in our simulations, HCl(g) is the dominant Cl-bearing gas under all conditions. We note, however, the lack of experimentally determined solubilities for Cl, N, and S in H-bearing systems\added{, as well as the limited range of temperatures and melt compositions over which they are calibrated \citep[though see][]{thompson2025water}}. The mixing ratios in the 10\% molten cases show generally similar behavior, though the more soluble elements (Cl and S) exhibit higher atmospheric mixing ratios compared to those determined in the 100\% molten cases, owing to the declining mass of molten material capable of storing them in the mantle. The corollary is that the more insoluble species (e.g., N$_2$(g)) have relatively lower mixing ratios. 

\subsubsection{Cooled atmosphere and elemental surface inventory}
\label{sect:trappist1e_cool}

%%%
\begin{figure}
\includegraphics[width=1\textwidth]{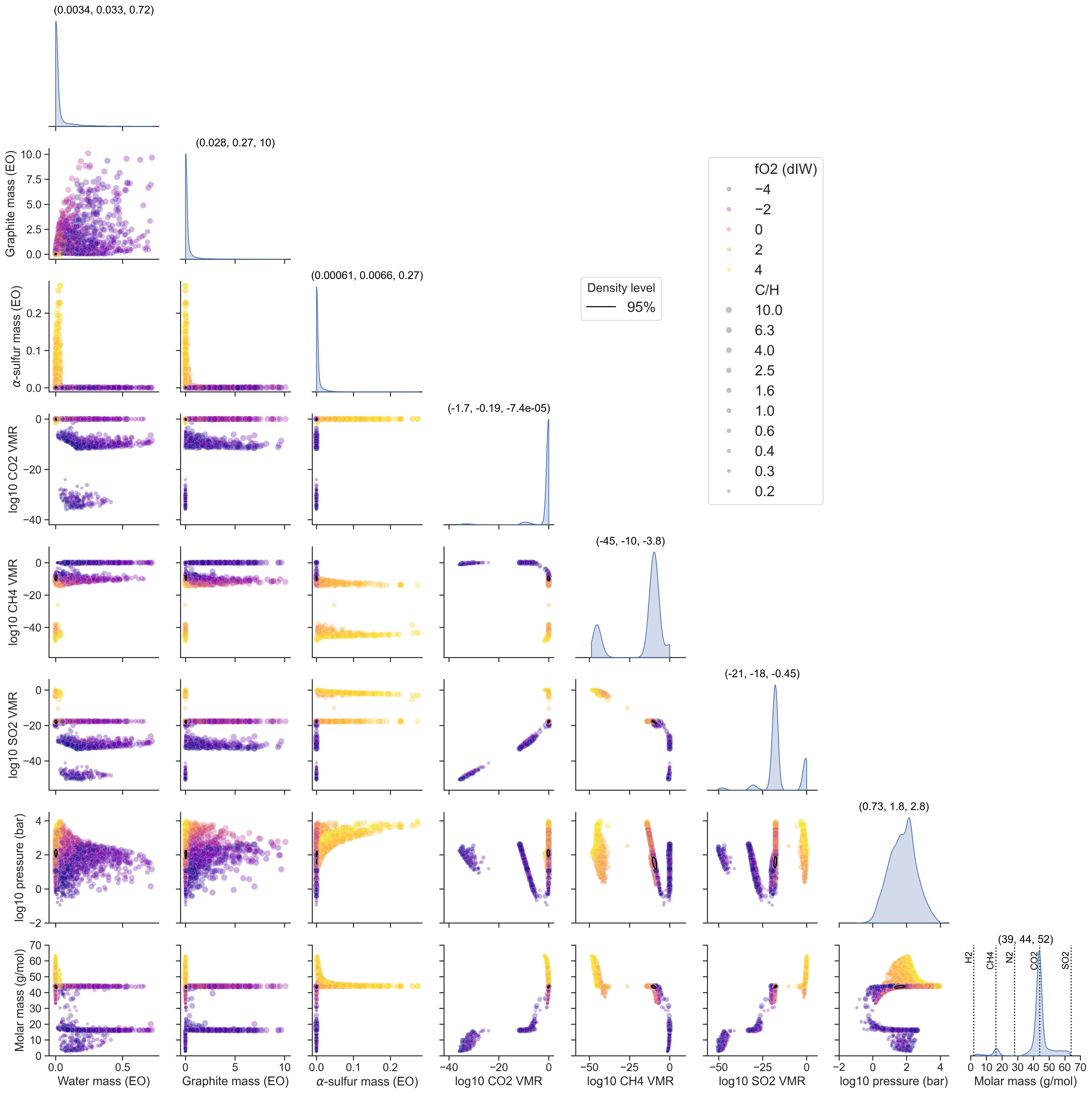}
\caption{Cooled atmospheres at 280 K for TRAPPIST-1e, starting with the elemental abundances in the atmosphere for a fully molten mantle and a surface temperature of 1800 K. Points are colored by oxygen fugacity of the initial, high-temperature atmosphere expressed relative to the IW buffer, and sized in proportion to log$_{10}$ C/H for the fully molten starting state. Condensed inventories of water, graphite, and $\alpha$-sulfur are normalized by the Earth's present-day ocean mass (EO). Other scatter plots show the VMR of major species in the atmosphere, total pressure, and molar mass. Density levels indicate areas with high likelihood in the scatter plots, while marginal distributions are shown on the diagonal. The 10th, 50th (median), and 90th percentiles of these distributions are also annotated above the marginal distributions as (p10, p50, and p90), respectively. Compare with the cool atmospheres derived from a partially molten mantle in Figure~\ref{fig:T1e_partially_molten_condensed}.}
\label{fig:T1e_fully_molten_condensed}
\end{figure}
%%%

The magma ocean stage establishes the initial volatile inventories in the atmosphere (Section~\ref{sect:early}), and as the atmosphere cools and condenses it distributes materials at the planetary surface that can participate in geochemical cycles and fuel a potential biosphere. To this end, we begin with the calculated atmospheric elemental abundances for each of the fully molten mantle cases presented in Section \ref{sect:early} and cool the atmospheres, isochemically (i.e., assuming a constant bulk atmosphere composition inherited from the magma ocean stage), to 280 K. This is a few kelvin above the freezing point of water at 1 bar and lies between the estimates of the equilibrium temperatures of TRAPPIST-1d and TRAPPIST-1e assuming a null albedo \citep{DGT18}. We include the contribution of any graphite present in the abundance of carbon of the high-temperature atmosphere, given that graphite remains close to the planetary surface \citep{KG19} and in equilibrium with the atmosphere. In addition to the aforementioned gas species and graphite (Section~\ref{sect:T1e_context}), we further allow for the presence of liquid water (H$_2$O), rhombic sulfur (also known as $\alpha$-sulfur), and ammonium chloride (ClH$_4$N) depending on thermodynamic stability, since these condensed phases can be major carriers of elements in the system. The same analysis is applied to the 10\% molten mantle cases in Appendix~\ref{app:trappist1e}.

\begin{deluxetable}{lr|rrr|rrr}
\tablecaption{Summary statistics of the TRAPPIST-1e models at 280 K, after cooling from equilibration with a fully molten or partially molten (10\% melt fraction) mantle at 1800 K. \label{table:models_with_condensates}
}
\tablewidth{0pt}
\tablehead{
\multicolumn{1}{c}{Dominant Species} & \multicolumn{1}{c}{} & \multicolumn{3}{c}{Total Pressure (bar)} & \multicolumn{3}{c}{Count with Condensates}\\
\colhead{(VMR $>$ 50\%)} & \colhead{Count} & \colhead{Min} & \colhead{Max} & \colhead{Median} & \colhead{H$_2$O} & \colhead{H$_2$O+C} & \colhead{H$_2$O+C+S+ClH$_4$N}}
\startdata
\multicolumn{8}{c}{At 280 K, following equilibration with a fully molten mantle at 1800 K}\\
\hline
H$_2$ & 150 & 12.25 & 268.41 & 65.20 & 150 & 0 & 0\\
CO$_2$ & 8336 & 0.45 & 8991.13 & 58.73 & 7709 & 6549 & 4406\\
CH$_4$ & 546 & 0.21 & 436.59 & 32.96 & 546 & 491 & 0\\
N$_2$ & 92 & 0.11 & 2.38 & 1.89 & 40 & 40 & 1\\
SO$_2$ & 823 & 6.51 & 231.68 & 133.86 & 580 & 0 & 0\\
ClH & 9 & 0.16 & 0.62 & 0.40 & 9 & 9 & 0\\
None & 44 & 1.74 & 234.88 & 17.17 & 19 & 1 & 1\\
Total & 10,000 & {} & {} & {} & 9053 & 7090 & 4408\\
\hline
\multicolumn{8}{c}{At 280 K, following equilibration with a partially molten mantle at 1800 K}\\
\hline
H$_2$ & 1634 & 0.76 & 675.95 & 66.50 & 1634 & 0 & 0\\
CO$_2$ & 5230 & 0.53 & 9547.52 & 104.16 & 5230 & 4002 & 4002\\
CH$_4$ & 1403 & 0.23 & 1009.14 & 62.98 & 1403 & 1172 & 0\\
N$_2$ & 100 & 0.27 & 2.13 & 1.06 & 100 & 100 & 83\\
SO$_2$ & 1180 & 6.65 & 374.21 & 229.39 & 1180 & 0 & 0\\
ClH & 33 & 0.27 & 7.05 & 2.65 & 33 & 33 & 20\\
H$_2$S & 212 & 0.48 & 135.60 & 26.96 & 212 & 212 & 94\\
None & 208 & 0.53 & 630.70 & 21.64 & 208 & 53 & 7\\
Total & 10,000 & {} & {} & {} & 10,000 & 5572 & 4206\\
\enddata
\tablecomments{Stable condensates are defined by an activity of unity.}
\end{deluxetable}

Figure~\ref{fig:T1e_fully_molten_condensed} shows the atmospheres at 280 K that were previously in equilibrium with a fully molten mantle and summary statistics are presented in Table~\ref{table:models_with_condensates}. The presence of condensates buffers the atmospheric composition and produces well-defined linear trends and clusters in the data. Therefore, and unlike at high temperatures, buffering of gas species fugacities by condensates means that atmospheres at 280 K can be grouped into discontinuous "families," of which we identify four types (see plots of log$_{10}$SO$_2$ VMR in Figure~\ref{fig:T1e_fully_molten_condensed}). Among these families, CO$_2$-rich atmospheres are dominant (83\% of the total) as revealed by the sharp peak in the frequency distribution at $\bar{\mu}=44$ g mol$^{-1}$. These develop from cooling of high temperature atmospheres produced across most of the initial $f$O$_2$ range, except in the most reduced ($<\Delta$IW = -3) and oxidized ($>\Delta$IW = +3) cases. Initially reduced atmospheres result in a subsidiary peak associated with CH$_4$(g)-rich types (5\% of the total) and a tail toward H$_2$(g)-rich atmospheres (1.5\% of the total). Another abundant gas in these atmospheres is NH$_3$(g). In general, however, there is a notable dearth of atmospheres with molar masses below about 40 g mol$^{-1}$ due to the (1) high C/H starting state caused by H$_2$O(g) dissolution in the magma ocean, and (2) the stability of CO$_2$(g) relative to CO(g) at low temperatures\added{, which attends the condensation of graphite, allowing the excess O in the atmosphere to oxidize the remaining CO(g) to form CO$_2$(g) \citep[see also][]{SBB20}}. Finally, oxidized starting conditions (i.e., $\Delta$IW = +3 to +4, see Section \ref{sect:early}) extend the tail of the distribution to high molar masses, defining the fourth family associated with CO$_2$--SO$_2$-bearing atmospheres, which represent 8\% of all cases. 

These atmospheres can also produce up to 0.3 Earth-ocean equivalents of $\alpha$-sulfur (Figure~\ref{fig:T1e_fully_molten_condensed}). Crucially, the starting conditions that favor large amounts of sulfur precipitation (oxidized) are different from the conditions that favor significant water, graphite, and ammonium chloride precipitation (IW and below). This arises because S is more abundant in high-temperature atmospheres at oxidizing conditions, while C and particularly H are more abundant at more reducing conditions (see Section \ref{sect:early}, Figure~\ref{fig:T1e_fully_molten_ratios}). There is also a trade-off between graphite and water, since high C/H in the starting conditions promotes larger graphite inventories, whereas low C/H forms larger water oceans. In the family of atmospheres with the lowest molar masses (${\bar\mu}<15$) produced by initially low C/H ratios and high H mass fractions, graphite saturation is not reached. The median total atmospheric pressure is roughly 65 bar and the maximum is a few kbar for oxidized starting conditions with high C/H and high initial number of water oceans.

Consideration of cases in which the mantle is only 10\% molten leads to cooled atmospheres that have greater mixing ratios of H-bearing species. This leads to a more continuous distribution in $\bar{\mu}$ across the simulation space (Figure \ref{fig:T1e_partially_molten_condensed}), in which the heights of peaks associated with H-bearing species (notably H$_2$(g) and CH$_4$(g), given that H$_2$O is always condensed) compared to that of CO$_2$(g) are relatively greater than for the 100\% molten cases. Nevertheless, the four families of atmospheres previously identified for the fully molten case are again evident, albeit slightly less distinct. CO$_2$ atmospheres are similarly dominant, with a tail extending to SO$_2$ atmospheres. Table~\ref{table:models_with_condensates} also reveals the formation of H$_2$S-rich atmospheres for the molten case, although they are low in number (2\% of the total).

\subsection{K2-18b and similar sub-Neptune-sized planets}
\label{sect:gasdwarf}
\subsubsection{Context and parameters}
Many sub-Neptunes likely harbor massive magma oceans beneath large insulating atmospheres, a configuration that corresponds to so-called "gas dwarfs." Such a structure has been proposed for the canonical sub-Neptune K2-18b \citep{SJN24,WBZ24}, although an alternative hypothesis suggests it might support clement conditions at the base of its atmosphere \citep{MSC23}. Other sub-Neptune-sized exoplanets are currently being observed with JWST (e.g., TOI-270 d, GJ 1214 b), and preliminary observations reveal absorption features of various gas species \added{\citep[e.g., H$_2$O, CO$_2$, CH$_4$, SO$_2$,][]{KZB23, BRC24, beatty2024, mukherjee2025, davenport2025}}. However, whether or not these planets possess a magma ocean surface, a temperate surface that could support liquid water, or no defined surface at all \cite[e.g.,][]{young2024phase}, remains unknown. Distinguishing between these scenarios requires self-consistent modeling frameworks that adequately capture the governing physics and chemistry of both the interior and the atmosphere. To this end, \atmodeller{} allows us to test the gas dwarf model with an important addition that has often been neglected in previous work on sub-Neptunes \citep{SY22, misener2023, young2023earth, SJN24, rogers2024fleeting, rigby2024}, though see \cite{TH24, seo2024}: the role of nonideality, in which the volatile species at the conditions of the magma ocean--atmosphere interface do not obey the ideal gas law. Rather, real gases decouple their partial pressure (which is related to molar abundance, mass balance, and mechanical pressure) from their thermodynamic pressure, which regulates equilibrium chemistry and solubility (Section~\ref{sect:eos}).

We simulate K2-18 b as a gas dwarf and demonstrate the effects of nonideality on sub-Neptune atmospheres. K2-18b has a total mass of $5.15 \times 10^{25}$ kg (8.63 $M_{\oplus}$) and an estimated mantle mass of $3.63 \times 10^{25}$ kg (assuming the same core/mantle mass ratio as the Earth) \citep{BWP19}. K2-18b's surface radius, assuming it has an optically thick atmosphere overlying a magma ocean surface, is estimated to be 11225 km \citep[1.76 $R_{\oplus}$; assuming an Earth-like interior composition and the super-Earth \replaced{equation of state}{mass--radius relation} taken from Table 2 of][]{HRH18}. The surface temperature at the magma ocean--atmosphere interface is set at 3000 K, \added{which is in broad agreement with formation models that predict a temperature of 3000 K at a pressure of 1 kbar after $\sim$3 Gyr of planetary evolution \citep{Tang2025,Heng2025,Rogers2021}.} We further note that this temperature is at or above the upper limit of the calibration for current solubility laws and real gas EOS. The mantle melt fraction is set to 100\%. We consider \replaced{5}{six} gas species in the atmosphere, each with their corresponding solubility laws, necessarily assuming the mantle composition is basaltic to remain internally consistent: H$_2$O \citep{DSH95}, H$_2$ \citep{HWA12}, CO \citep{YNN19}, CO$_2$ \citep{DSH95}, and CH$_4$ \citep{AHW13}. O$_2$ is also included as a gas species.  For the cases with real gases, the CORK model is used for H$_2$O, an ab initio EOS model for H$_2$ \citep{CD21}, and a corresponding states model for CO, CO$_2$, and CH$_4$ \citep{SS92} (Table~\ref{table:realgas}). O$_2$ is assumed to behave as an ideal gas in all cases, which does not introduce any uncertainty owing to its vanishingly low fugacities under the range of conditions studied.

To explore the effects of nonideality, we perform two series of simulations in which volatiles can partition between the silicate mantle and the atmosphere. The total hydrogen inventory is varied from 0.1\% to 3\% of K2-18b's mass (Figures \ref{fig:k218b_massfrac} and \ref{fig:k218b_idealvsreal}) assuming both real and ideal gas behavior. In these simulations, all other parameters are held constant: $T_{\mathrm{ surface}} = 3000$ K, C/H (by mass) = 0.32 ($\sim100\times$ the solar ratio, as recently inferred for K2-18b, \cite{WBZ24}) and $f$O$_2$ at $\Delta$IW = -3. The equivalent cases, in which each of the other two variables ($\Delta$IW and C/H ratio) are varied while the others are held constant, are shown in Figures~\ref{fig:k218b_fO2} and \ref{fig:k218b_CtoH}, respectively. All simulation data are available to download \citep{atmodeller_data}.

\subsubsection{Volatile storage in the mantle}

\begin{figure}
\includegraphics[width=1\textwidth]{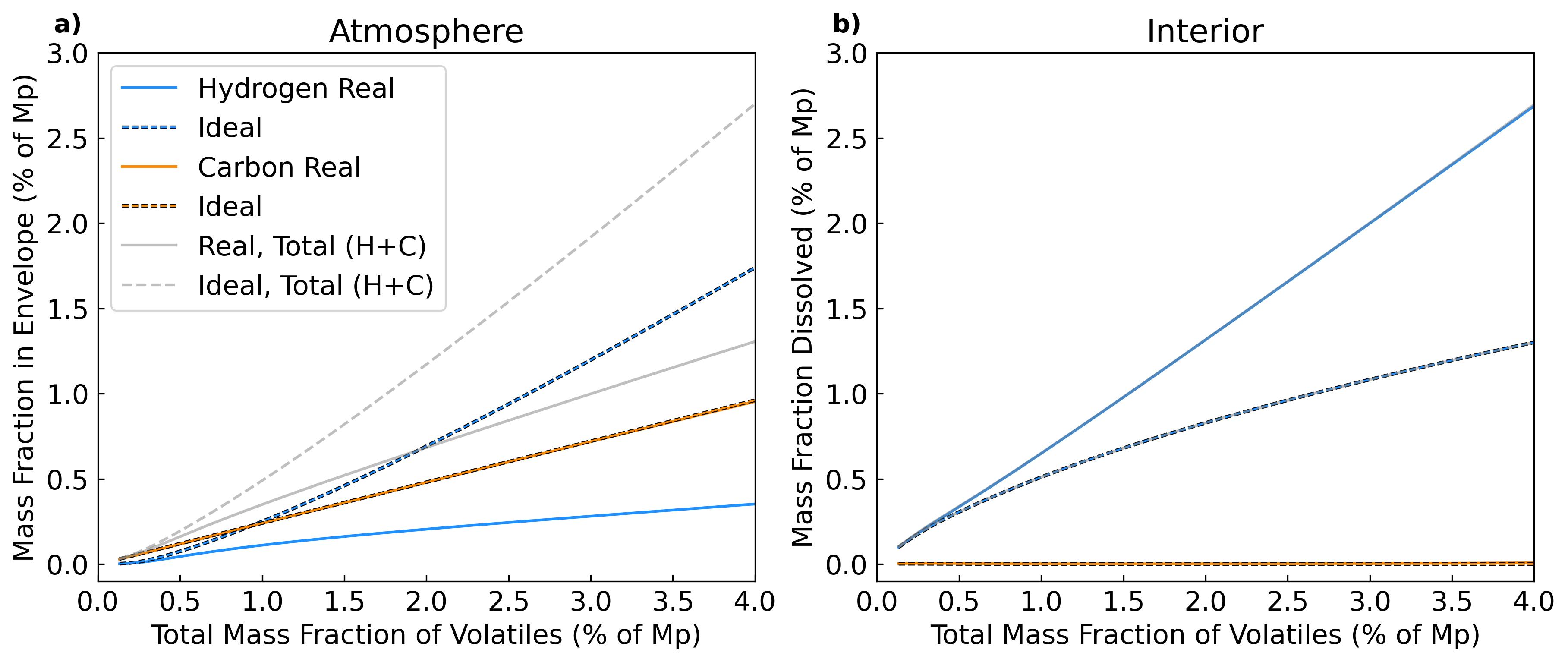}
\caption{Comparison between the mass fraction of volatile elements in the (a) atmosphere and (b) interior, of K2-18b for real (solid lines) and ideal (dashed lines) gas behavior for hydrogen (blue), carbon (orange), and their sum (gray). In (b), the sum of H+C overlaps with H, since it is the dominant dissolved volatile element in both the real and ideal gas cases. The total hydrogen budget expressed as a fraction of planet's mass ($M_p$) varies from 0.1\% to 3\%, while other input parameters are fixed; $T_{\mathrm{surface}} = 3000$ K, C/H (by mass) = 0.32 (which is $\sim$100$\times$ the solar ratio) and $f$O$_2$ at $\Delta$IW = -3.}
\label{fig:k218b_massfrac}
\end{figure}

\begin{figure}
\includegraphics[width=1\textwidth]{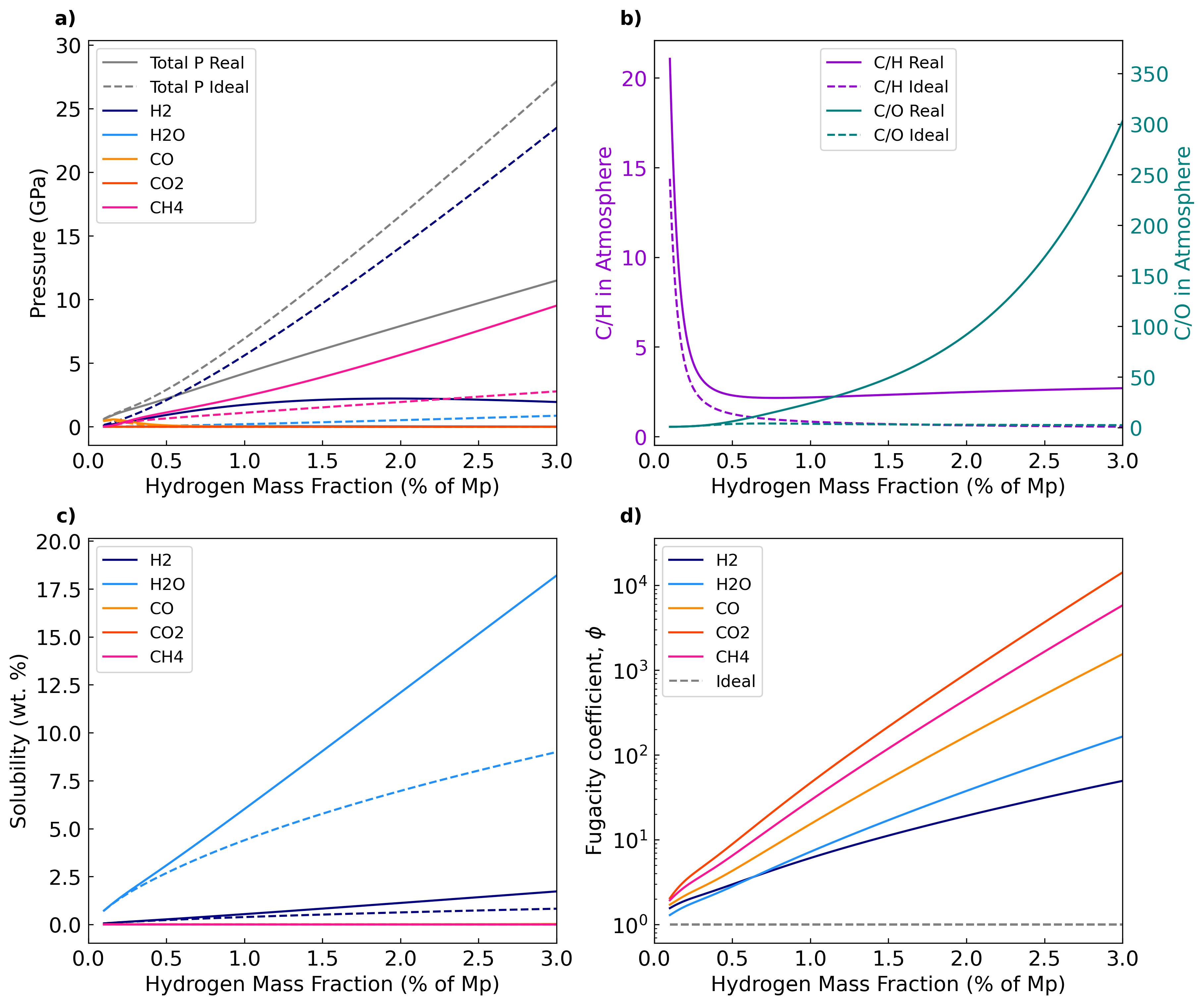}
\caption{Simulations at the magma ocean--atmosphere interface of K2-18b assuming real versus ideal gas behavior. As in Figure \ref{fig:k218b_massfrac}, the total hydrogen budget expressed as a fraction of the planet's mass ($M_p$) varies from 0.1\% to 3\%, while other input parameters are fixed: $T_{\mathrm{surface}} = 3000$ K, C/H (by mass) = 0.32 (which is $\sim$100$\times$ the solar ratio) and $f$O$_2$ at $\Delta$IW = -3. Unlike Figure~\ref{fig:k218b_massfrac}, here the $x$-axis corresponds to the total hydrogen mass fraction of the planet, expressed as a percentage of K2-18b's mass. The solid lines correspond to the cases assuming real gases and the dashed lines are for cases with ideal gas behavior. (a) Pressures of volatile species in the atmosphere (GPa) and the total atmospheric pressure (gray). (b) C/H (purple, left $y$-axis) and C/O (teal, right $y$-axis) in the atmosphere. (c) Solubility of each volatile species (wt\%, i.e., fraction of the planet's mantle mass). (d) Fugacity coefficient ($\phi$) of each volatile species. The coefficient for ideal behavior (gray dashed line) is unity for all volatiles.}
\label{fig:k218b_idealvsreal}
\end{figure}

The key result is that for planets with large volatile inventories, the mass fraction of volatile elements (H+C) retained in the envelope (up to 4\% of the planet mass initially added, $M_p$) decreases as solubility increases, and decreases even further when real gas behavior is taken into account. In Figure~\ref{fig:k218b_massfrac}a, a case without solubility (not shown) would produce a 1:1 relationship between the envelope volatile mass fraction ($y$-axis) and the total volatile mass fraction ($x$-axis). When ideal gas behavior and solubility are considered (gray dashed line), the relationship is approximately 2:3, and when both real gas behavior and solubility are included (gray solid line), the relationship is closer to 1:3. The ideal with solubility case overestimates the volatile mass fraction in the envelope (Figure \ref{fig:k218b_massfrac}a), and hence the total atmospheric pressure by a factor of $\geq$2 compared to the real with solubility case (Figure \ref{fig:k218b_idealvsreal}a). Relative to a case without solubility, the envelope volatile mass fraction is reduced by 2/3 when considering both H- and C-bearing species solubilities and real gases. For cases that do consider solubility, assuming a total hydrogen mass equivalent to $\sim$3\% of K2-18b's mass results in a $>$50\% decrease of the total atmosphere pressure for the real case ($\sim$12 GPa) compared to the ideal case (27 GPa). Therefore, these cases demonstrate that both solubility and nonideality have comparable influences in determining the first-order characteristics of sub-Neptune atmospheres. 

The mass fraction of H+C in the envelope predicted by the ideal and real cases with solubility begins to diverge above 0.2--0.3 total volatile mass fraction (compare gray dashed and solid lines, Figure~\ref{fig:k218b_massfrac}). This is because, as the hydrogen mass fraction of the planet increases, the atmospheric surface pressure increases, causing the gas fugacity coefficients, $\phi$, to deviate more strongly from ideality (Figure \ref{fig:k218b_idealvsreal}d). Because the temperatures and pressures far exceed the critical points of the gas species considered here, repulsive forces between the molecules predominate, and volumes ($V$), and hence $\phi$ exceed those of ideal gases. For a constant mass budget of each element, this property increases the fugacity, and hence solubility in silicate liquids, since solubility is proportional to thermodynamic pressure (i.e., fugacity) not partial pressure (Figure \ref{fig:k218b_idealvsreal}c). Consequently, the enhanced solubility of volatile elements, primarily hydrogen species (H$_2$ and H$_2$O), limits the amount of H and O (and to a lesser extent C) in the atmosphere and acts to buffer increases in the total atmospheric pressure (Figure \ref{fig:k218b_idealvsreal}c,d). 

At fixed $f$O$_2$($\Delta$IW = -3) and C/H (0.32 by mass; 100$\times$ solar), for low H mass fractions, CO(g) is the dominant species, but is exceeded by H$_2$(g) across all hydrogen mass fractions $>$0.2\% in the ideal case, whereas $p$CH$_4$ exceeds $p$H$_2$ above $\sim$0.3\% H mass fraction in the real case. The predominance of CH$_4$(g) results from the interplay between total pressure and the fugacity coefficients of the major atmospheric species, as described by the homogeneous equilibrium:

\begin{equation}
    \mathrm{CO}_2\mathrm{(g)} + \mathrm{2H}_2\mathrm{(g)} = \mathrm{CH}_4\mathrm{(g)} + \mathrm{O}_2\mathrm{(g)},
    \label{eq:CH4-H2}
\end{equation}

where, at equilibrium, and at constant $f$O$_2$:

\begin{equation}
    K_{(\ref{eq:CH4-H2})} = \frac{f\mathrm{CH}_4 \cdot f\mathrm{O}_2}{f\mathrm{CO}_2 \cdot(f\mathrm{H}_2)^2} = \frac{\phi \mathrm{CH}_4}{\phi \mathrm{CO}_2 \cdot (\phi \mathrm{H}_2)^2} \cdot \frac{x \mathrm{CH}_4}{x \mathrm{CO}_2 \cdot (x \mathrm{H}_2)^2} \cdot \frac{f\mathrm{O}_2}{P^2} .
    \label{eq:CH4-H2-K}
\end{equation}

Therefore, $x$CH$_4$ is proportional to $(P^2 \cdot \phi \mathrm{H}_2^2) / \phi \mathrm{CH}_4$, whereas $x\mathrm{H}_2$ scales with $(\phi \mathrm{CH}_4)^{0.5} / (\phi \mathrm{H}_2 \cdot P)$. Based on this reasoning, at constant bulk composition, temperature and $f$O$_2$, CH$_4$(g) should become dominant as $P$ increases. However, this is not observed in the ideal case, where $p$H$_2$ exceeds $p$CH$_4$ across the entire pressure range (Figure~\ref{fig:k218b_idealvsreal}a). In ideal cases, H$_2$ remains stable because the addition of gas with C/H (by mass) = 0.32 gives rise to C/H ratios in the atmosphere that are still too low (i.e., $\sim$1 by mass; Figure~\ref{fig:k218b_idealvsreal}b) to form significant quantities of methane  (Figure~\ref{fig:k218b_idealvsreal}a). On the other hand, for real cases, the increase in $\phi$H$_2$ results in its increased dissolution relative to the ideal case (Figure~\ref{fig:k218b_idealvsreal}c) while CH$_4$ never dissolves to any substantial extent, causing a commensurate increase in C/H of the atmosphere ($\sim$ 2.5 by mass, Figure~\ref{fig:k218b_idealvsreal}b) that is closer to the stoichiometry of CH$_4$ (C/H = 3 by mass). Although $\phi$CH$_4$ is in the range 10$^3$ and $\phi$H$_2$ $\sim$ 20--30, since $x$H$_2$ depends upon $(\phi$H$_2)^2$ and $1/\phi$CH$_4$, the effect of relative differences in $\phi$ essentially cancel one another out.

\begin{figure}
\includegraphics[width=1\textwidth]{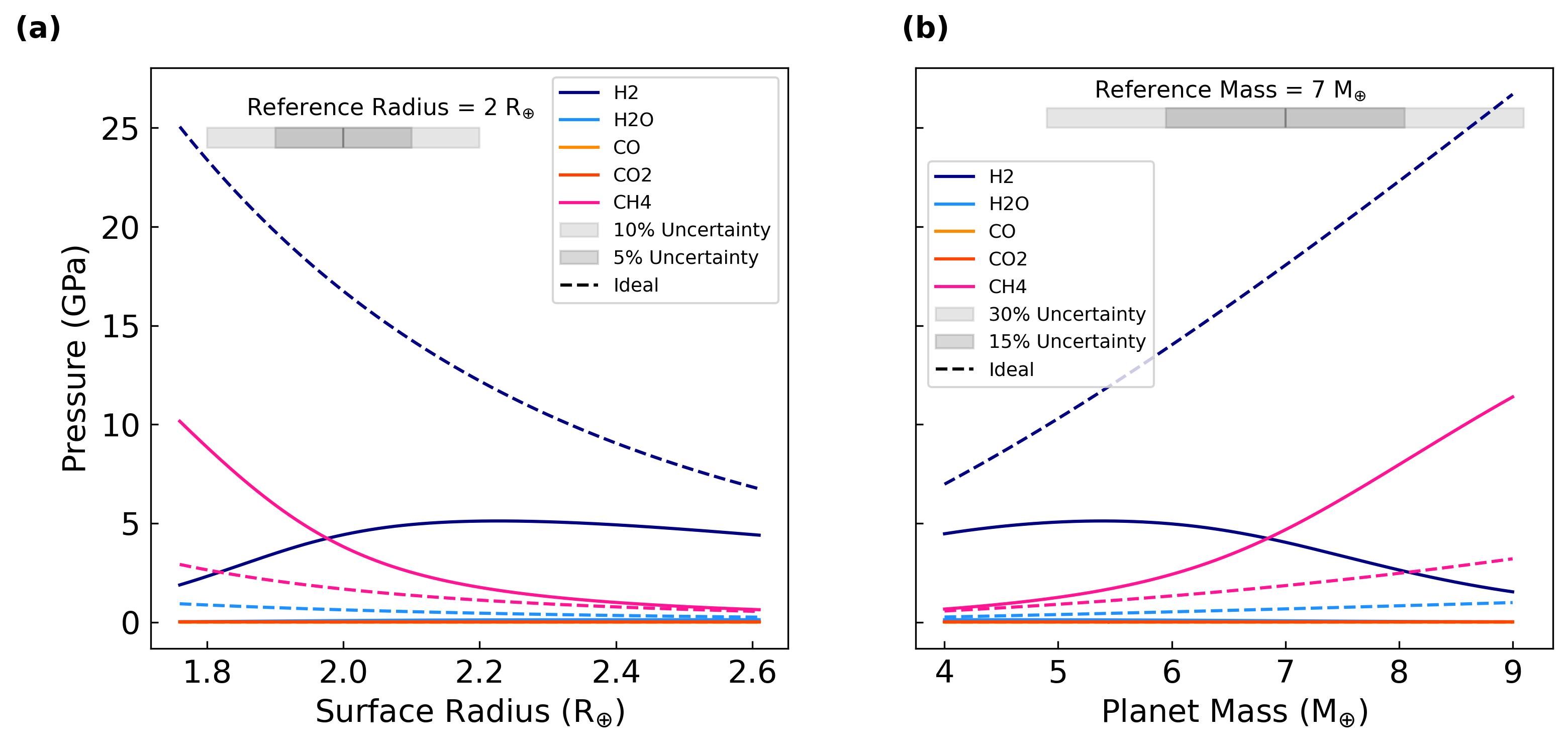}
\caption{The effect of planetary surface radius and total mass on the atmospheric speciation of K2-18b at the magma ocean--atmosphere interface: $T_{\mathrm{surface}} = 3000$ K, the oxygen fugacity is fixed at $\Delta$IW = -3, the total hydrogen mass of the planet is 3\% of the planet's mass, and C/H is 100$\times$ solar. (a) Pressures (in GPa) of the gas species assuming real gas EOS (solid lines) and ideal behavior (dashed lines) as a function of the surface radius of the planet which varies from 1.76 to 2.61 $R_{\oplus}$, where the planet mass is constant at 8.63 $M_{\oplus}$. The gray horizontal bar shows typical uncertainties on planetary radius measurements of 5\% and 10\% for a 2 $R_{\oplus}$ planet. (b) Pressures as a function of planet mass varying from 4 to 9 $M_{\oplus}$, where the planetary surface radius is constant at 1.76 $R_{\oplus}$. The gray horizontal bar shows typical uncertainties on planetary mass measurements of 15 and 30\% for a 7 $M_{\oplus}$ planet.}
\label{fig:k218b_radiusmass}
\end{figure}

The aforementioned calculations were performed for a fixed planetary mass and radius. However, these quantities are only determined within limits for sub-Neptunes \citep[at best 15\% relative for mass and 3.5\% for radius,][]{luque2022precis}. Hence, we now determine, for a fixed hydrogen mass fraction, C/H, $f$O$_2$ and surface temperature, how variations in a planet's surface radius or mass can influence the atmospheric speciation we infer when employing real gas EOS. As shown in Figure \ref{fig:k218b_radiusmass}a, in the real case, for a fixed mass of 8.63 $M_{\oplus}$, the atmosphere will be dominated by CH$_4$ below $\sim$2 $R_{\oplus}$ and by H$_2$ above 2 $R_{\oplus}$. This contrasts with the ideal scenario, in which H$_2$ is always the dominant species. Similarly, at constant radius (1.76 R$_{\oplus}$), the atmosphere is H$_2$ dominated below $\sim$7 M$_{\oplus}$ and CH$_4$ dominated above 7 M$_{\oplus}$ for real gases (Figure \ref{fig:k218b_radiusmass}b). Hence, typical uncertainties in mass and radius (Figure~\ref{fig:k218b_radiusmass}, gray horizontal bars) can bracket the transition regime between H$_2$ and CH$_4$ dominated atmospheres, highlighting the importance of improving mass and radius measurement uncertainties to accurately interpret atmospheric observations. Mass and radius both influence the atmospheric pressure at the surface, which in turn impact fugacity coefficients, through the relationship:

\begin{equation}
    P = \frac{M_ag}{4 \pi R_s^2},
    \label{eq:mass-pressure}
\end{equation}

where $M_a$ is the total mass of the atmosphere, $g$ is the acceleration due to gravity, and $R_s$ is the surface radius. Because pressure is proportional to 1/$R_s^2$, even small changes in surface radius have tangible effects on total pressure, for a constant atmospheric mass budget. Conversely, the linear dependence of $P$ on $g$ (and hence on the planetary mass, $M_p$) results in a more modest mutual dependence. \added{However, because solubility is included, $M_a$ is not constant}, and increasing pressure for a constant total mass budget results in higher dissolution (since $x_v \propto p_v$, where $v$ is any given gas species), giving rise to the monotonic decrease in $p$H$_2$ (and hence total $P$) with increasing $R_s$ or with decreasing $M_p$ in the ideal case. Conversely, increasing $P$ leads to increases in $\phi$ (analogous to the mechanism discussed with respect to Equation~\ref{eq:CH4-H2}, above), promoting additional H dissolution as molecular H$_2$. The net result is sub-Neptunes are expected to be CH$_4$-rich for cases in which mean densities are roughly $\geq$6500 kg m$^{-3}$ at $\Delta$IW = -3, H=3\% $M_p$, and bulk C/H=$\sim$100$\times$ solar. Figures~\ref{fig:k218b_fO2} and \ref{fig:k218b_CtoH} illustrate the effects of varying oxygen fugacity from -6 < $\Delta$IW < 0 (at constant C/H = 0.32 and H mass fraction = 1\% $M_p$) and total C/H mass ratio (at constant $f$O$_2$ = $\Delta$IW = -3 and H mass fraction = 1\% $M_p$), respectively, on volatile speciation in the atmosphere and dissolution in the interior assuming real and ideal gas behavior, for K2-18b-like cases.

%%%

\section{Discussion}
\subsection{Earth-sized planets}
For Earth-sized planets, our results show that the dominant atmosphere at temperate conditions that emerges after planet formation is CO$_2$-rich with a total surface pressure around 100 bar. Such an atmosphere forming upon cooling on the early Earth from a magma ocean state was first hypothesized by \cite{holland1984}, computed by \cite{zahnle2007}, and demonstrated by \cite{SBB20} for a magma ocean initially equilibrated at its estimated $f$O$_2$ close to the IW buffer. CO$_2$-rich atmospheres are compatible with that of Venus at the present day, both in composition and total pressure. The composition is also similar to that of modern Mars, hinting that this mode of atmosphere formation may have been commonplace on rocky planets in our solar system. That the Earth does not currently possess a 97:3 CO$_2$--N$_2$ atmosphere points to the operation of geochemical cycles that drew down carbon from the atmosphere through weathering by the Urey reaction to form carbonates \citep{sleep2001,zahnle2007} that are subsequently recycled into the interior, thus evolving its atmosphere away from CO$_2$-rich over geological timescales. This is because carbon is abundant, CO$_2$ is strongly favored at 280 K, and water condensation buffers the emergence of hydrogen-dominated atmospheres even when total C/H is low (as with atmospheres derived from a partially molten starting state, Appendix~\ref{app:trappist1e}).

\begin{figure}
%\figurenum{2}
\includegraphics[width=1\textwidth]{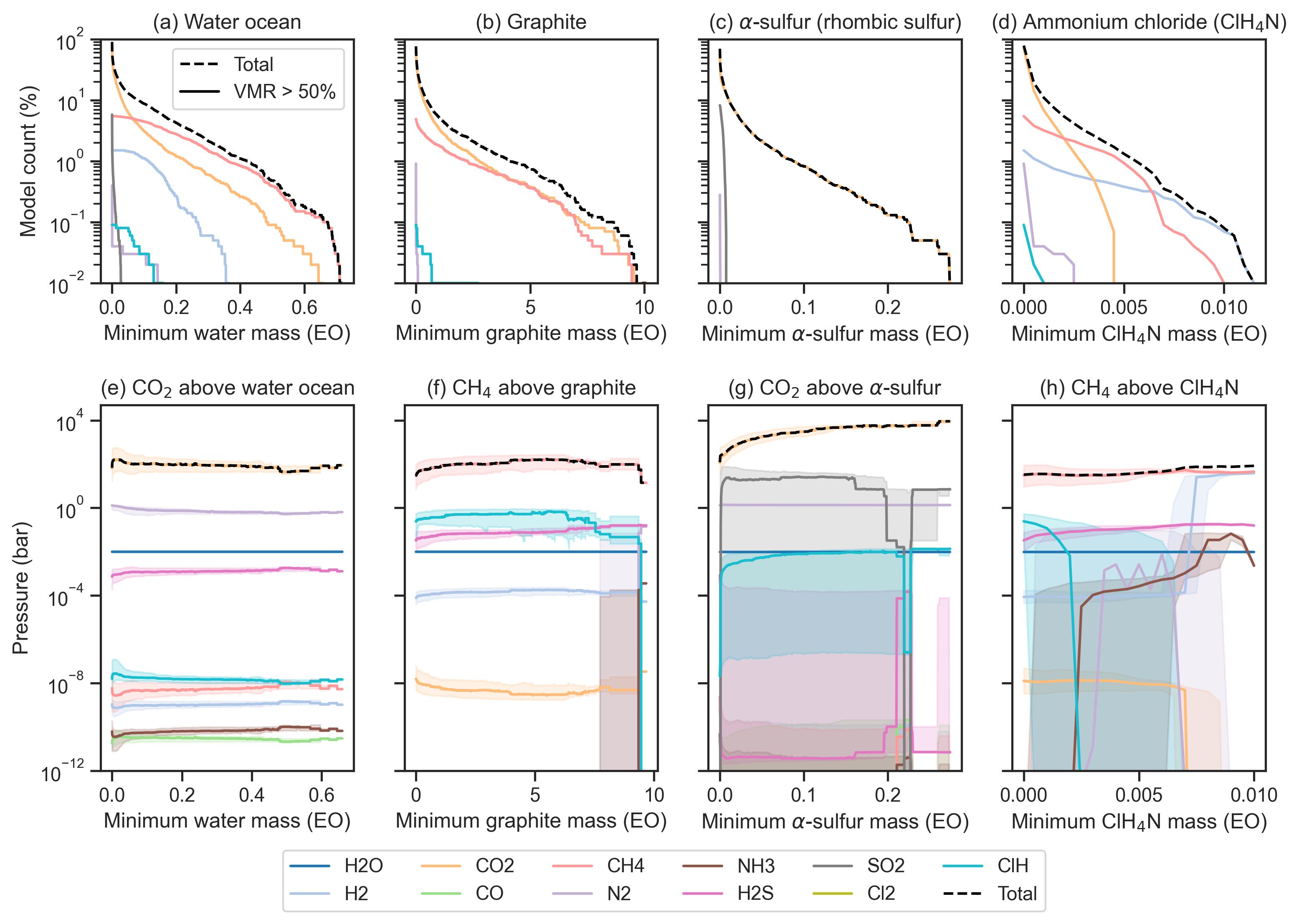}
\caption{Atmospheric speciation of TRAPPIST-1e at 280 K above a planetary surface with stable condensates for atmospheres originally in equilibrium with a fully molten mantle. In all panels, the curve colors correspond to the gas species listed in the legend and condensate masses are relative to Earth oceans (EO). Upper panels show the percentage of models by dominant species (VMR $>$ 50\%) that satisfy the requirement of a minimum mass of (a), water, (b), graphite, (c) $\alpha$-sulfur, and (d) ammonium chloride. Lower panels illustrate the composition and total pressure of the atmosphere for (e) CO$_2$-rich above a water ocean, (f) CH$_4$-rich above graphite, (g) CO$_2$-rich above $\alpha$-sulfur, and (h) CH$_4$-rich above ammonium chloride. Median values are indicated by lines and shaded regions bracket the first and third quartiles. Compare to the atmospheric speciation derived from a partially molten mantle in Figure~\ref{fig:T1e_10_condensates}.}
\label{fig:T1e_100_condensates}
\end{figure}

% Elemental abundances
Condensates at planetary surfaces, such as water, graphite, $\alpha$-sulfur, and ammonium chloride, are of primary interest for the establishment of geochemical cycles \citep{hirschmann2018comparative,dasgupta2024} and a habitable environment \citep{benner2020,mrnjavac2023moon,rimmer2024}. Figures~\ref{fig:T1e_100_condensates} and \ref{fig:T1e_10_condensates} show atmospheres at 280 K above surfaces of an Earth-like planet together with stable condensates with which they are in equilibrium, derived from a fully molten and partially molten mantle, respectively. Each panel mandates that a particular condensate is stable with a minimum mass (shown on the $x$-axis), where the upper row reveals the number of models associated with a different dominant (VMR $>$ 50\%) species as a function of the number of Earth-ocean equivalents by mass (EO) condensed. The lower row shows the variety of atmospheric species for dominant classes of atmospheres that support condensates, where either CO$_2$ or CH$_4$ is chosen as the dominant species (Table~\ref{table:models_with_condensates}). The conditions for water oceans and graphite to be stable are comparable, with two classes of atmospheres emerging as determined by a VMR greater than 50\%: CO$_2$-rich and CH$_4$-rich, with the former representing 83\% of all atmospheres, whereas the latter only around 5\% (e.g., Figure~\ref{fig:T1e_100_condensates}). For both classes, the atmospheres have median total surface pressures of a few tens of bar, and due to buffering imposed by C(cr) and H$_2$O(l) condensates, the partial pressures of other C- and H-bearing species are relatively constant, regardless of the condensate reservoir mass.

Small amounts of water can be stable beneath atmospheres dominated by SO$_2$, N$_2$, H$_2$, and ClH, although these are all less prevalent than CO$_2$ and CH$_4$ (Figure~\ref{fig:T1e_100_condensates}a). Conditions that permit the condensation of solid ammonium chloride, ClH$_4$N(cr) mirror those for water and graphite, reflecting the greater availability of NH$_3$ and HCl in reduced atmospheres relative to their oxidized counterparts (Figure~\ref{fig:T1e_fully_molten_minor}). Ammonium chloride most commonly condenses in small mass fractions ($<$0.0025 EO) beneath CO$_2$-rich atmospheres, but the highest ClH$_4$N(cr) mass fractions ($>$0.006 up to 0.012 EO) occur, albeit rarely ($\sim$1\% of all models), in equilibrium with H$_2$. Conversely, $\alpha$-sulfur stability occurs almost exclusively beneath CO$_2$-rich atmospheres, in which the partial pressure of CO$_2$ is on average higher than in equilibrium with water and graphite (compare the orange lines in Figure~\ref{fig:T1e_100_condensates}e and g). The CO$_2$-rich atmospheres in equilibrium with $\alpha$-sulfur also frequently have several tens of bar of SO$_2$ and only 10$^{-11}$ bar of H$_2$S. Note that SO$_2$(g) is almost never the major gas phase above $\alpha$-sulfur because most of the S budget is hosted in the condensed phase. In total, water, graphite, $\alpha$-sulfur, and ammonium chloride coexist in around 40\% of the models, albeit at potentially low abundance. Nevertheless, our results show that many early atmospheres are conducive to the formation of condensates on temperate planetary surfaces (here 280~K), should these atmospheres be able to cool isochemically and reach chemical equilibrium with stable condensates.

\begin{figure}
\gridline{\fig{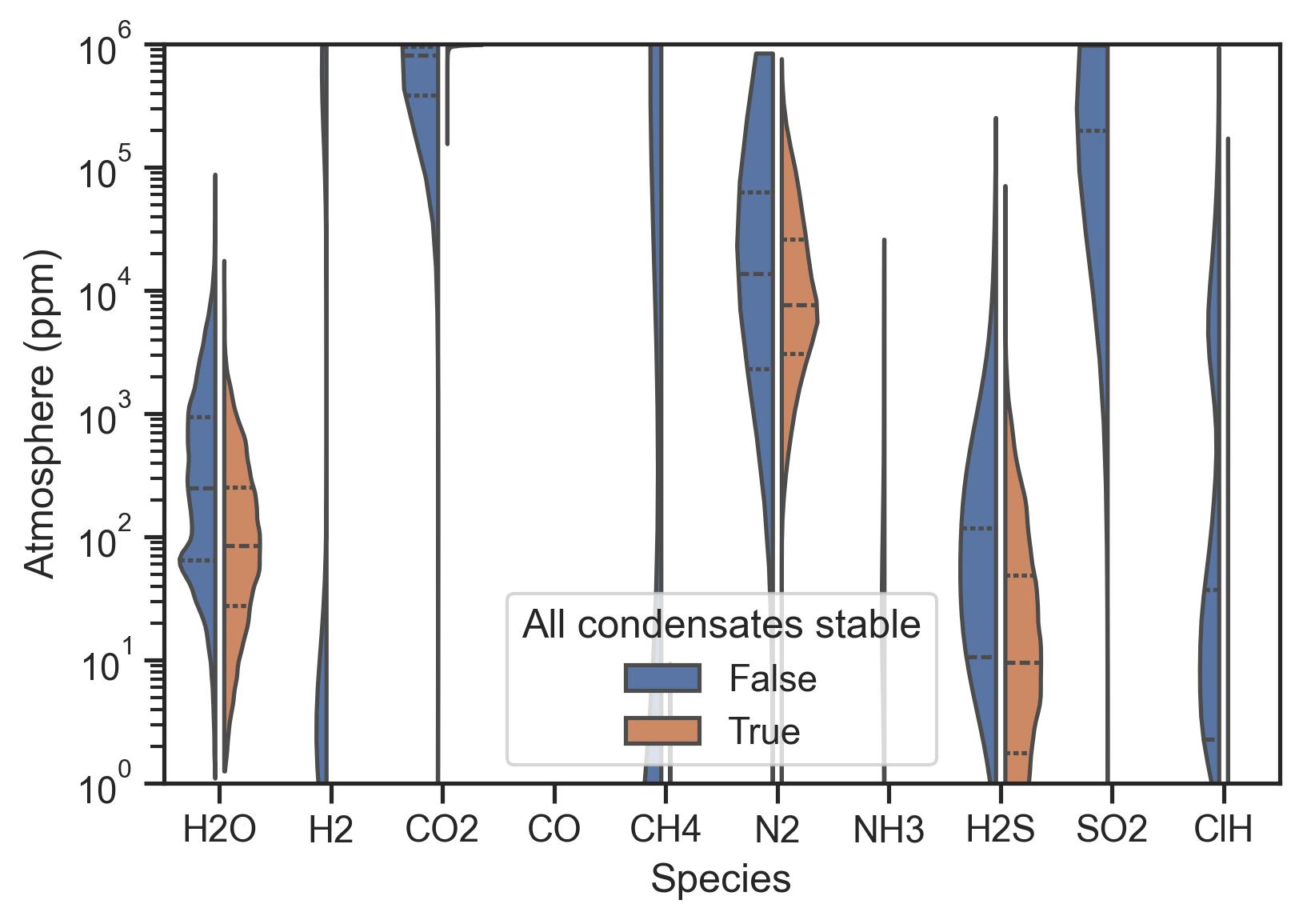}{0.475\textwidth}{(a) Atmospheres derived from a fully molten starting state from 1 to 10$^6$ ppm.}
\fig{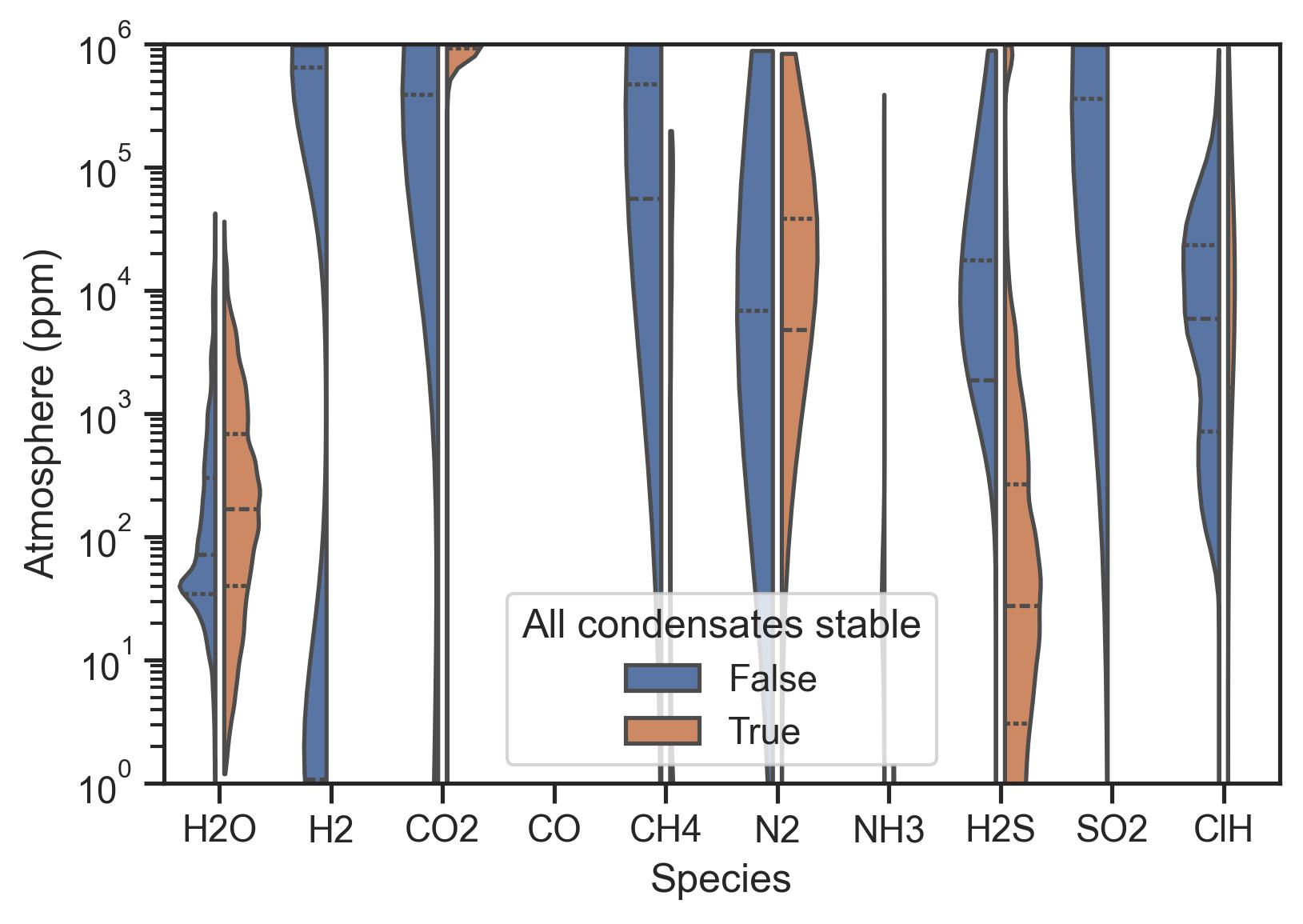}{0.475\textwidth}{(b) Atmospheres derived from a partially molten starting state from 1 to 10$^6$ ppm.}
}
\gridline{\fig{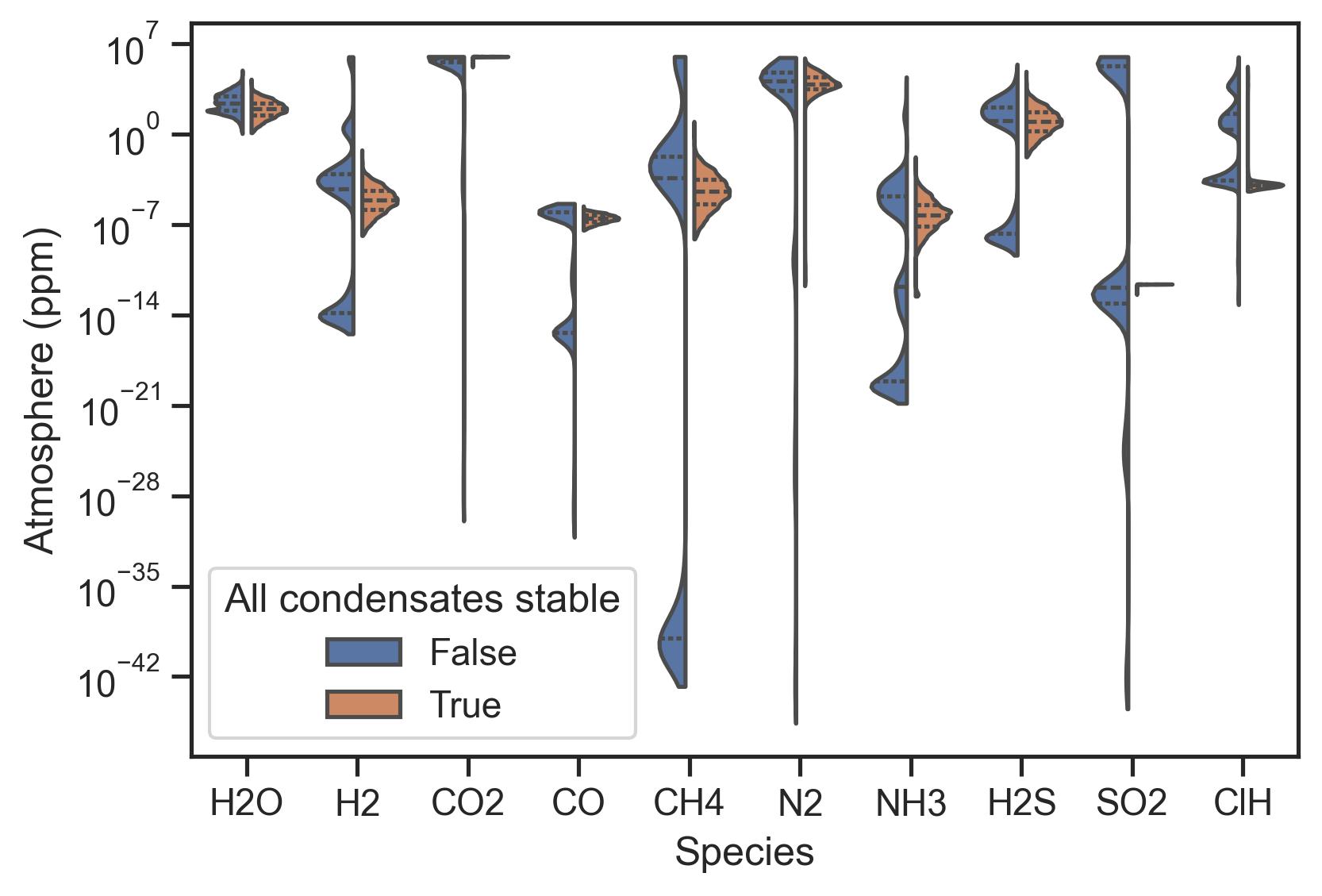}{0.475\textwidth}{(c) As for (a) and showing the full data range.}
\fig{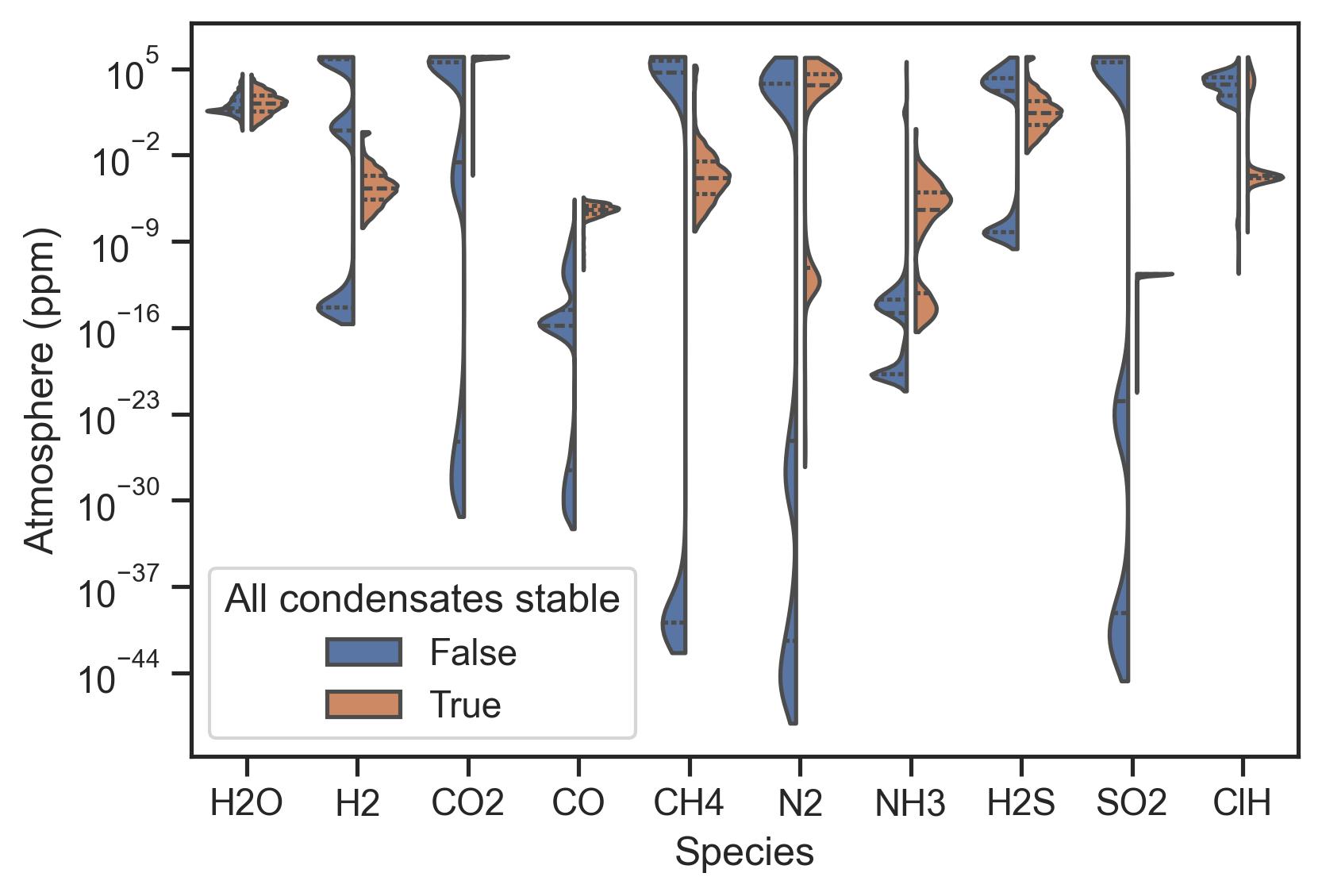}{0.475\textwidth}{(d) As for (b) and showing the full data range.}
}

\caption{Atmospheric speciation of TRAPPIST-1e at 280 K in equilibrium with stable reservoirs of water, graphite, $\alpha$-sulfur, and ammonium chloride. "True" indicates atmospheres above planetary surfaces where all four condensates coexist. Conversely, "False" means that at least one of the four condensates are not stable. From this classification, atmospheres that are more amiable to support surfaces primed for habitable environments can be distinguished. The width of each violin plot is proportional to the number of models, where interior dotted lines represent the first and third quartiles and dashed lines represent the second quartile (median).}
\label{fig:T1e_observe}
\end{figure}

Figure~\ref{fig:T1e_observe} summarizes the atmospheric abundances (ppm) above planetary surfaces that simultaneously support water, graphite, $\alpha$-sulfur, and ammonium chloride. The objective is to identify characteristic atmospheric signatures that may be used to identify planetary surfaces with a higher propensity for habitability \citep{krissansen2018detectability, schwieterman2019rethinking,konrad2024,tokadjian2024detectability,borges2024,young2024inferring}. To satisfy joint condensate stability, reduced species, such as H$_2$, NH$_3$ and CH$_4$ are low in abundance---almost always below the ppm level---because otherwise $\alpha$-sulfur does not form. This comes from the fact that these species are present in the atmosphere only at very low $f$O$_2$ at the high-temperature initial conditions, at which S is highly soluble. Therefore, S is never present in sufficient amounts in these atmospheres to be able to condense upon cooling to a low temperature. Conversely, oxidized initial conditions that produce significant SO$_2$ lead to final, elevated $f$O$_2$s of these atmospheres that prevent the condensation of graphite via CO$_2$(g) = C(cr) + O$_2$(g), even at high $p$CO$_2$. Consequently, any detection of SO$_2$ in an atmosphere is incompatible with the presence of graphite. On the other hand, SO$_2$ is highly unlikely to be the dominant atmospheric gas in the presence of H$_2$O(l) or $\alpha$-S(cr). Species whose fugacities are less redox sensitive cannot readily be used to discriminate between stable versus unstable condensates (e.g., H$_2$O and N$_2$). Figure~\ref{fig:T1e_observe_WG} illustrates the same analysis but excludes $\alpha$-sulfur and ammonium chloride, and hence only considers the joint stability of water and graphite. In this scenario, reduced species can be accommodated at higher abundances (e.g. H$_2$, NH$_3$), and notably CH$_4$-rich atmospheres are compatible with both water oceans and graphite at planetary surfaces.

Our analysis provides a geological baseline for the expected states of planetary atmospheres after they emerge from the magma ocean stage, postformation but prior to modification by geochemical cycling, atmospheric (photo)chemistry and escape, and later, possibly life. It should be noted that these computations assume chemical equilibrium is reached at 280 K in the absence of chemical modification of the atmosphere following the magma ocean stage. In practice, the timescales over which the atmospheres cool may be sufficiently short so as to permit chemical kinetics and photochemical processes \citep{arney2016, zahnle2020, drant2025}, in addition to continued volcanic outgassing \citep{ONS20,guimond2021low,liggins2022} and/or atmospheric escape/erosion \citep[e.g.,][]{krissansen2024erosion} to invalidate our dual assumptions of chemical equilibrium at a constant bulk composition. Hence, these results should be regarded as an end-member case, or to provide the initial atmospheric composition to inform the subsequent time evolution of cooling. We anticipate that our analysis can inform a prior expectation in modeling studies and statistical analyses, in which observations can be used to quantify whether a planetary atmosphere has evolved away from its earliest outgassed state.

The establishment of geological cycling may require an early reservoir of water at the surface or in the upper mantle \citep[e.g.,][]{TT17} to alter the rheology and partial melting of rocks \citep{K16}. In this regard, surfaces that emerge postformation with condensates, notably liquid water, may already be primed to cycle volatiles with the interior and thereby modify the size and composition of the atmosphere. Conversely, for surfaces that emerge without condensates, there could be a higher propensity to remain locked in an inactive state since geochemical cycling cannot initiate, at least not immediately. Table~\ref{table:models_with_condensates} reveals that most (91\%) Earth-like planets that emerge from the magma ocean stage have the prerequisite conditions for surficial water, with at least 56\% of planets also accommodating graphite. Since condensates buffer the atmospheric speciation (due to the equivalence of chemical potentials, an intensive variable), finding atmospheres that are compatible with stable condensates will not directly allow the size of those condensate reservoirs to be determined.

%%%
%%%
\subsection{Sub-Neptunes}
Our analysis permits new insights on two key aspects of the nature of sub-Neptunes: (1) the influence of atmospheres on their mass-radius relationships and demographics \citep[e.g.,][]{Bean2021, luque2022density} and (2) the identity and abundance of gas species in equilibrium with magma oceans, providing grounds for comparison with spectroscopic observational constraints \citep[e.g.,][]{MSC23,BRC24}.

\subsubsection{Nonideality and the "Fugacity Crisis"}
Regarding point 1, \cite{KFS19} proposed that a "fugacity crisis," in which the enhanced dissolution of H$_2$ at high surface pressures ($\sim$1--8 GPa) hinders atmospheric growth, can explain the drop-off in exoplanet abundance above $\sim$3 $R_{\oplus}$ observed in exoplanet demographics studies (i.e., the "radius cliff") \citep{FP18}; though see also \cite{SY22} for a different interpretation. Our work goes beyond the H$_2$ nonideality case considered in \cite{KFS19}, by incorporating additional volatile species (e.g., H$_2$O, CH$_4$, CO, and CO$_2$) along with their dissolution into the interior and real gas EOS. Despite the relatively large mass of envelopes (2\%--3\% of the planetary mass) required to explain the mass-radius relationships of sub-Neptune exoplanets in the absence of interior dissolution, these quantities are modest, on a molar basis, with respect to the proportion of condensed mass in the planet. As such, in the end-member case of a solar gas interacting with a rocky sub-Neptune interior, the equilibrium H/O and C/O ratios will be modified from their solar values \citep[$\sim$2000 and $\sim$0.55 by moles, respectively,][]{asplund2009} via the mutual solubility of O in the envelope and C and H in the mantle. Since the H/O and C/O ratios of nominally rocky material are $<<$1 \citep[H/O and C/O ratios in the BSE are $\sim$0.005 and $\sim$3.6 $\times$ 10$^{-4}$ by moles, respectively,][]{palmeoneill2014}, chemical equilibrium will result in a decrease in H/O and C/O in the envelope relative to solar values \citep[see also][]{young2023earth}.

Here, because only volatile element mass balance is considered (i.e., there is no notion of the chemical composition of the condensed phase other than for C, H, O, and N), this process is simulated by considering a higher $f$O$_2$ ($\Delta$IW = -3) in our sub-Neptune models than that imposed by the solar nebula gas \cite[$\Delta$IW = -6.5,][]{grossman2008redox}. Should $f$O$_2$ at the magma ocean--envelope interface fall below $\Delta$IW = -3.5, then H$_2$ predominates in the gas phase (envelope), even for C/H ratios 100$\times$ solar, though CH$_4$ remains a minor species (Figure~\ref{fig:k218b_fO2}). As noted by \cite{KFS19,SY22}, however, the experiments underpinning the solubility law for H$_2$ dissolution in silicate liquids were performed over a limited temperature (1673--1773~K) and pressure (up to 3 GPa) range \citep{HWA12}; the temperature range is notably much lower than the 3000 K assumed for equilibrium at the interface. Therefore, how the equilibrium constant of the reaction H$_2$(g) = H$_2$(l) depends on temperature, and pressures above 3 GPa, is unconstrained, yet is essential to assess the "fugacity crisis" hypothesis.

\subsubsection{Role of Nonideality in Inferring Atmospheric Chemistry}

A key feature of spectroscopic determinations of the speciation of gases in the atmospheres of temperature sub-Neptunes with equilibrium temperatures in the range 250--400 K, such as K2-18b ($T_\mathrm{eq}$ = 255~K) and TOI-270d ($T_\mathrm{eq}$ = 354 K), is the ubiquity of CO$_2$(g) and CH$_4$(g) \citep{MSC23,BRC24}. Mixing ratios of these species are of the order 1--2\% each, and their ratio, $x\mathrm{CH}_4/x\mathrm{CO}_2$ is roughly unity \citep{MSC23,BRC24}, with evidence for H$_2$O(g) at similar levels for TOI-270d but unresolved (below 10$^{-3}$) in K2-18b. In all cases, the background gas is inferred to be H$_2$. Although these transmission spectroscopy measurements probe the upper atmosphere \citep[near 10$^{-2}$--10$^{-4}$ bar,][]{BRC24,madhusudhan2025} and the speciation cannot therefore be directly compared with that predicted in our models, they provide first-order constraints on the chemical degrees of freedom permitted in sub-Neptune atmospheres. \added{Processes above the planetary surface, such as the stabilization of a supercritical water ocean \citep{Luu2024}, chemical kinetics in the gas phase \citep{glein2025deciphering}, photochemical reactions \citep{tsai2024,WBZ24,drant2025}, and production of aerosols \citep{jaziri2025}, can also modify the composition of the observable atmosphere from that expected at thermochemical equilibrium.}

Inferred metallicities in the atmosphere required to fit the observed CH$_4$ and CO$_2$ abundances are of the order 100--200$\times$ solar \citep{WBZ24,yang2024chemical}, which are consistent with 100$\times$ solar composition used in our simulations (Section \ref{sect:gasdwarf}). At these metallicities, we highlight that CH$_4$(g) is expected to be a major species. Consequently, the C/H of the bulk atmosphere is expected to be $\sim$2.5 by mass (Figure~\ref{fig:k218b_idealvsreal}), and its variation is antithetic with that of C/O, which increases monotonically from near zero to 300 as the fraction of H present as a function of planetary mass increases from zero to 3\% (at constant $\Delta$IW = -3). Inferred C/H and C/O ratios (by mass) from the measurements of \cite{MSC23,BRC24} are of the order 0.1--0.3 ($\sim$30--100$\times$ solar) and 0.02--0.5, respectively. Hence, both are lower than those predicted in our simulations for a real gas in equilibrium with a magma ocean (Figure~\ref{fig:k218b_idealvsreal}). Partitioning of carbon into a metal phase, which we do \added{not} explicitly model, could reduce the atmospheric C/O ratio \citep{WDS25}. \added{However, since metallicity (and hence C/O) is inferred directly from atmospheric observations and our models reproduce these metallicities, any additional C sequestered into the core is immaterial to the comparison.}

The lower C/O ratios observed than predicted by our models leads to another discrepancy---the absence of CO$_2$(g) in our simulations. As highlighted for terrestrial planetary atmospheres (Section \ref{sect:trappist1e}), however, any CO(g) that is stable at high temperatures (here, 3000~K) is partitioned between CO$_2$(g), C(cr). and/or CH$_4$(g) at low temperatures ($\sim$300--400 K, the temperatures in the photosphere recovered from transmission spectra) depending on the prevailing gas chemistry and total pressure. This implies substantial amounts of CO$_2$(g) could form in the upper atmosphere upon cooling. Therefore, future work should concentrate on how gas-condensed phase equilibria act to modify the speciation of sub-Neptune atmospheres throughout the atmospheric column. 

A simplified version of this exercise was carried out by \cite{yang2024chemical}, in which speciation at the pressures of observation in the atmospheres of K2-18b and TOI-270d were extrapolated to 200~bar. In order to match the CH$_4$/CO$_2$ proportions in the upper atmosphere, these authors inferred H$_2$/H$_2$O ratios at 200~bar to be roughly unity for K2-18b and 0.5 for TOI-270d. Assuming these ratios are representative of those at any magma ocean/atmosphere interface, this would correspond to $\Delta$IW in the range 0 to $-$1. These $f$O$_2$s are similar to, or more oxidized than, those inferred for small, differentiated rocky bodies in the solar system \citep{wadhwa2008}. Such high oxygen fugacities would seem at odds with the substantial quantities (1\%--3\% by mass) of solar-like gas inferred to have been accreted by sub-Neptunes. Indeed, we predict C/H ratios that are too high ($>$2.5) at these $f$O$_2$s, and H$_2$ would no longer be a predominant constituent of the atmosphere (Figure~\ref{fig:k218b_fO2}). Therefore, preliminary indications suggest that either (1) bulk planetary C/H ratios should be relatively low ($<$10$^{-1}$ by mass), either by processes that fractionate C from other atmosphere-forming elements during accretion relative to solar or its preferential sequestration into sub-Neptune cores, and/or (2) the lower atmosphere has a distinct chemical composition with higher C/H and C/O than does the upper atmosphere sensed by transmission measurements (e.g., by inefficient mixing or condensation of graphite in the atmospheric column), in order to maintain simultaneously low C/H (0.1--0.3) and C/O (0.02--0.5) ratios observed in temperate sub-Neptunes.

\section{Conclusions}

For Earth-sized rocky planets such as TRAPPIST-1e, the atmosphere during the magma ocean stage is likely CO-rich due to two key factors. First, CO(g) is thermodynamically favored at 1800 K across a broad range of oxygen fugacities. Second, carbon-bearing species are significantly less soluble than hydrogen-bearing species---particularly H$_2$O(g)---,leading to an elevated atmospheric C/H, often exceeding the bulk C/H by several orders of magnitude. As a result, hydrogen-bearing species are depleted in the atmosphere, even when the mantle is only partially molten. In a fully molten mantle, the further depletion of hydrogen by the sequestration of H$_2$O(g) into the liquid leads to atmospheres that are defined by entirely continuous mixing ratios of their constituent species, favoring atmospheres dominated by (1) heavy carbon species (predominantly CO(g)), (2) CO$_2$(g) $\pm$ SO$_2$(g)-bearing varieties at high $f$O$_2$s ($\Delta$IW = +3 to +4), or (3) H$_2$(g) $\pm$ CH$_4$(g)-bearing types at low $f$O$_2$ (below $\Delta$IW = -3). 

Based on the equilibrium results of atmospheres with molten mantles, we cool the atmospheres to 280 K and permit (if thermodynamically stable) the formation of water (oceans), graphite, $\alpha$-sulfur, and ammonium chloride. Unlike at high temperatures, the presence of condensates buffers gas mixing ratios to produce discrete "families" of atmospheres. Notably, CO$_2$-rich atmospheres are not only widespread (83\% of cases) but also uniquely capable of supporting a wider variety of surface condensates compared to those dominated by other species. A small fraction (5\%) are CH$_4$-rich atmospheres that can sustain both water and graphite, while roughly 8\% are CO$_2$--SO$_2$-rich in equilibrium with native sulfur. The remainder ($\sim$1\%) are H$_2$-rich. Depending on the conditions during early equilibration with a magma ocean, atmospheres rich in N$_2$, HCl, or H$_2$S can also host diverse surface condensates; however, such cases comprise only $\sim$1\% of all cases. In short, water condensation buffers the partial pressures of hydrogen-bearing species in the atmosphere, such that even low C/H does not produce substantial hydrogen atmospheres except at the most reduced conditions where H$_2$ can dominate. Thus, for both high and low temperature atmospheres, carbon always wins.

Our simulations can be considered as geological reference cases---the expected state of a rocky planet after its earliest evolution. From this state, subsequent processes such as geological outgassing and atmospheric (photo)chemistry, dynamics, and escape can modify the atmosphere over geological timescales. Hence, our results can be used to inform the priors and initial conditions of other models, as well as provide an observational baseline for rocky planets to determine if they have evolved away from their birth state. In this regard, the simulations can inform observational studies and instrument requirements to investigate the atmospheres of temperate rocky planets.

For sub-Neptune-sized planets, we performed nonideal gas-phase chemistry simulations including dominant C- and H-bearing species at conditions relevant to the magma ocean--envelope interface. The combination of solubility and nonideality conspire to stabilize CH$_4$ as a major gas species for a range of parameter choices that are reasonable given current observational constraints for sub-Neptunes. As for Earth-sized rocky planets, this result is a natural outcome of the high solubility of hydrogen-bearing species (notably molecular hydrogen for sub-Neptunes) compared to carbon-bearing species, a result that is exacerbated due to nonideality at the high-temperature (3000~K) and pressure (several GPa) conditions assumed to occur at the interface. Furthermore, the transition between a H$_2$ and CH$_4$ atmosphere can occur within the range of observational uncertainties in mass and radius, even for a fixed volatile budget and $f$O$_2$, thereby providing a new challenge to infer atmospheric composition. Given detections of carbon-bearing species in the atmospheres of sub-Neptunes, such as K2-18b and TOI-270d, our findings emphasize the need to include and realistically model gas envelopes throughout their extent.

To achieve these results, we devised an open-source code \atmodeller{}, built on JAX, to probe the nature of atmospheres above rocky planets from Earth size to sub-Neptune size. The code includes solubility laws, real gas EOS, and thermodynamic data. Leveraging the functionality of the JAX Python library allows the code to be both user friendly and highly performant, and community development of the code is encouraged.

\section*{acknowledgments}
We thank Daniel Kitzmann, Patrick Kidger, and Johanna Haffner for discussions. D.J.B., M.A.T., and P.A.S. were supported by the Swiss State Secretariat for Education, Research and Innovation (SERI) under contract No. MB22.00033, a SERI-funded ERC Starting grant ``2ATMO." P.A.S. also thanks the Swiss National Science Foundation (SNSF) through an Eccellenza Professorship (\#203668). M.A.T. is supported by NASA through the NASA Hubble Fellowship grant \#HST-HF2-51545 awarded by the Space Telescope Science Institute, which is operated by the Association of Universities for Research in Astronomy, Inc., for NASA, under contract NAS5-26555. K.H. acknowledges the FED-tWIN research program STELLA (Prf-2021-022) funded by the Belgian Science Policy Office (BELSPO) and the FWO research grant G014425N. M.T. acknowledges financial support from the Chair of Theoretical Astrophysics of Extrasolar Planets at LMU Munich. \added{We thank the reviewer for their observant and constructive comments, which have enhanced the clarity of the manuscript.}


%\vspace{5mm}

%\facilities{HST(STIS), Swift(XRT and UVOT), AAVSO, CTIO:1.3m,


%\software{astropy \citep{2013A&A...558A..33A,2018AJ....156..123A},  
  %        Cloudy \citep{2013RMxAA..49..137F}, 
  %        Source Extractor \citep{1996A&AS..117..393B}
  %        }
\software{\atmodeller{} v1.0.0 (\url{https://github.com/ExPlanetology/atmodeller}). In addition, the data used in this study is issued under the GNU General Public License (GPL) 3.0 and available to download \citep{atmodeller_data}.}



% Force appendices content to appear in the appendices (i.e. tables etc.)
\clearpage

\appendix

\section{Sampling Strategy and Parameter Choices}
\label{app:sampling}
%\explain{New appendix A added in response to reviewer main comment 1}

The logarithmic sampling strategy adopted to explore the atmospheric diversity of TRAPPIST-1e is physically motivated as follows. Oxygen fugacity ($f$O$_2$) exerts a fundamental control on atmospheric composition, particularly through its influence on redox speciation; for example, determining the relative abundances of reduced versus oxidized species such as H$_2$ and H$_2$O. Thermodynamically, $f$O$_2$ is defined by an equilibrium reaction, and when referenced to Fe--FeO (IW) equilibrium (Equation~\ref{eq:IW}) is conventionally expressed as a $\log_{10}$ offset relative to the IW buffer. This convention naturally supports the use of a log-uniform prior, especially within a thermochemical framework where partial pressures, fugacities, and equilibrium constants combine multiplicatively (and thus additively in log-space). Observationally, the present-day $f$O$_2$s of rocky bodies in the solar system spans several orders of magnitude from roughly 5 $\log_{10}$ units below IW (Mercury) to 2 $\log_{10}$ units above IW (carbonaceous chondrites) \citep[e.g.,][]{Grewal2024}, with several extrasolar rocky bodies showing comparable redox states \citep{Doyle19}. Early Earth conditions during the magma ocean stage were close to IW \citep{SBB20}.

Therefore, to capture this diversity, we center our log-uniform prior on zero offset relative to IW (anchored to early Earth) and allow sampling across a broad range of $\pm 5 \log_{10}$ units. While no prior is without limitations, we adopt this scale-invariant formulation because it more faithfully represents ignorance over the parameter’s order of magnitude. It offers the dual advantages of anchoring to a well-motivated reference point (early Earth) while permitting exploration beyond the present-day solar system constraints on $f$O$_2$, which is warranted by the potentially greater diversity of exoplanetary environments. For consistency with the log-uniform prior, we apply the same sampling strategy to the other two parameters: the total hydrogen inventory and the C/H mass ratio are each varied across one order of magnitude on either side of an Earth-like reference value. The upper bound of the total hydrogen inventory is motivated by planet formation simulations, which show that terrestrial planets can accrete several times Earth's water content \citep{RQL07b}. The lower bound, by contrast, reflects scenarios of limited delivery or substantial early loss.

The corner plots display bivariate scatter plots (e.g., Figure~\ref{fig:T1e_fully_molten}), which directly reflect the model outcomes and are therefore insensitive to the choice of prior, depending only on the sampled parameter ranges. In contrast, the likelihood contours, marginal distributions, and derived summary statistics (e.g., Table~\ref{table:models_with_condensates}) are prior dependent. A uniform distribution might appear to be a simple alternative prior, but its inadequacy for $f$O$_2$ is evident: when sampled uniformly between $10^{-5}$ and $10^5$, the mean value in $\log_{10}$ space is $\sim$4.7, which is heavily skewed toward the upper bound and far outside the range characteristic of the early Earth or any known rocky body in the solar system. New observations and models may eventually reveal the true distributions of the parameters that govern atmospheric diversity, but, in the absence of such knowledge, a logarithmic sampling strategy remains the most reasonable and defensible choice.

% Ensure tables and figures in the appendix are labeled correctly
% Might not be required for the ApJ style?
\renewcommand{\thetable}{\Alph{section}\arabic{table}}
\renewcommand{\thefigure}{\Alph{section}\arabic{figure}}

\section{Partially molten mantle}
\label{app:trappist1e}
\setcounter{figure}{0}
\setcounter{table}{0}
\setcounter{equation}{0}

\begin{figure}
%\figurenum{1}
\includegraphics[width=1\textwidth]{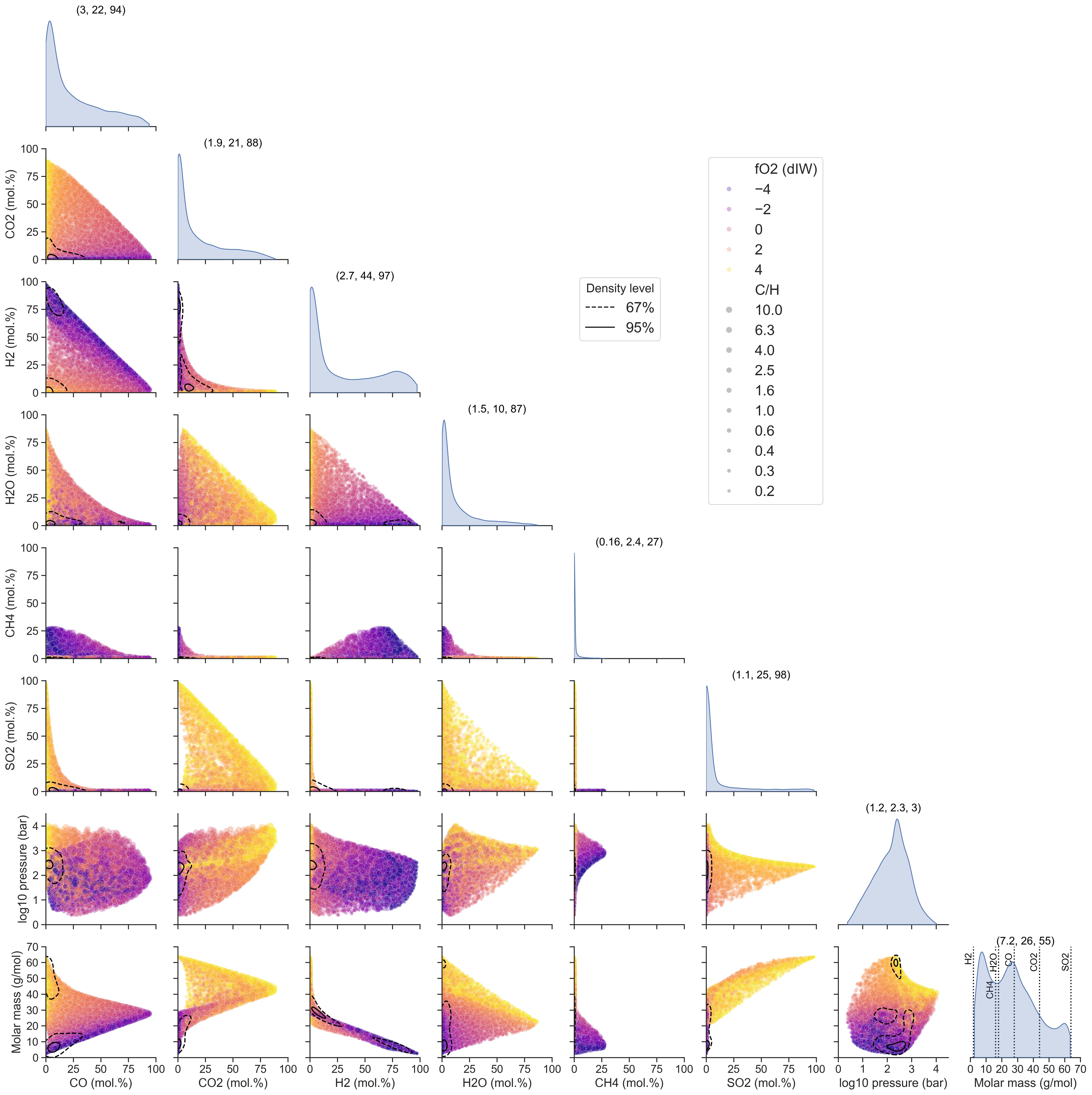}
\caption{For an early TRAPPIST-1e with a partially molten mantle (10\% melt fraction, surface temperature of 1800 K), scatter plots of atmospheric molar abundance, total pressure, and molar mass for major species. Points are colored by oxygen fugacity expressed relative to the IW buffer and sized in proportion to log$_{10}$ C/H. Density levels indicate areas with high likelihood in the scatter plots while marginal distributions are shown on the diagonal. The 10th, 50th (median), and 90th percentiles of these distributions are also annotated above the marginal distributions as (p10, p50, and p90), respectively. Compare to a fully molten mantle in Figure~\ref{fig:T1e_fully_molten}, noting that the extent of the axes for H$_2$O and CH$_4$ is different.}
\label{fig:T1e_partially_molten}
\end{figure}

\begin{figure}
\includegraphics[width=1\textwidth]{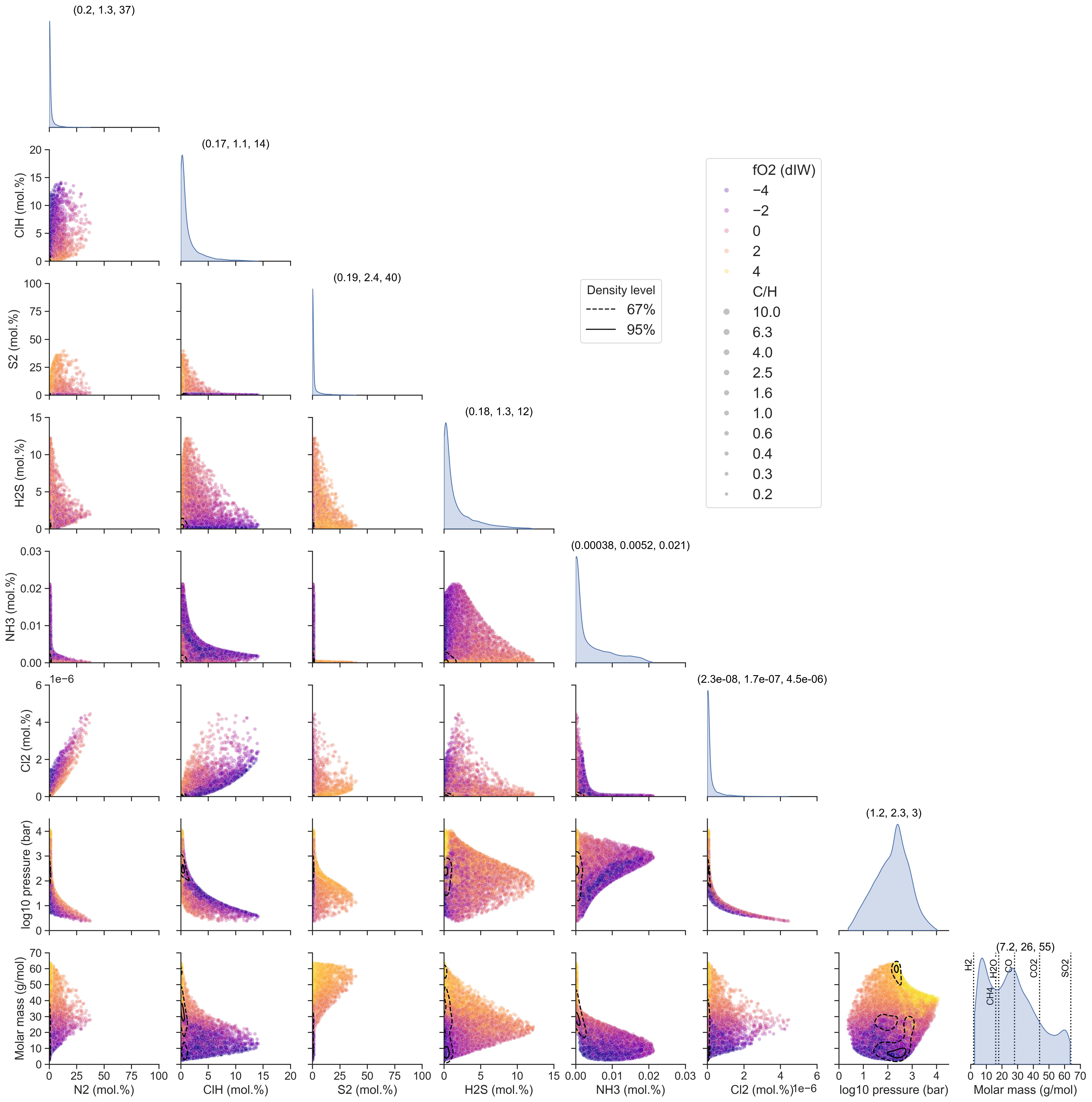}
\caption{For an early TRAPPIST-1e with a partially molten mantle (10\% melt fraction, surface temperature of 1800 K), scatter plots of atmospheric molar abundance, total pressure, and molar mass, for N$_2$ and minor species: ClH, S$_2$, H$_2$S, NH$_3$, and Cl$_2$. Points are colored by oxygen fugacity expressed relative to the IW buffer and sized in proportion to log$_{10}$ C/H. Density levels indicate areas with high likelihood in the scatter plots, while marginal distributions are shown on the diagonal. The 10th, 50th (median), and 90th percentiles of these distributions are also annotated above the marginal distributions as (p10, p50, and p90), respectively. Compare to a fully molten mantle in Figure~\ref{fig:T1e_fully_molten_minor}.}
\label{fig:T1e_partially_molten_minor}
\end{figure}

\begin{figure}
%\figurenum{3}
\includegraphics[width=1\textwidth]{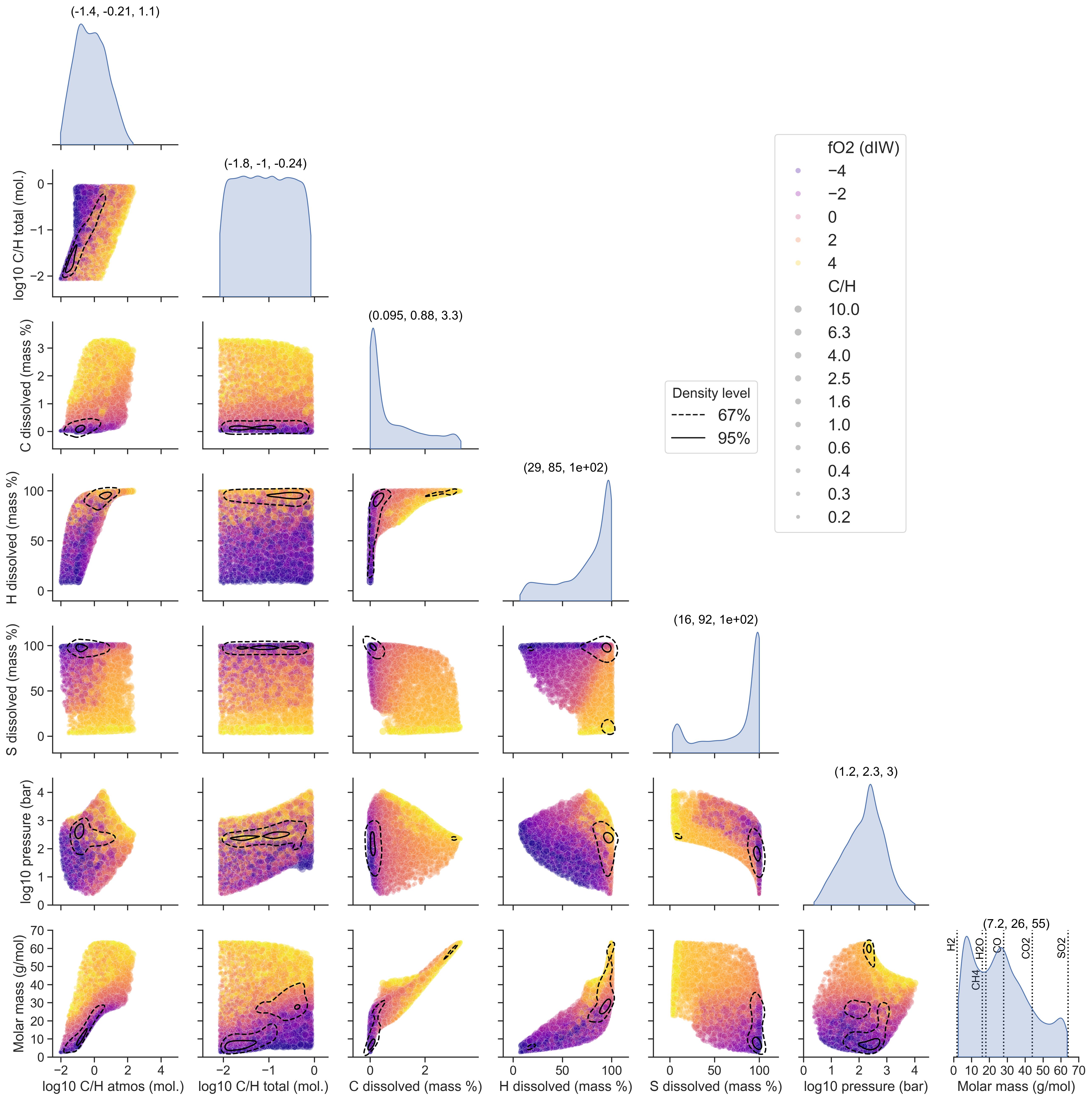}
\caption{For an early TRAPPIST-1e with a partially molten mantle (10\% melt fraction, surface temperature of 1800 K), scatter plots of atmospheric and total C/H, C, H, and S dissolved in the interior relative to the total inventory of the element by mass, total pressure, and molar mass. Points are colored by oxygen fugacity and sized in proportion to log$_{10}$ C/H. Density levels indicate areas with high likelihood in the scatter plots, while marginal distributions are shown on the diagonal. The 10th, 50th (median), and 90th percentiles of these distributions are also annotated above the marginal distributions as (p10, p50, and p90), respectively. Compare to a fully molten mantle in Figure~\ref{fig:T1e_fully_molten_ratios}.}
\label{fig:T1e_partially_molten_ratios}
\end{figure}

\begin{figure}
%\figurenum{1}
\includegraphics[width=1\textwidth]{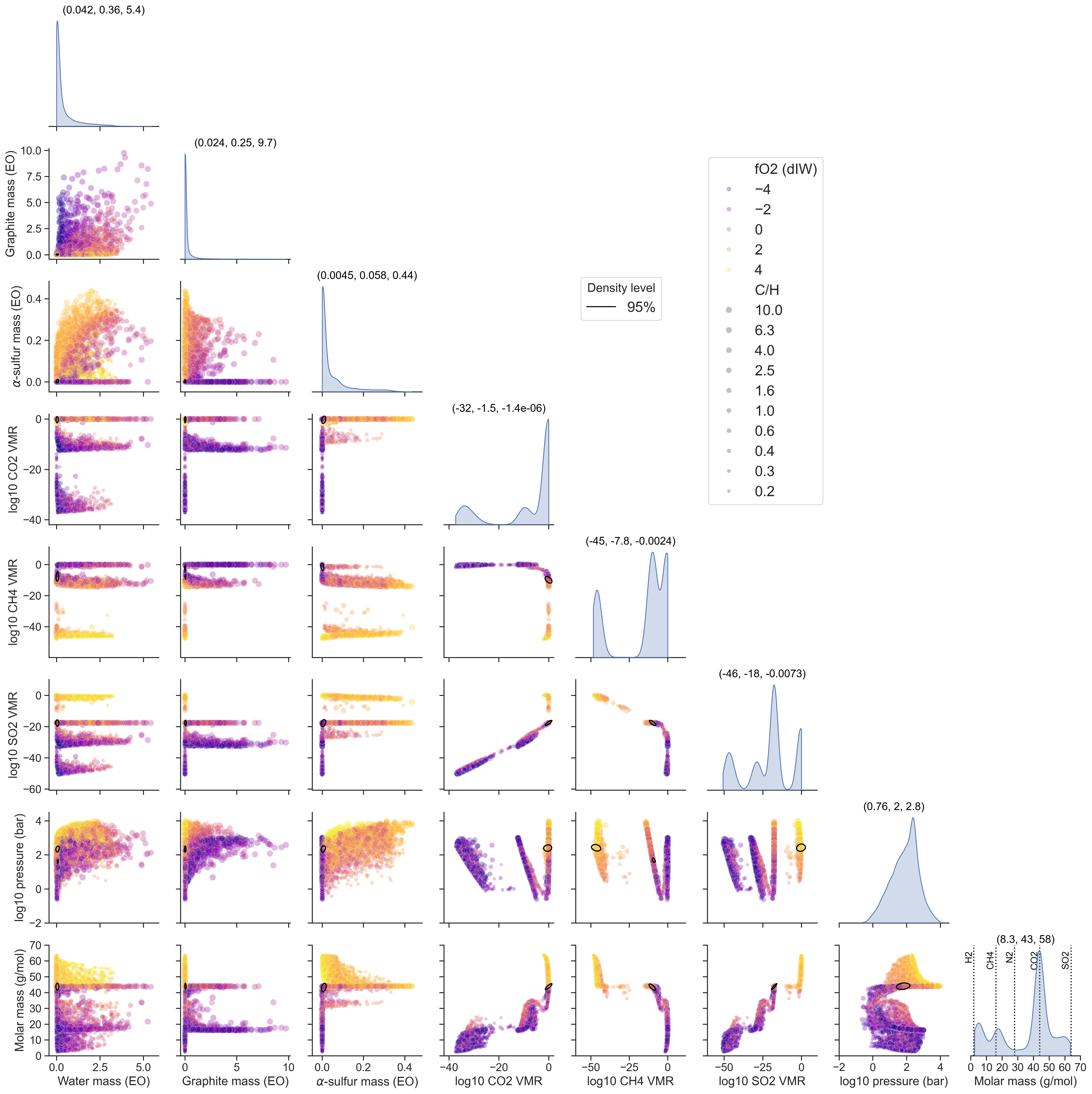}
\caption{Cooled atmospheres at 280 K for TRAPPIST-1e, starting with the elemental abundances in the atmosphere for a partially molten mantle (10\% melt fraction) and a surface temperature of 1800 K. Points are colored by oxygen fugacity of the initial, high-temperature atmosphere expressed relative to the IW buffer and sized in proportion to log$_{10}$ C/H for the partially molten starting state. Condensed inventories of water, graphite, and $\alpha$-sulfur are normalized by the Earth's present-day ocean mass (EO). Other scatter plots show the VMR of major species in the atmosphere, total pressure, and molar mass. Density levels indicate areas with high likelihood in the scatter plots, while marginal distributions are shown on the diagonal. The 10th, 50th (median), and 90th percentiles of these distributions are also annotated above the marginal distributions as (p10, p50, and p90), respectively. Compare to the cool atmospheres derived from a fully molten mantle in Figure~\ref{fig:T1e_fully_molten_condensed}.}
\label{fig:T1e_partially_molten_condensed}
\end{figure}

We repeat the analysis presented in Section~\ref{sect:trappist1e} for a partially molten TRAPPIST-1e with a mantle melt fraction of 10\%, whereby the influence of solubility is lessened compared to a fully molten mantle (compare Figures~\ref{fig:T1e_fully_molten}, \ref{fig:T1e_fully_molten_minor}, \ref{fig:T1e_fully_molten_ratios}, \ref{fig:T1e_fully_molten_condensed} to Figures~\ref{fig:T1e_partially_molten}, \ref{fig:T1e_partially_molten_minor}, \ref{fig:T1e_partially_molten_ratios}, \ref{fig:T1e_partially_molten_condensed}, respectively). Even with just 10\% melt fraction, C/H in the atmosphere is boosted and its distribution broadened compared to total C/H (Figure~\ref{fig:T1e_starting_CH_ratio}). Compared to a fully molten mantle, a partially molten mantle permits a greater diversity of atmospheres since a broader range of gas mixtures comprised of hydrogen and carbon species are possible. Although larger H$_2$O atmospheres can occur, the median H$_2$O in the atmosphere is still only 11 mol\% versus 1 mol\% for a fully molten mantle. The mean molar mass (Figure~\ref{fig:T1e_partially_molten}, bottom right) reflects this greater atmospheric diversity, where a subsidiary peak around 5 g mol$^{-1}$ represents H$_2$-dominated atmospheres at reduced conditions in a mixture with CH$_4$ and CO. The CO peak at 28 g mol$^{-1}$ is evident, albeit now supplemented by mixtures of H$_2$O (18 g mol$^{-1}$) and CO$_2$ (44 g mol$^{-1}$) in approximately equal proportion. The molar mass distribution also has an extended tail, driven by the prevalence of CO across a broad range of oxygen fugacity as well as the emergence of CO$_2$ (44 g mol$^{-1}$) and SO$_2$ (64 g mol$^{-1}$) as prevalent oxidized species. Regardless, the thermodynamic stability of CO over a range of $f$O$_2$ ensures that it remains a dominant species.

%%%
%%%

%%%
%%%

%%%
%%%

For a partially molten mantle, dissolved H is more than 60\% for models above the IW buffer due to abundant H$_2$O which is soluble. For reduced conditions below the IW buffer, H$_2$ is more abundant than H$_2$O, and, since H$_2$ is not as soluble as H$_2$O, less total hydrogen is dissolved, between about 10\% and 60\%. For comparison, dissolved C never exceeds around 3\% regardless of whether the reduced (CO) or oxidized (CO$_2$) species is most abundant. We recall that total C/H is an input parameter, which is varied between -1 and 1 log$_{10}$ units by mass, which corresponds approximately to between -2 and 0 log$_{10}$ units by moles. This gives rise to a median log$_{10}$ C/H in the atmosphere of -0.2 and C/H can reach several hundred for the most oxidized conditions. Hence even with a small mantle melt fraction (10\%) total C/H can differ from atmospheric C/H by around two orders of magnitude.

%%%
%%%

Figure~\ref{fig:T1e_partially_molten_condensed} shows the atmospheres at 280 K derived from the partially molten starting state. Compared to those derived from the fully molten starting state, the atmospheres exhibit the same complex relationships, albeit the solution space is enlarged for certain quantities. This is because the smaller initial C/H allows H$_2$ and CH$_4$ atmospheres to be more prevalent at reduced starting conditions which can produce mixtures of hydrogen and carbon species. Nevertheless, CO$_2$ atmospheres are dominant in similarity with atmospheres derived from the fully molten starting state. H$_2$O is only present in the atmosphere below the percent level since water condenses to produce a median surficial reservoir of 0.4 Earth oceans by mass (EO) with a maximum inventory around 5.5 EO, which is almost a tenfold increase compared to the fully molten starting condition (Figure~\ref{fig:T1e_fully_molten_condensed}). $\alpha$-sulfur condensation can produce up to around 0.4 EO, although an expected amount is closer to 0.1 EO. In contrast, the total graphite reservoir is usually around 0.3 EO, but for reduced starting conditions can reach up to 10 EO. Large reservoirs of all three condensates can coexist, usually for starting conditions around the IW buffer.

\begin{figure}
%\figurenum{2}
\includegraphics[width=1\textwidth]{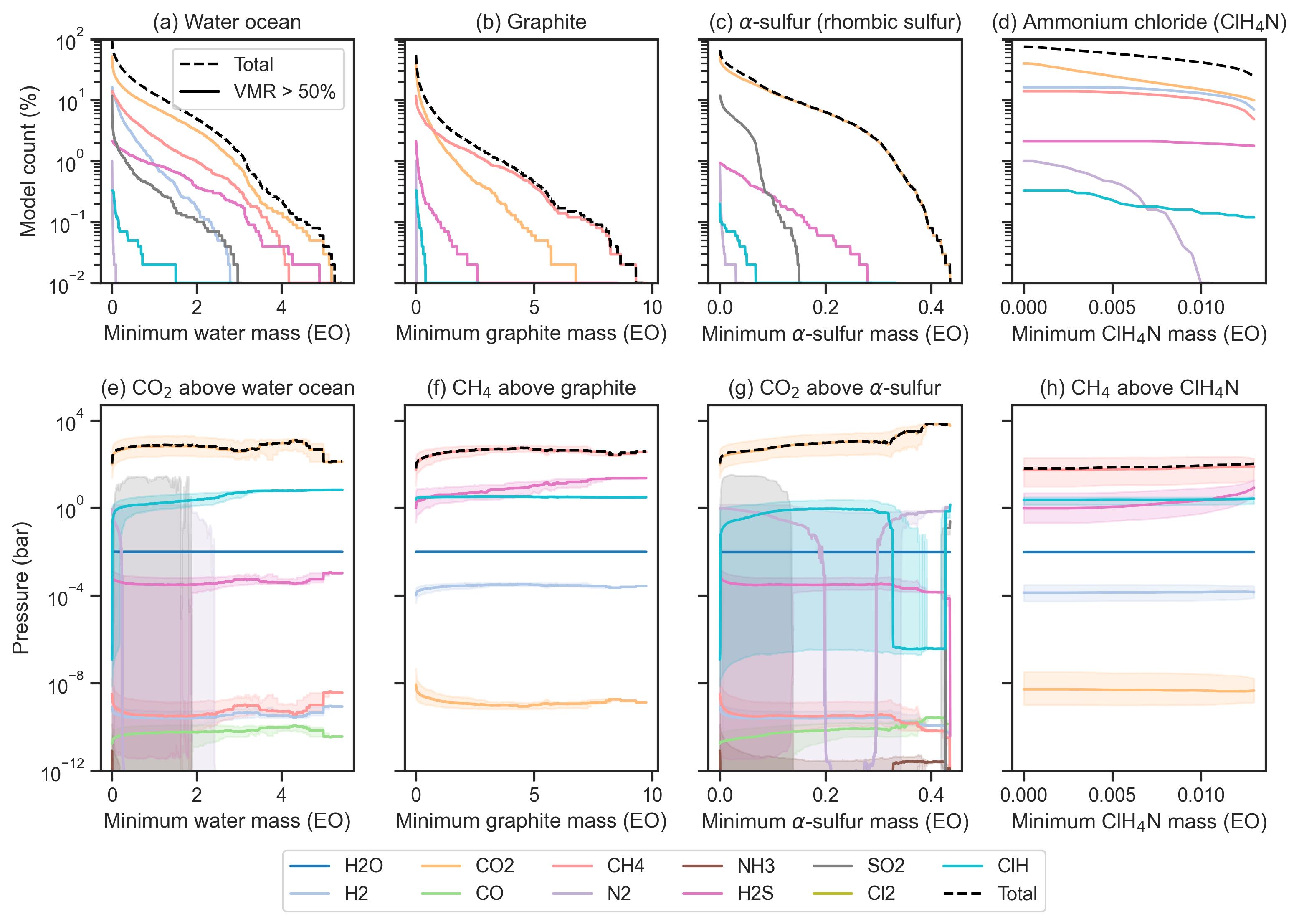}
\caption{Atmospheric speciation of TRAPPIST-1e at 280 K above a planetary surface with stable condensates for atmospheres originally in equilibrium with a partially molten mantle (10\% melt fraction). In all panels, the curve colors correspond to the gas species listed in the legend and condensate masses are relative to Earth oceans (EO). The upper panels show the percentage of models by dominant species (VMR $>$ 50\%) that satisfy the requirement of a minimum mass of (a) water, (b) graphite, (c) $\alpha$-sulfur, and (d) ammonium chloride. Lower panels illustrate the composition and total pressure of the atmosphere for (e) CO$_2$-rich above a water ocean, (f) CH$_4$-rich above graphite, (g) CO$_2$-rich above $\alpha$-sulfur, and (h) CH$_4$-rich above ammonium chloride. Median values are indicated by lines and shaded regions bracket the first and third quartiles. Compare to the atmospheric speciation derived from a fully molten mantle in Figure~\ref{fig:T1e_100_condensates}.}
\label{fig:T1e_10_condensates}
\end{figure}

\begin{figure}
\gridline{\fig{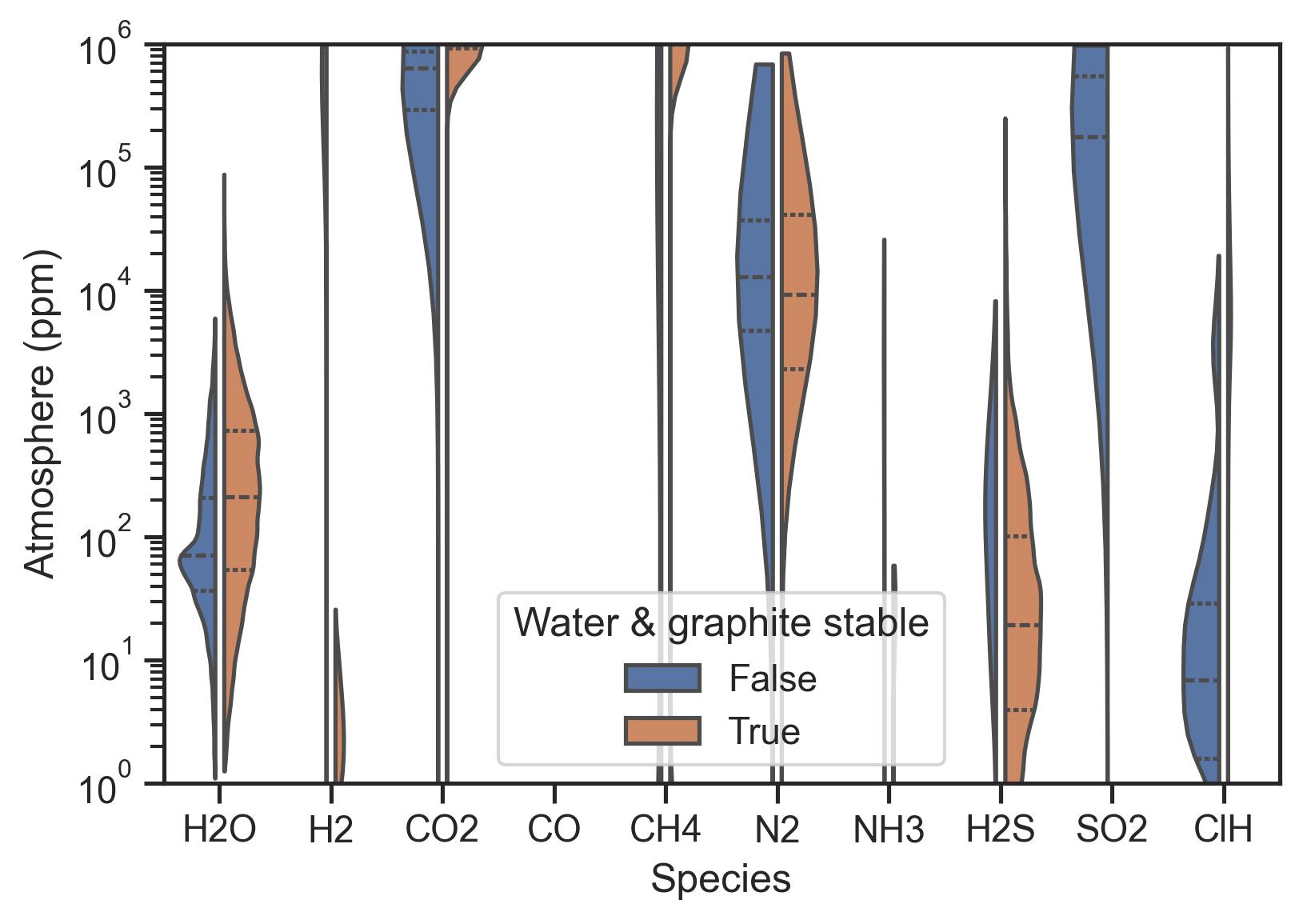}{0.475\textwidth}{(a) Atmospheres derived from a fully molten starting state from 1 to 10$^6$ ppm.}
\fig{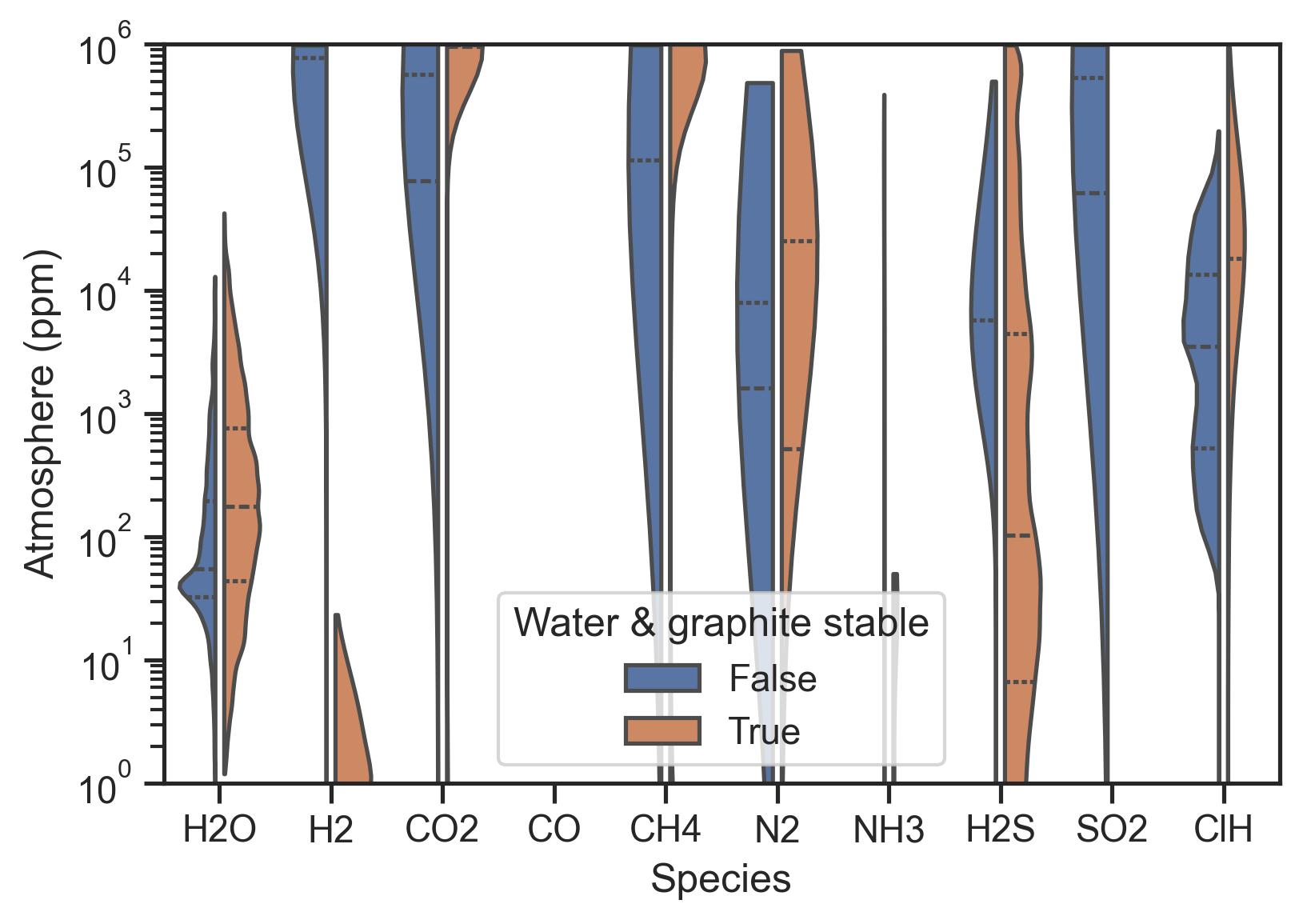}{0.475\textwidth}{(b) Atmospheres derived from a partially molten starting state from 1 to 10$^6$ ppm.}
}
\gridline{\fig{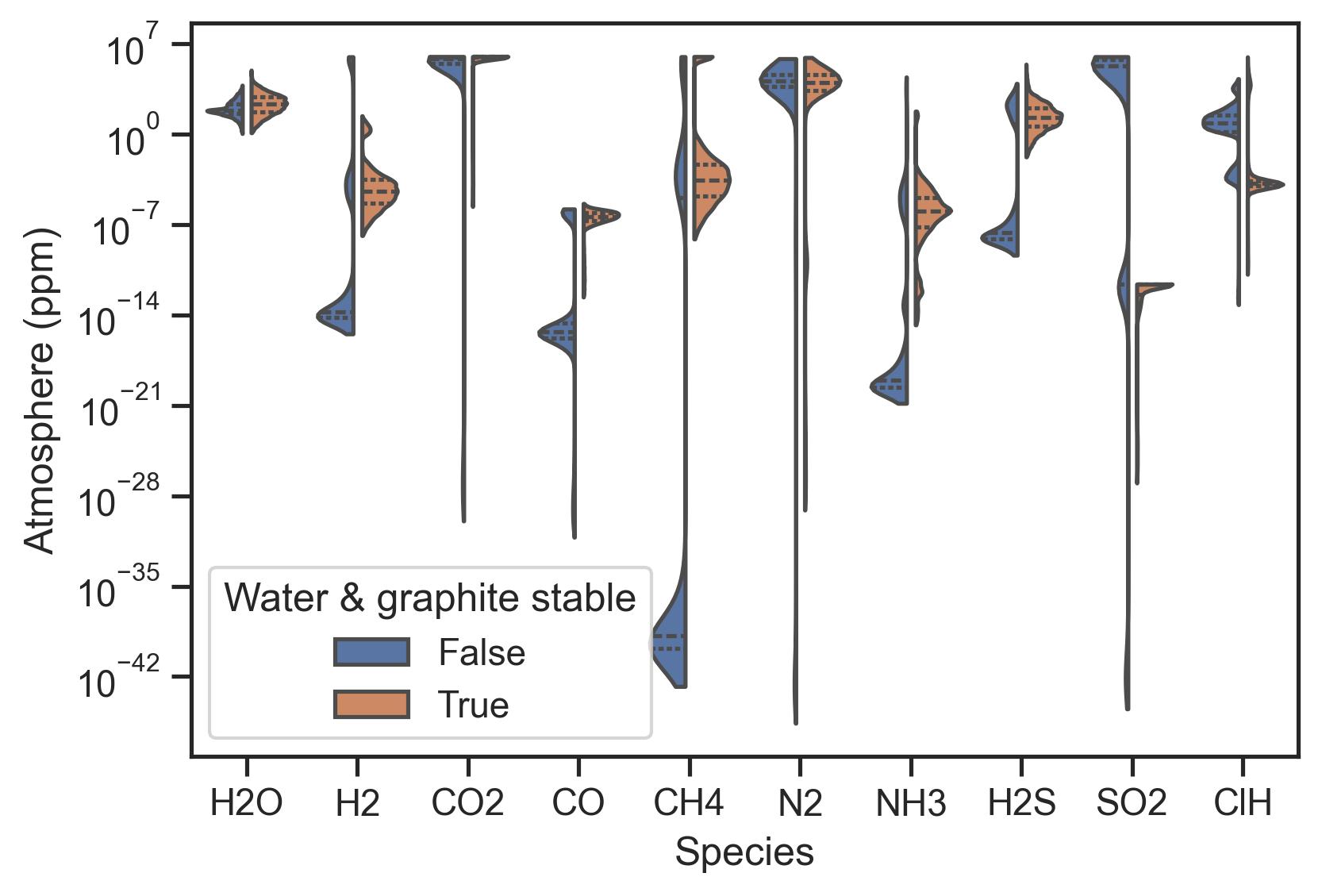}{0.475\textwidth}{(c) As for (a) and showing the full data range.}
\fig{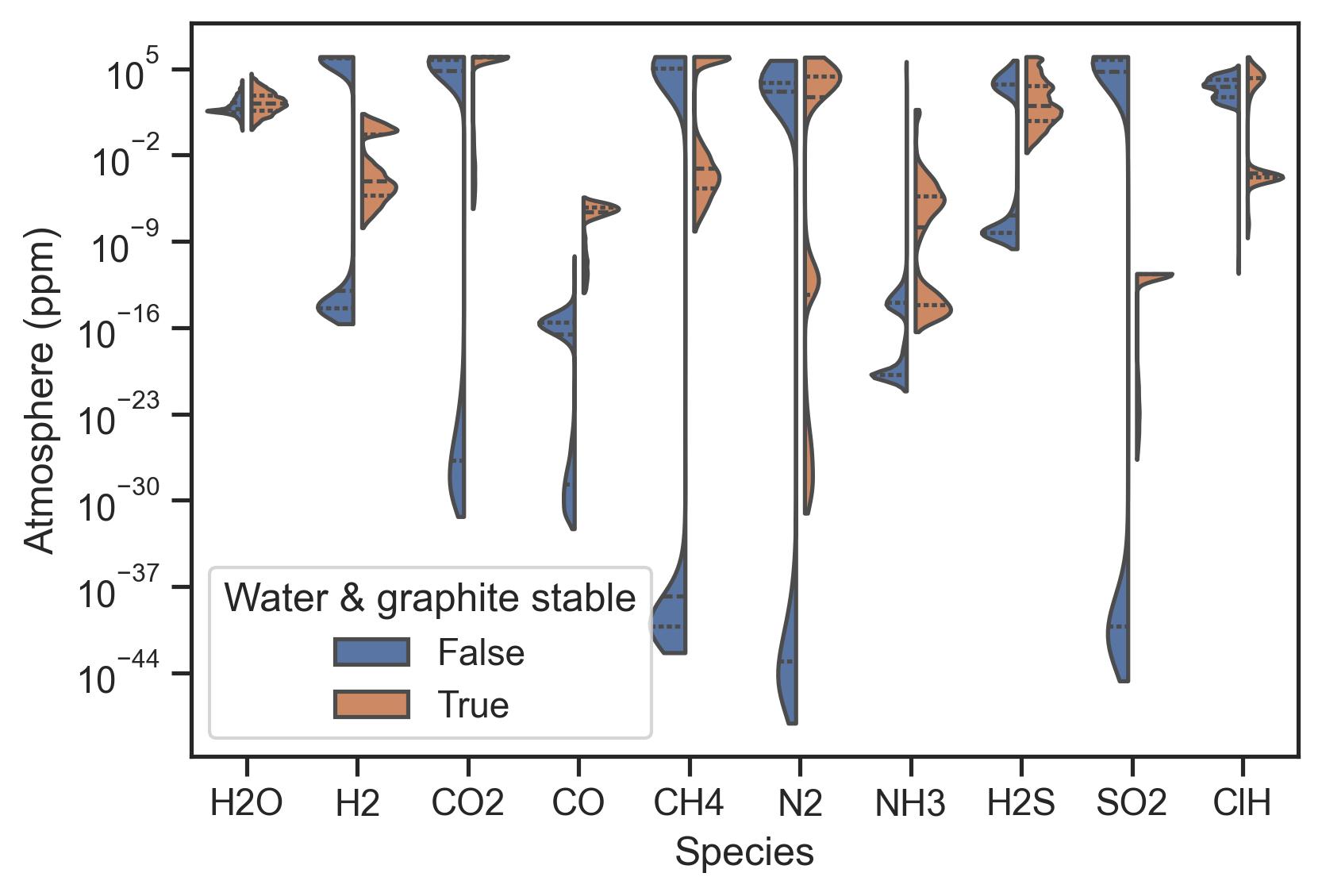}{0.475\textwidth}{(d) As for (b) and showing the full data range.}
}
\caption{Atmospheric speciation of TRAPPIST-1e at 280 K in equilibrium with stable reservoirs of water and graphite. "True" indicates atmospheres above planetary surfaces where both condensates coexist. Conversely, "False" means that at least one of the two condensates is not stable. From this classification, atmospheres that are more amiable to support surfaces primed for habitable environments can be distinguished. The width of each violin plot is proportional to the number of models, where interior dotted lines represent the first and third quartiles and dashed lines represent the second quartile (median). A scenario additionally including $\alpha$-sulfur and ammonium chloride is presented in Figure~\ref{fig:T1e_observe}.}
\label{fig:T1e_observe_WG}
\end{figure}

The overall trends of atmospheres above planetary surfaces with stable condensates are comparable to the fully molten starting state in terms of the dominance of CO$_2$ and CH$_4$ atmospheres (compare Figures~\ref{fig:T1e_10_condensates} and ~\ref{fig:T1e_100_condensates}). However, there are a few notable differences. First, most water oceans are stable beneath CO$_2$ atmospheres and most graphite reservoirs are stable beneath CH$_4$ atmospheres. This trend is reversed compared to the atmospheres derived from a fully molten starting state. Second, H$_2$S atmospheres are absent for the fully molten starting state, but now feature above all three condensate reservoirs. Third, SO$_2$ atmospheres can support up to 3 EO of water as well as 0.15 EO of $\alpha$-sulfur. Finally, N$_2$ atmospheres, although low in number, are also compatible with large reservoirs of water and $\alpha$-sulfur. Figure~\ref{fig:T1e_observe_WG} shows the atmospheric speciation in equilibrium with stable reservoirs of water and graphite only, which excludes $\alpha$-sulfur and ammonium chloride compared to Figure~\ref{fig:T1e_observe} \deleted{in the main text}.

Tables~\ref{table:atmos_ppm_fully_molten} and \ref{table:atmos_ppm_partially_molten} summarize the atmosphere speciation for TRAPPIST-1e at 280 K for atmospheres with stable water, graphite, and $\alpha$-sulfur, for fully molten and partially molten starting states, respectively (Figure~\ref{fig:T1e_observe}).

\begin{deluxetable*}{lccccc|ccccc}
\tablecaption{Atmosphere speciation (ppm) of TRAPPIST-1e at 280 K where all considered condensates are stable\label{table:atmos_ppm_fully_molten}, following cooling and collapse after equilibrating with an early mantle that was fully molten.}
\tablewidth{0pt}
\tablehead{
\colhead{Species} & \multicolumn{5}{c}{All Condensates Stable (ppm)} & \multicolumn{5}{c}{At Least One Condensate Unstable (ppm)} \\
\cline{2-6} \cline{7-11} 
& \colhead{Min} & \colhead{Max} & \colhead{Q1} & \colhead{Q2} & \colhead{Q3} 
& \colhead{Min} & \colhead{Max} & \colhead{Q1} & \colhead{Q2} & \colhead{Q3} 
}
\startdata
H$_2$O  & 1.3e+00 & 1.8e+04 & 2.8e+01 & 8.6e+01 & 2.5e+02 & 1.1e+00 & 8.8e+04 & 6.5e+01 & 2.5e+02 & 9.4e+02 \\
H$_2$   & 1.4e-08 & 5.8e-02 & 1.4e-06 & 7.8e-06 & 4.0e-05 & 3.4e-16 & 9.5e+05 & 1.6e-14 & 5.5e-05 & 8.5e-04 \\
CO$_2$  & 1.6e+05 & 1.0e+06 & 9.7e+05 & 9.9e+05 & 1.0e+06 & 1.1e-30 & 1.0e+06 & 3.8e+05 & 8.1e+05 & 9.6e+05 \\
CO      & 3.6e-08 & 2.7e-06 & 1.7e-07 & 3.0e-07 & 5.1e-07 & 6.2e-32 & 4.2e-06 & 4.4e-16 & 6.8e-10 & 9.3e-07 \\
CH$_4$  & 7.5e-09 & 9.4e+00 & 3.7e-06 & 3.5e-05 & 3.1e-04 & 1.7e-43 & 1.0e+06 & 9.5e-40 & 4.3e-04 & 1.8e-02 \\
N$_2$   & 1.8e-12 & 7.6e+05 & 3.1e+03 & 7.7e+03 & 2.6e+04 & 2.5e-46 & 8.4e+05 & 2.3e+03 & 1.4e+04 & 6.3e+04 \\
NH$_3$  & 2.8e-13 & 1.7e-02 & 7.9e-08 & 5.6e-07 & 3.4e-06 & 1.4e-21 & 2.6e+04 & 7.2e-20 & 1.6e-12 & 1.5e-05 \\
H$_2$S  & 1.7e-02 & 7.1e+04 & 1.8e+00 & 9.6e+00 & 4.9e+01 & 4.2e-10 & 2.5e+05 & 1.9e-08 & 1.1e+01 & 1.2e+02 \\
SO$_2$  & 3.8e-13 & 2.4e-12 & 2.3e-12 & 2.4e-12 & 2.4e-12 & 3.2e-45 & 9.7e+05 & 8.3e-14 & 1.4e-12 & 2.0e+05 \\
ClH     & 3.8e-05 & 1.7e+05 & 8.3e-05 & 1.1e-04 & 1.5e-04 & 6.2e-14 & 9.3e+05 & 2.6e-04 & 2.3e+00 & 3.8e+01 \\
\enddata
\tablecomments{Q1, Q2, and Q3 represent the first (25th percentile), second (median, 50th percentile), and third (75th percentile) quartiles, respectively. These data are represented in Figure~\ref{fig:T1e_observe}(a,c), where interior dotted lines show Q1 and Q3 and dashed lines show Q2 (median).}
\end{deluxetable*}

\begin{deluxetable*}{lccccc|ccccc}
\tablecaption{Atmosphere speciation (ppm) of TRAPPIST-1e at 280 K where all considered condensates are stable\label{table:atmos_ppm_partially_molten}, following cooling and collapse after equilibrating with an early mantle that was partially molten (10\% melt fraction).}
\tablewidth{0pt}
\tablehead{
\colhead{Species} & \multicolumn{5}{c}{All Condensates Stable (ppm)} & \multicolumn{5}{c}{At Least One Condensate Unstable (ppm)} \\
\cline{2-6} \cline{7-11} 
& \colhead{Min} & \colhead{Max} & \colhead{Q1} & \colhead{Q2} & \colhead{Q3} 
& \colhead{Min} & \colhead{Max} & \colhead{Q1} & \colhead{Q2} & \colhead{Q3} 
}
\startdata
H$_2$O  & 1.2e+00 & 3.7e+04 & 4.0e+01 & 1.7e+02 & 6.9e+02 & 1.0e+00 & 4.3e+04 & 3.5e+01 & 7.2e+01 & 3.0e+02 \\
H$_2$   & 1.3e-08 & 7.9e-01 & 2.5e-06 & 2.2e-05 & 2.2e-04 & 2.1e-16 & 9.8e+05 & 5.0e-15 & 1.1e+00 & 6.5e+05 \\
CO$_2$  & 2.3e-04 & 1.0e+06 & 9.3e+05 & 9.8e+05 & 1.0e+06 & 4.7e-32 & 1.0e+06 & 6.1e-26 & 2.7e-03 & 3.9e+05 \\
CO      & 4.9e-12 & 3.9e-06 & 1.9e-07 & 3.9e-07 & 7.8e-07 & 4.8e-33 & 2.6e-06 & 3.0e-28 & 1.6e-16 & 3.0e-15 \\
CH$_4$  & 6.8e-09 & 2.0e+05 & 7.7e-06 & 1.4e-04 & 3.1e-03 & 4.2e-43 & 9.9e+05 & 1.2e-40 & 5.6e+04 & 4.8e+05 \\
N$_2$   & 5.2e-28 & 8.4e+05 & 7.4e-12 & 4.8e+03 & 3.8e+04 & 8.0e-49 & 8.8e+05 & 4.1e-42 & 7.4e-26 & 6.9e+03 \\
NH$_3$  & 4.7e-17 & 1.4e+00 & 6.5e-14 & 3.6e-07 & 9.9e-06 & 7.3e-22 & 3.9e+05 & 1.9e-20 & 1.7e-15 & 2.1e-14 \\
H$_2$S  & 1.6e-02 & 9.8e+05 & 3.1e+00 & 2.8e+01 & 2.7e+02 & 2.5e-10 & 8.9e+05 & 6.1e-09 & 1.9e+03 & 1.8e+04 \\
SO$_2$  & 5.6e-22 & 2.4e-12 & 2.2e-12 & 2.4e-12 & 2.4e-12 & 2.2e-45 & 9.8e+05 & 7.1e-40 & 1.1e-22 & 3.6e+05 \\
ClH     & 5.8e-09 & 9.4e+05 & 1.5e-04 & 2.2e-04 & 1.7e+03 & 2.5e-12 & 9.0e+05 & 7.3e+02 & 6.0e+03 & 2.4e+04 \\
\enddata
\tablecomments{Q1, Q2, and Q3 represent the first (25th percentile), second (median, 50th percentile), and third (75th percentile) quartiles, respectively. These data are represented in Figure~\ref{fig:T1e_observe}(c,d), where interior dotted lines show Q1 and Q3 and dashed lines show Q2 (median).}
\end{deluxetable*}

\section{Comparison with FactSage 8.2}
\label{app:factsage}
\setcounter{figure}{0}
\setcounter{table}{0}
\setcounter{equation}{0}

We independently verified the results of our model by comparison to FactSage calculations \citep[version 8.2,][]{bale2016}, assuming ideal gas behavior (Table~\ref{table:factsage}). This comparison reveals that the partial pressure of a given volatile differs by at most a few percent. Additional comparison cases are available in the open-source code including cases with condensates.

\begin{deluxetable*}{llll|lllll|lllll}
\centerwidetable
\tablecaption{Comparison of select outgassed atmospheres calculated using \atmodeller{} compared to FactSage 8.2 calculations\label{table:factsage}}
\tablewidth{0pt}
\tablehead{
\multicolumn{4}{c}{Case} & \multicolumn{5}{c}{Atmodeller} & \multicolumn{5}{c}{FactSage 8.2}\\
\hline
\colhead{Hydrogen} & \colhead{C/H} & \colhead{$f \rm O_2$} & \colhead{Temp} & \colhead{CO} & \colhead{CO$_2$} & \colhead{H$_2$} & \colhead{H$_2$O} & \colhead{CH$_4$} & \colhead{CO} & \colhead{CO$_2$} & \colhead{H$_2$} & \colhead{H$_2$O} & \colhead{CH$_4$} \\
\colhead{Oceans} & \colhead{kg/kg} & \colhead{$\Delta$IW$^\dagger$} & \colhead{K} & \multicolumn5c{bar} & \multicolumn5c{bar}}
\decimals
\tablecolumns{14}
\startdata
3 & 1 & -2.0 & 1400 & 6.2 & 0.2 & 175 & 13.8 & 38.1 & 6.2 & 0.2 & 176 & 13.8 & 38\\
3 & 1 & +0.5 & 1400 & 46.6 & 30.7 & 237 & 337 & 28.7 & 46.4 & 30.9 & 237 & 337 & 28.7\\
1 & 0.1 & +2.0 & 1400 & 0.9 & 3.2 & 27.8 & 218 & 0 & 0.9 & 3.3 & 27.4 & 218 & 0\\
1 & 5 & +4.0 & 1400 & 9.6 & 358 & 5.4 & 432 & 0 & 10.21 & 357 & 5.8 & 432 & 0\\
1 & 1 & 0 & 873 & 0 & 0 & 59 & 18.3 & 19.5 & 0 & 0 & 59 & 18.3 & 19.5
\enddata
\tablecomments{$^\dagger$ Log$_{10}$ shift relative to the IW buffer as defined by \citet{OP93,HGD08}, whereas previous calculations \cite[][Table E1]{BHS21} were performed relative to the $f$O$_2$ buffer defined by \cite{OE02}.}
\end{deluxetable*}

\section{Bounded real gas equation of state}
\label{app:boundedeos}
\setcounter{figure}{0}
\setcounter{table}{0}
\setcounter{equation}{0}

Real gas EOS are usually calibrated with experimental data, and are therefore valid within a certain range of temperature and pressure. Even if the parameters of a real gas EOS are within the calibrated range, numerical solvers perform better when the function is both bounded and smooth because these properties help avoid instability and ensure reliable convergence. Boundedness ensures that the solution does not experience unphysical behaviors, such as division by zero or infinite values, which can lead to numerical errors or divergence. Smoothness, on the other hand, helps prevent large, abrupt changes in the EOS that can cause solvers to struggle with step-size control or lead to inaccurate approximations, ultimately improving the robustness and accuracy of the numerical methods.

For a given EOS, a pressure range is defined based on calibration data to bracket where the EOS is physically reasonable and well behaved. Below the minimum calibration pressure, often around 1 bar, the gas species is set to obey the ideal gas law. Above the maximum calibration pressure $P_{\text{maxc}}$, we let the compressibility factor $Z$ (e.g., Equation~\ref{eq:fugacity}) follow an empirical virial-like equation \citep[e.g.,][]{SF87,SF87a}:
\begin{equation}
Z(P,T) = \frac{PV_m}{RT} = A + B(P-P_{\text{maxc}}), \quad \text{for } P > P_{\text{maxc}}
\end{equation}
where $R$ is the gas constant, $T$ temperature, $V_\text{m}$ molar volume, $A$ the compressibility factor  of the EOS evaluated at ($P_{\text{maxc}}$, T), and $B$ the gradient of $Z$ with respect to $P$ above $P_{\text{maxc}}$, which could be ignored (set to zero) for simplicity. The volume integral between $P_{\text{maxc}}$ and $P$ can be expressed analytically
\begin{equation}
\int_{P_{\text{maxc}}}^P V_m dP = RT \left[ \ln\left(\frac{P}{P_{\text{maxc}}}\right) \left( A - B P_{\text{maxc}} \right) + B \left(P - P_{\text{maxc}}\right) \right].
\label{eq:realgasbound}
\end{equation}
Equation~\ref{eq:realgasbound} can be included as part of a volume integral calculation when $P>P_{\text{maxc}}$, from which the fugacity can be determined using
\begin{equation}
RT\ln f = \int V dP.
\end{equation}

\section{Treatment of oxygen}
\label{app:oxygen}
\setcounter{figure}{0}
\setcounter{table}{0}
\setcounter{equation}{0}

The condensed reservoir (molten/solid rock) of a rocky planet dominates its gaseous reservoir in terms of mass or number of moles. During the magma ocean stage, in which the molten mantle and atmosphere communicate due to similar dynamic timescales, it is therefore expected that mantle chemistry plays a role in regulating the chemistry of the combined atmosphere-mantle system. In particular, oxygen fugacity ($f$O$_2$) is a crucial parameter in mantle geochemical systems because it governs the oxidation state of elements, influencing mineral stability, phase equilibria, and the speciation of volatiles such as hydrogen, carbon, and sulfur. It is reasonable to assume that oxygen fugacity ($f$O$_2$) is buffered by the IW equilibrium in the early history of a rocky planet because planetary accretion and differentiation involve metal-silicate equilibration under reducing conditions. \added{During core formation, metallic iron segregates from the silicate mantle, and the prevailing redox state is largely controlled by Fe--FeO equilibrium, which defines the IW buffer \citep[e.g.,][]{H21}:}

\begin{equation}
    \underset{\text{metal}}{\text{Fe}} + 1/2\ {\text O}_2 \Longleftrightarrow \underset{\text{w\"ustite}}{\text{FeO}}
    \label{eq:IW}
\end{equation}

Rock analyses confirm the redox state of Earth's magma ocean was close to the IW buffer \citep{SBB20}. Hence, for the high-temperature calculations of both TRAPPIST-1e and K2-18b the control of mantle chemistry on atmospheric speciation is encapsulated in the constraint on $f$O$_2$, which is defined relative to the IW buffer that exclusively determines (absolute) $f$O$_2$ (in bar) at a given temperature and a reference pressure of 1 bar. Representing $f$O$_2$ relative to the IW buffer does not necessitate that the atmosphere of K2-18b is actually buffered by the Fe--FeO reaction at present day. Furthermore, it should be appreciated that imposing a buffered oxygen constraint does not in itself constrain the total oxygen abundance. This is evident, for example, by considering the following redox reaction:
\begin{equation}
\mathrm{H_2O} = \mathrm{H_2} + \frac{1}{2} \mathrm{O_2},
\end{equation}
where the equilibrium constant of the reaction $K(T)$ is defined as
\begin{equation}
K(T) = \frac{f{\mathrm{H}_2} \cdot f{\mathrm{O}_2}^{1/2}}{f{\mathrm{H_2O}}},
\label{eq:redoxfO2}
\end{equation}
where $T$ is temperature and $f$ fugacity, in which the later can be considered equivalent to partial pressure for this discussion. If $f$O$_2$ is constrained relative to the IW buffer, then both $f$O$_2$ and $K$ are solely functions of temperature and these collectively govern the ratio of $f$H$_2$ to $f$H$_2$O but importantly not their absolute values. Imposing an additional hydrogen abundance constraint (or alternatively pressure constraint) is required to break the degeneracy to allow a unique determination of $f$H$_2$ and $f$H$_2$O. That is, two equations (Equation~\ref{eq:redoxfO2} and hydrogen mass balance) can be solved to determine two unknowns ($f$H$_2$ and $f$H$_2$O), implicitly constraining the total oxygen abundance. Hence, if $f$O$_2$ is prescribed as a constraint, \atmodeller{} solves for the total oxygen abundance self-consistently as part of the solution process. For the cooled atmosphere calculations of TRAPPIST-1e, the elemental abundance of species in the atmosphere (including oxygen) are prescribed as constraints, in which case \atmodeller{} then ensures a self-consistent determination of $f$O$_2$.

\section{Interplay of Solubility and Nonideality}
\label{app:interplay}
\cite{CS18} and \cite{KFS19} present models for H$_2$ dissolution in sub-Neptunes, for ideal and nonideal scenarios, respectively. Here, we develop a toy model to demonstrate how these previous results are also obtained with the formulation of \atmodeller{}. We start with a simplified mass balance equation used by \atmodeller{} \citep[e.g., Appendix C,][]{BHS21}, where the first term is the dissolved mass of volatile, the second term the mass of volatile in the atmosphere, and the third term the total mass of volatile
\begin{equation}
X_v M_m + 4 \pi R_s^2 \left( \frac{\mu_v}{\bar{\mu}} \right) \frac{p_v}{g}  = m_v,
\label{eq:volmasssingle}
\end{equation}
where:
\begin{itemize}
\item $X_v$ is the volatile abundance in the melt;
\item $M_m$ is the mass of melt;
\item $R_s$ is the surface radius;
\item $g$ is gravity, positive by definition;
\item $\mu_v$ is the molar mass of the volatile;
\item $\bar{\mu}$ is the mean molar mass of the atmosphere;
\item $p_v$ is the surface partial pressure of the volatile; and
\item $m_v$ is the total volatile mass.
\end{itemize}

We employ a general expression for solubility with a power-law dependence on fugacity
\begin{equation} 
X_v = \alpha_v \exp{(-T_0/T)} f^{1/\beta_v},
\label{eq:simplesol}
\end{equation}
where $\alpha_v$, $\beta_v$, and $T_0$ are constants specific to a particular equilibrium between a gas species and its dissolved component and $T$ is temperature at the atmosphere-interior interface. We define a functional form for the fugacity coefficient (Equation~\ref{eq:fugacity}) that captures the general tendency of fugacity coefficients to increase exponentially with pressure

\begin{equation}
    \phi = \exp{\left(\frac{P-P_{\rm ideal}}{P_0}\right)},
    \label{eq:simplefugacity}
\end{equation}
where $P$ is total pressure, $P_0$ is a constant that regulates nonideality and $P_{\rm ideal}=1$ bar is the pressure at which the volatile behaves ideally. Hence ideal behavior is recovered as $P_0 \rightarrow \infty$ or $P \rightarrow P_{\rm ideal}$. For simplicity, temperature dependence can also be accounted for through the choice of P$_0$. Substituting Equations~\ref{eq:simplesol} and \ref{eq:simplefugacity} into \ref{eq:volmasssingle} gives

\begin{equation}
    A \exp{\left(-\frac{T_0}{T}\right)}  \exp{\left(\frac{P}{\beta_v P_0}\right)} \zeta_m P^{1/\beta_v} + \Gamma \left( \frac{\mu_v}{\bar{\mu}} \right) P  = \zeta_v,
    \label{eq:fullsimple}
\end{equation}

where
\begin{equation}
    A = \alpha_v \exp{\left(-\frac{P_{\rm ideal}}{\beta_v P_0} \right)},\quad \zeta_m = \frac{M_m}{M_p}, \quad \Gamma = \frac{4 \pi R_s^2}{M_p g}, \quad \zeta_v = \frac{m_v}{M_p},
\end{equation}

and $M_p$ is the mass of the rocky planet, i.e. silicate mantle plus iron core. Equation~\ref{eq:fullsimple} can be simplified for a single species atmosphere ($\mu_v = \bar{\mu}$) and for $\beta_v=1$, appropriate for H$_2$ \citep{HWA12}:

\begin{equation}
    \left[ A \exp{\left(-\frac{T_0}{T}\right)} \exp{\left(\frac{P}{\beta_v P_0}\right)} \zeta_m + \Gamma \right] P  = \zeta_v,
    \label{eq:fullsinglespecies}
\end{equation}

The ratio of the terms in the square brackets corresponds to the ratio of the dissolved volatile mass $M_{\rm diss}$ and the atmospheric volatile mass $M_{\rm atm}$

\begin{equation}
    \frac{M_{\rm diss}}{M_{\rm atm}} = \frac{A G M_p^2 \exp{\left(-\frac{T_0}{T}\right)} \exp{\left(\frac{P}{\beta_v P_0}\right)} \zeta_m }{4 \pi R_s^4}
    \label{eq:chachan}
\end{equation}

where $G$ is the gravitational constant. For the limit of ideal behavior ($P_0 \rightarrow \infty$) and a fully molten core ($\zeta_m=1$) the scaling in Equation 12 of \cite{CS18} is recovered. Hence, the dissolved volatile mass fraction relative to that of the atmosphere increases with planet size and higher surface temperature \citep[Figure 1,][]{CS18,gillmann2024}. \cite{KFS19} additionally include the nonideal dissolution of H$_2$ into magma driven by the nonlinear increase of fugacity with pressure (quantified by $P_0$), which again gives rise to enhanced volatile storage in the interior.

\section{Additional K2-18b Figures}
\label{app:gasdwarf}
\setcounter{figure}{0}
\setcounter{table}{0}
\setcounter{equation}{0}
Figures~\ref{fig:k218b_fO2} and \ref{fig:k218b_CtoH} show how atmospheric pressures (a), atmospheric C/H and C/O ratios by mass (b), solubilities (c), and fugacity coefficients (d) vary with oxygen fugacity and total C/H mass ratio, respectively, between the real and ideal cases. In these simulations, the planet mass, surface radius, and temperature are the same as those in Section~\ref{sect:gasdwarf} and Figures~\ref{fig:k218b_massfrac} and \ref{fig:k218b_idealvsreal}, and only one input parameter is varied (either oxygen fugacity or C/H mass ratio), while the others are kept constant. 

\begin{figure}
\includegraphics[width=1\textwidth]{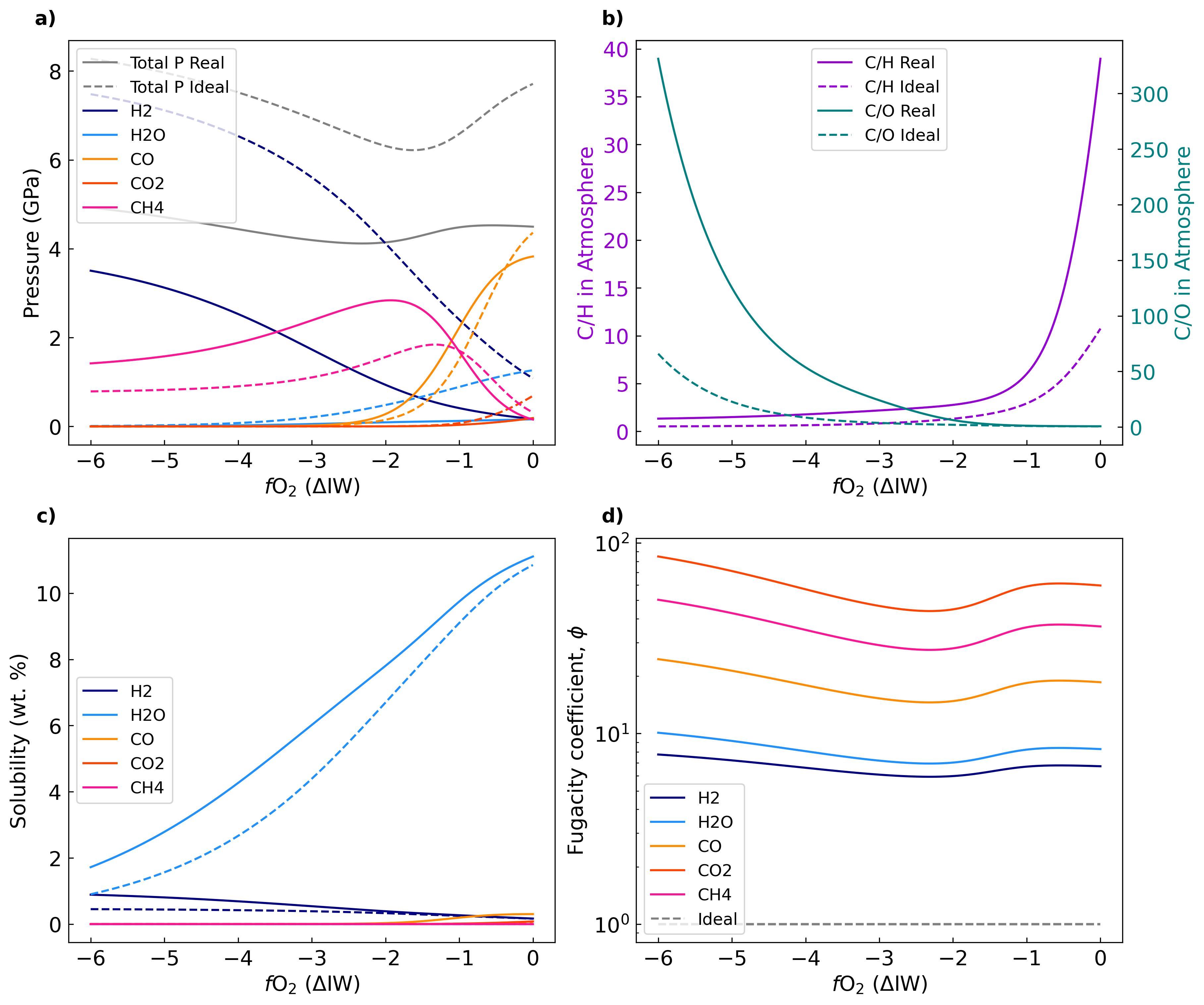}
\caption{Simulations at the magma ocean--atmosphere interface of K2-18b assuming real versus ideal gas behavior under various oxygen fugacity conditions. The oxygen fugacity ($f$O$_2$) of the planet varies from $\Delta$IW = -6 to 0, while other input parameters are fixed, with $T_\mathrm{surface} = 3000$ K, a total hydrogen mass fraction of 1\% of the planet's mass and C/H by mass of 100$\times$ solar (i.e., $\sim$0.3). The $x$-axis corresponds to $f$O$_2$, expressed as log$_{10}$ units relative to the IW buffer. The solid lines correspond to the cases assuming real gases, and the dashed lines assume ideal gas behavior. (a) Pressures of volatile species in the atmosphere (GPa) and the total atmospheric pressure (gray). (b) C/H (purple, left $y$-axis) and C/O (teal, right $y$-axis) in the atmosphere. (c) Solubility of each volatile species (wt\%, i.e., fraction of the planet's mantle mass). (d) Fugacity coefficient ($\phi$) for each volatile species. The coefficient for ideal behavior (gray dashed line) is unity for all volatiles.}
\label{fig:k218b_fO2}
\end{figure}

\begin{figure}
\includegraphics[width=1\textwidth]{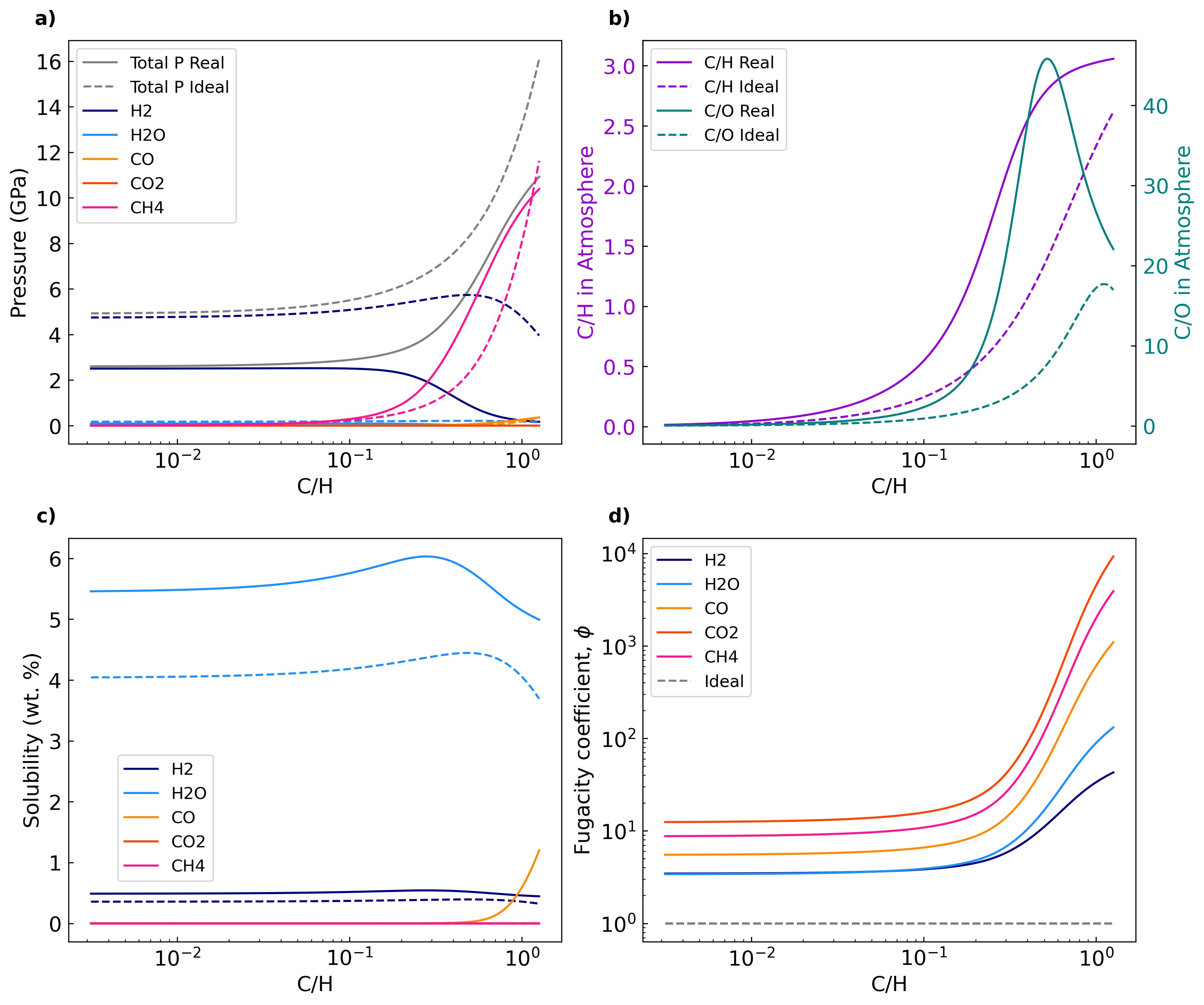}
\caption{Simulations at the magma ocean--atmosphere interface of K2-18b assuming real versus ideal gas behavior for different C/H by mass. C/H by mass varies from solar ($\sim$3E-3) to that of the BSE ($\sim$1.3), while other input parameters are fixed, with $T_\mathrm{surface} = 3000$ K, a total hydrogen mass fraction of 1\% of the planet mass, and oxygen fugacity ($f$O$_2$) at $\Delta$IW = -3. The $x$-axis corresponds to C/H by mass on a log$_{10}$-scale. The solid lines correspond to the cases assuming real gases, and the dashed lines assume ideal gas behavior. (a) Pressures of volatile species in the atmosphere (GPa) and the total atmospheric pressure (gray). (b) C/H (purple, left $y$-axis) and C/O (teal, right $y$-axis) in the atmosphere. (c) Solubility of each volatile species (wt\%, i.e., fraction of the planet's mantle mass). (d) Fugacity coefficient ($\phi$) for each volatile species. The coefficient for ideal behavior (gray dashed line) is unity for all volatiles.}
\label{fig:k218b_CtoH}
\end{figure}


%\bibliography{refs}{}
%\bibliographystyle{aasjournal}


\listofchanges

\end{document}